\documentclass[12pt]{iopart}
\expandafter\let\csname equation*\endcsname\relax
\expandafter\let\csname endequation*\endcsname\relax
\usepackage{epsfig}
\usepackage{graphicx}
\usepackage{amssymb}
\usepackage{amsmath}
\usepackage{mathtools}
\usepackage{multicol}
\usepackage[english]{babel}
\usepackage{enumitem}
\usepackage{gensymb}
\usepackage{geometry}
\usepackage{cancel}
\usepackage[titletoc,title]{appendix}
\usepackage{cite}
\usepackage{epstopdf}
\usepackage{soul}
\usepackage{epstopdf, epsfig}
\usepackage{xcolor}

\newcommand{\uvec}[1]{\hat{\vec{#1}}} 
\renewcommand\vec\mathbf

\begin{document}

\title{Sheath constraints on turbulent magnetised plasmas}
\author{A Geraldini$^{1}$, S. Brunner$^{1}$, F. I. Parra$^{2}$}
\address{$^{1}$ Swiss Plasma Center, \'Ecole Polytechnique F\'ed\'erale de Lausanne (EPFL), CH-1015 Lausanne, Switzerland}
\address{$^2$  Princeton Plasma Physics Laboratory, Princeton, NJ 08540, USA }
\ead{alessandro.geraldini@epfl.ch}    

\begin{abstract}
A solid target in contact with a plasma charges (negatively) to reflect the more mobile species (electrons) and thus keep the bulk plasma quasineutral. 
To shield the bulk plasma from the charged target, there is an oppositely (positively) charged sheath with a sharp electrostatic potential variation on the Debye length scale $\lambda_{\rm D}$.
In magnetised plasmas where the magnetic field is inclined at an oblique angle $\alpha$ with the target, some of the sheath potential variation occurs also on the ion sound gyroradius length scale $\rho_{\rm S} \cos \alpha$, caused by finite ion gyro-orbit distortion and losses.
We consider a collisionless and steady-state magnetised plasma sheath whose thickness $l_{\rm ms} \sim \max (\lambda_{\rm D}, \rho_{\rm S} \cos \alpha)$ is smaller than the characteristic length scale $L$ of spatial fluctuations in the bulk plasma, such that the limit $l_{\rm ms}  /  L \rightarrow 0$ is appropriate.
Spatial structures are assumed to be magnetic field-aligned. 
In the case of small magnetic field angle $ \alpha \sim  \delta \equiv \rho_{\rm S} / L \ll 1$, electric fields tangential to the target transport ions towards the target via ExB drifts at a rate comparable to the one from parallel streaming. 
A generalised form of the kinetic Bohm-Chodura criterion at the sheath entrance is derived by requiring that the sheath electric field have a monotonic spatial decay far from the target. 
The criterion depends on tangential gradients of potential and ion distribution function, with additional nontrivial conditions.
\end{abstract}

\section{Introduction} 
 
Sheaths arising at the interface between a plasma and a solid target --- also referred to as a wall --- have been studied for as long as plasmas themselves have been \cite{Tonks-1929, Bohm-1949}.
Despite this, there remain fundamental unanswered questions about the mathematical formulation of the magnetised plasma-sheath transition beyond fluid models.
The boundary conditions that sheaths impose on turbulent magnetised plasmas, such as the ones present in fusion devices, are also still a subject of active research \cite{Loizu-2012, Krashenninikov-2017, Shi-2017}.

Considering an electrostatic plasma, the electric field $\vec{E}$ satisfies $\vec{E} = - \nabla \phi$ and the magnetic field $\vec{B}$ is constant in time.
Thus, one of Maxwell's equations reduces to Poisson's equation for the electrostatic potential,
\begin{align} \label{Poisson-3D}
\varepsilon_0 \nabla_{\vec{x}}^2 \phi (\vec{x})  = en_{\rm e}(\vec{x}) - Zen_{\rm i} (\vec{x}) \rm .
\end{align}
Here $n_{\rm i}$ and $n_{\rm e}$ are the ion and electron densities, respectively, $e$ is the proton charge, $Z$ is the charge state of the ions, $\vec{x}$ is the position, $\nabla_{\vec{x}}^2 $ is the Laplacian and $\varepsilon_0$ is the permittivity of free space.
The length scale associated with the electrostatic potential variation in (\ref{Poisson-3D}) is the Debye length, $\lambda_{\rm D} = \sqrt{\varepsilon_0 T_{\rm e} / e^2 n_{\rm ref}}$, where $n_{\rm ref}$ is a reference electron density and $T_{\rm e}$ is the electron temperature.
Indeed, (\ref{Poisson-3D}) is readily re-expressed to
\begin{align} \label{Poisson-3D-2}
\frac{ \lambda_{\rm D}^2 }{L^2} \nabla_{\vec{X}}^2 \varphi (\vec{X})  = N_{\rm e}(\vec{X}) - ZN_{\rm i} (\vec{X}) \rm ,
\end{align}
where $L$ is the length scale of the physical processes we observe in a plasma, $\vec{X} = \vec{x}/L$, $N_s = n_s/n_{\rm ref}$ for $s=$ i and e, and $\varphi = e\phi/ T_{\rm e}$ is the electrostatic potential normalised to the electron temperature.
The very high degree of accuracy to which plasmas satisfy the quasineutrality equation,
\begin{align} \label{quasi-3D}
0  = N_{\rm e}(\vec{X}) - ZN_{\rm i} (\vec{X}) \rm ,
\end{align}
is related to the smallness of the ratio of the Debye length relative to the bulk length scale, $\lambda_{\rm D} / L \ll 1$.
The plasma effectively adjusts itself (almost instantanously relative to the plasma processes occurring on the length scale $L$) to satisfy quasineutrality.
We thus normally use equation (\ref{quasi-3D}) to solve for the electrostatic potential in the bulk.

The formation of a sheath is mathematically related to the fact that (\ref{Poisson-3D-2}) is a higher order differential equation than (\ref{quasi-3D}).
The particle densities appearing in (\ref{quasi-3D}) and on the right hand side of (\ref{Poisson-3D-2}) depend on the electric field $\vec{E} = - \nabla \phi$.
Hence, the quasineutrality equation fully specifies the variation of the electrostatic potential in the quasineutral plasma.
However, (\ref{Poisson-3D-2}) requires boundary conditions for $\varphi$. 
It is unsurprising that the boundary conditions become necessary close to a solid target: in practice, the electrostatic potential at the target can be biased with respect to that of the quasineutral plasma at any arbitrary value. 
In fact, even without a wall bias, the plasma naturally sets itself at a potential higher than the wall potential.
This is so that a large enough number of electrons, which are much faster than ions due to their significantly smaller mass, are reflected before reaching the target, which ensures that the loss rate of electrons is equal to that of the slower ions.
The quasineutral plasma potential cannot smoothly join to the wall potential via equation (\ref{quasi-3D}) alone, but requires the Laplacian term in (\ref{Poisson-3D-2}). 
Since this term is negligible at the length scales of the bulk plasma, there must be a very thin region near the target where it becomes critical in allowing the potential to reach its value at the target. 
This region is known as the Debye sheath, and its size is the Debye length. 
The treatment of the Debye sheath as a boundary layer is an example of a singular perturbation theory in which the largest derivative of a differential equation is multipled by a small parameter \cite{Bender-Orszag}.
By using perturbation theory in $\lambda_{\rm D} / L \ll 1$, the Debye sheath has been thoroughly studied in many different situations, and the necessary condition for the sheath to be stationary in the asymptotic limit $\lambda_{\rm D} / L \rightarrow 0$ has been derived: the well-known Bohm condition.
Riemann's review \cite{Riemann-review} remains to this date a complete, instructive and profound analysis of the Bohm condition in unmagnetised plasmas. 

In magnetised plasmas where the magnetic field is obliquely incident with the wall, the \emph{magnetised plasma sheath} ensures that the plasma potential smoothly reaches the wall potential value.
Here there is another length scale of potential variation in addition to the Debye length, which is the projection of the ion sound gyroradius, $\rho_{\rm S} = c_{\rm S} / \Omega_{\rm i} = \sqrt{(T_{\rm i} + ZT_{\rm e})m_{\rm i}}/(ZeB)$, onto the direction normal to the target, $\rho_{\rm S} \cos \alpha$, where $\alpha$ is the magnetic field angle at the wall \cite{Chodura-1982}.
To define the ion sound gyroradius, we need the ion sound speed $c_{\rm S} = \sqrt{(ZT_{\rm e} + T_{\rm i})/m_{\rm i}}$, the ion temperature $T_{\rm i}$, the ion cyclotron frequency $\Omega = ZeB/m_{\rm i}$, the ion mass $m_{\rm i}$ and the magnetic field strength $B = |\vec{B}|$.
The ion sound gyroradius is often much larger than the Debye length, although we adopt a maximal ordering $\lambda_{\rm D} \sim \rho_{\rm S}\cos \alpha$ for the sake of generality (it includes the special case $\alpha \approx 90^{\circ}$).
The characteristic thickness of the magnetised sheath is thus $l_{\rm ms} = \max (\lambda_{\rm D}, \rho_{\rm S} \cos \alpha)$.
There have been a number of other important studies on this region, including but not limited to the following ones considering:
the effect of the presence of a high-energy tail in the electron distribution function on the total sheath potential drop \cite{Tskhakaya-2002b}; the effect of magnetic field strength, inclination angle and collisionality on the magnetised sheath structure \cite{Riemann-1994, Kim-1995}; the effect of the inclination angle and other plasma parameters on electron emission \cite{Tskhakaya-2000, Komm-2020}; the effect of electron losses on the Bohm condition and on the collapse of the Debye sheath at shallow angles of incidence \cite{Ewart-2021, Castillo-2024}.

The formation of the electric field on the second length scale $\rho_{\rm S}$ in the magnetised sheath is the consequence of two different mechanisms \cite{Sato-1994}.
The ions in their Larmor orbits (gyro-orbits) are absorbed by the target at a characteristic distance from the target of  the order of $\rho_{\rm i} \cos\alpha$, where $\rho_{\rm i} = \sqrt{T_{\rm i} m_{\rm i}}/(ZeB)$ is the thermal ion Larmor radius.
This first mechanism reduces the ion density close to the target and causes a non-uniform potential.
The self-consistent spatial profile of the electrostatic potential is, however, determined by a balance between polarisation and guiding center charge densities.
The distortion of ion gyro-orbits is thus the second mechanism determining the shape of the potential profile.
Consider a distance far enough from the target that the potential variation is still weak, and the scale length of the potential variation is much longer than the ion gyroradius.
The ion density at a given position is approximately the density of ions whose guiding center lie at that position, $N^{\rm gc}_{\rm i} (\vec{X})$, plus several higher order corrections small in $\rho_{\rm i}^2 / L^2$ and $\rho_{\rm B}^2 / L^2$.
One of these corrections is the polarisation density arising from the non-uniformity of the potential on the scale of gyro-orbits, $( \rho_{\rm B}^2 / L^2) \nabla_{\perp}^2 \varphi (\vec{X})$ (see e.g. equation (2.4) in \cite{Shi-2017} and (7) in \cite{Pan-2018}).
Here, $\nabla_{\perp}^2$ is the Laplacian in the plane perpendicular to the magnetic field, while $\rho_{\rm B} = v_{\rm B} / \Omega_{\rm i} = \sqrt{ZT_{\rm e} m_{\rm i}}/(ZeB)$ is referred to as the Bohm gyroradius, with $v_{\rm B} = \sqrt{ZT_{\rm e} / m_{\rm i}}$ known as the Bohm speed (or cold-ion sound speed).
The ion density can thus be expressed as $N_{\rm i} (\vec{X}) = N^{\rm gc}_{\rm i} (\vec{X}) + ( \rho_{\rm B}^2 / L^2) \nabla_{\perp}^2 \varphi (\vec{X}) + \ldots $.
Under the assumption that the solid target is planar, the gradients in the direction normal to the target are the largest ones in the sheath.
Thus, Poisson's equation sufficiently far from the target so that the sheath potential variation is still weak takes the form
\begin{align} \label{Poisson-3D-3}
\frac{  \rho_{\rm B}^2 \cos^2 \alpha + \lambda_{\rm D}^2 }{L^2} \partial_{X}^2 \varphi = N_{\rm e}(\vec{X}) - ZN^{\rm gc}_{\rm i} (\vec{X}) + \ldots  \rm ,
\end{align}
where $X = x / L$ and $x$ is the distance from the target.
If the magnetic field angle at the wall is oblique, then $\cos \alpha \neq 0$ and some
of the sheath potential variation is driven by the polarisation term on the left hand side of (\ref{Poisson-3D-3}) at distances of $\rho_{\rm B}\cos \alpha$ from the target.

This simplified picture illustrates that the magnetised and the unmagnetised sheaths have a similar mathematical structure when viewed as boundary layers of the bulk plasma arising from a singular perturbation theory. 
In a magnetised plasma, however, the magnetised sheath can only be a boundary layer for drift-reduced models where the polarisation density is either neglected or is a small correction arising from electrostatic potential variation on a length scale much larger than $\rho_{\rm B}$. 

In this paper, we generalise the kinetic plasma-sheath constraints to magnetised plasmas including the effect of field-aligned spatial fluctuations at shallow angles (particularly relevant to magnetic fusion plasmas).
We exclude ion reflection in the magnetised sheath, and calculate a threshold for the gradient of the tangential electric field below which ion reflection can be excluded.
The derivation proceeds as follows.
We first calculate the effect on the ion trajectories of the weak sheath electrostatic potential variation far from the target using perturbation theory.
By solving the Vlasov equation using these ion characteristics, we obtain an analytical expression for the ion density perturbation.
The perturbed Poisson's equation can then be used to derive a linear integro-differential equation for the electrostatic potential.
The electrostatic potential solution is argued to be an exponential function of the distance from the target.
Inserting this solution as an Ansatz into Poisson's equation, we derive two separate constraints which make the assumption of a monotonically decaying electrostatic potential solution self-consistent: the so-called kinetic Bohm-Chodura condition and a polarisation condition.

The kinetic Bohm-Chodura condition derived herein is in fact identical to that previously derived by Claassen and Gerhauser \cite{Claassen-Gerhauser-1996b}.
Their derivation, however, did not include the ion polarisation density and therefore has a rigorous mathematical grounding only if $\lambda_{\rm D} \gg \rho_{\rm S}$, which is not satisfied in typical fusion plasmas.
A similar cautionary remark had been made by Cohen and Ryutov \cite{Cohen-Ryutov-2004-sheath-boundary-conditions}.
The inclusion of the polarisation density makes the derivation substantially more involved, and leads to a non-trivial sheath polarisation condition: essentially the requirement that the ion polarisation density not reverse sign. 
The polarisation condition can be shown to be trivially satisfied in the limits of: normal incidence $\alpha \rightarrow \pi / 2$; no gradients tangential to the target; and cold ions. 

This paper is organised as follows.
In section~\ref{sec-orderings} the orderings are presented and discussed.
We then perturbatively solve, in section \ref{sec-traj}, for the ion trajectories and ion distribution function far from the target in the magnetised sheath (near the magnetised sheath entrance).
Then, in section~\ref{sec-Poisson} we analyse Poisson's equation at the magnetised sheath entrance and derive the Bohm-Chodura and polarisation conditions for the sheath electrostatic potential profile to monotonically decay far from the target in a turbulent magnetised plasma.
A higher order analysis which determines the electrostatic potential when the Bohm-Chodura condition is marginally satisfied is carried out in section~\ref{sec-higher}. 
We summarise and briefly discuss our findings in section~\ref{sec-conc}.

We remark that some more involved calculations are relegated to the appendices.
For a first lighter read, we suggest skipping section~\ref{sec-higher}, whose results may be considered non-essential, as well as all the appendices, which leaves just a little over one half of the length of the manuscript.

\section{Orderings} \label{sec-orderings}

\begin{figure}
\centering
\includegraphics[scale=1.0]{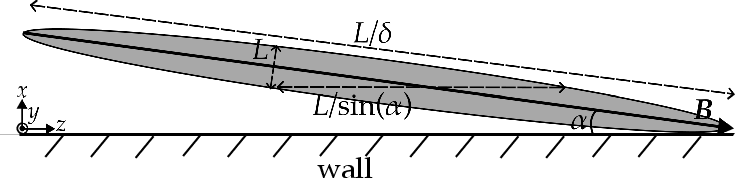}
\caption{The geometry of the system and the Cartesian axes are shown, together with a cartoon of a typical fluctuation in the $xz$ plane represented by a tilted ellipse which is strongly elongated in the direction of the magnetic field $\vec{B}$. The length scale of fluctuations in the direction perpendicular to the magnetic field is $L$, while parallel to the magnetic field it is $L/\delta$. 
The length scale in the $z$ (horizontal) direction is the distance $L/\sin \alpha$ measured obliquely through the fluctuation.
The magnetised sheath is a very thin layer next to the target, $l_{\rm ms} \ll L$ (not shown).} 
\label{fig-lzscale}
\end{figure}

The geometry of the system considered in this paper is shown in figure \ref{fig-lzscale}, where Cartesian $x$, $y$ and $z$ directions are depicted: the coordinate $x$ measures the distance from the wall,  $y$ measures displacements tangential to the target and perpendicular to the magnetic field $\vec{B}$, while $z$ measures displacements in the other direction tangential to the target.

The plasma is assumed to be electrostatic, so that the electric field $\vec{E}$ satisfies $\vec{E} = -\nabla \phi$ for an electrostatic potential $\phi$.
The constant magnetic field $\vec{B}$ is assumed to be uniform in space and entirely generated by currents far away from the region being considered \cite{Geraldini-2017}.
It is therefore assumed that any current present in the magnetised sheath does not affect the magnetic field.
In this paper, the smallest length scale of the bulk plasma, $L$, is assumed to be much larger than the magnetised sheath length scale, while being much smaller than the device size $a$,
\begin{align} \label{orderings-length}
 \rho_{\rm S} \lesssim l_{\rm ms} \sim \rho_{\rm S} \cos \alpha + \lambda_{\rm D} \ll L \ll a \rm .
\end{align}
The width of the magnetised sheath is typically a multiple, of order unity, of the length scale $l_{\rm ms}$.
The analysis in this paper is performed to lowest order in the small parameter $l_{\rm ms} / L \ll 1$.
Further taking $L/a \ll 1$ guarantees that the effects of the gradient and curvature of the magnetic field are negligible.
In fusion devices, the device size $a$ is the minor radius while the bulk length scale $L$ is the cross-field width of large-amplitude (order unity) turbulent fluctuations (of density and electrostatic potential) in the Scrape-Off Layer, as schematically represented in figure~\ref{fig-lzscale}. 
We restrict ourselves to long-wavelength turbulence because turbulent structures in the downstream region of the Scrape-Off Layer, in particular at the plasma-wall boundary, have been observed to be an order of magnitude larger than the ion gyroradius \cite{Carralero-2015}. 
We do not consider the small-wavelength, small-amplitude turbulent fluctuations that are important in the core of fusion devices but less important in the edge.
Such small-wavelength fluctuations would anyway not be scale-separable with respect to the magnetised sheath, which would disqualify a boundary layer treatment.
The curvature and gradients of the magnetic field can be ignored because both the characteristic width of the magnetised sheath and (as we will see) the characteristic distance that a typical ion moves tangentially to the wall within the magnetised sheath are much smaller than the device length scale. 

We will consider plasmas satisfying the ordering 
\begin{align} \label{ordering-T}
T_{\rm e} \gtrsim T_{\rm i} \rm .
\end{align}
This formally excludes the possibility of cold electrons, $T_{\rm i} \gg T_{\rm e}$.
This ordering is justified by the fact that it appropriately describes the order unity temperature ratio typical in the Scrape-Off Layer of fusion plasmas \cite{Mosetto-2015}, while encompassing the cold ion limit often used in fluid plasma modelling \cite{Ricci-2012}.
Moreover, the physics discussed in this paper becomes less relevant when (\ref{ordering-T}) is not satisfied.
Note that in the ordering (\ref{ordering-T}) the ion sound and Bohm gyroradii and speed become of a comparable size, $\rho_{\rm S} \sim \rho_{\rm B} \equiv \sqrt{m_{\rm i} ZT_{\rm e}}/ (ZeB)$ and $c_{\rm S} \sim v_{\rm B} \equiv \sqrt{ZT_{\rm e}/m_{\rm i}}$. 
For the remainder of this paper, we adopt $\rho_{\rm S}$ and $c_{\rm S}$ when referring to the ordering of a speed or length scale, while reserving $v_{\rm B}$ and $\rho_{\rm B}$ only for the exact quantity it denotes.
In some cases, particularly when referring to the scaling of the ion distribution function, it will be useful to use the ion thermal velocity $v_{\rm t,i} \equiv \sqrt{T_{\rm i} / m_{\rm i}}$.

\begin{figure}
\centering
\includegraphics[scale=0.7]{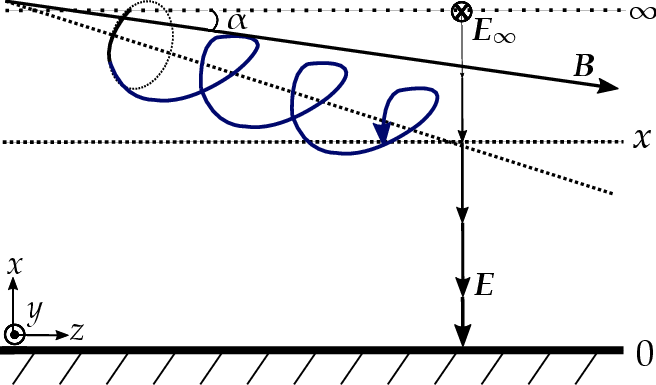}
\caption{Cartoon of an ion trajectory after it ``enters'' the magnetised sheath, crossing the dotted line labelled ``$\infty$'' (representing schematically a distance infinitely far away from the wall on the magnetised sheath scale $l_{\rm ms}$). 
Here, the electrostatic field at the sheath entrance $\vec{E}_{\infty} = - \uvec{e}_y \phi_{\infty}'(y)$ is small (the case $\phi_{\infty}'(y) > 0$ is illustrated). 
When the particle reaches the dotted line labelled ``$x$'', which is at a finite large distance from the target, the electrostatic potential is different only by a small amount relative to its value at infinity, \emph{viz} the ordering (\ref{phihat-def}). 
Thus, the electric field acting on the ion is weak, and the ion trajectory is approximately helical. Note the small offset between the axis of the helix and the magnetic field, caused by the drift $- \frac{\phi_{\infty}'(y) }{ B} \left[ \uvec{e}_x \cos \alpha + \uvec{e}_z \sin \alpha \right] $. When the magnetic field is at a small angle $\alpha$ with the target, as shown, the drift of the orbit alters the angle between the axis of the helix and the wall significantly. 
The ion trajectory can be further calculated perturbatively: at first order, this involves effectively integrating the electric force over the lowest order helix.}
\label{fig-geometry}
\end{figure}

Both in the plasma bulk and in the magnetised sheath, the electrostatic potential variations are ordered to be large in amplitude,
\begin{align} \label{phi-largeamp}
\frac{ e \phi }{T_{\rm e}}  \sim 1  \rm .
\end{align}
In the plasma bulk these variations occur most strongly in the direction perpendicular to the magnetic field, but on length scales $L$ much longer than the ion Larmor radius, $L \gg \rho_{\rm s}$. 
We define the small parameter $\delta  =  \rho_{\rm S} / L \ll 1$ satisfying the ordering
\begin{align} \label{alpha-delta}
\delta \equiv \frac{\rho_{\rm S}}{L} \lesssim \alpha \lesssim 1 \rm .
\end{align}
According to (\ref{alpha-delta}), we assume that the magnetic field angle $\alpha$ with the target surface cannot be much smaller than $\delta$, although it can be small and comparable to $\delta$.
Moreover, we assume that the length scales of electrostatic potential variation are longer than $L$ along the magnetic field (field-aligned turbulence), $\sim L / \delta$ (see figure~\ref{fig-geometry}).
The longer length scale in the parallel direction has been argued to be consistent with the turbulent fields expected in magnetised plasmas and with the steady-state fields expected due to the size and geometry of the Scrape-Off Layer (see Appendix B in \cite{Geraldini-2017}).
Considering the length scales of the large-amplitude spatial fluctuations, summarised in figure~\ref{fig-lzscale}, the components of the electrostatic field in the bulk are ordered to be
\begin{align} \label{bulkE-ordering}
\frac{ \uvec{b} \cdot (\nabla \phi)_{\rm bulk} }{B} \sim \delta^2 c_{\rm S} \ll \frac{ | \uvec{b} \times (\nabla \phi)_{\rm bulk} |}{B}  \sim \delta c_{\rm S} \ll c_{\rm S} \text{ for }  x~ \sim L / (\delta \sin \alpha + \cos \alpha)   \rm ,
\end{align}
where $\uvec{b} = \vec{B}/B$ is the unit vector in the magnetic field direction.
Note that ordering (\ref{bulkE-ordering}) for the bulk is valid for $x \sim L / (\delta \sin \alpha + \cos \alpha)  $. For $\alpha \ll 1$, this gives $x \sim L$, and for $\alpha = \pi / 2$, $x \sim L/\delta$, as one would expect from $x$ being mostly perpendicular or mostly parallel to $\vec{B}$. This estimate for $x$ can be obtained from $(\partial_x \phi )_{\rm bulk} / B = \uvec{e}_x \cdot \left( \uvec{b} \uvec{b} \cdot (\nabla \phi )_{\rm bulk} -  \uvec{b} \times (\uvec{b} \times (\nabla \phi)_{\rm bulk})  \right)/B \sim \delta c_{\rm S} \left( \delta \sin \alpha + \cos\alpha \right)$. 
Apart from the distance at which it is valid, ordering (\ref{bulkE-ordering}) for the electric field components does not depend on the magnetic field angle $\alpha$ because it refers to the parallel and perpendicular components of the electric field in the bulk, sufficiently far from the wall that the magnetised sheath electric field normal to the target is negligible.
In the magnetised sheath, however, where the electrostatic potential undergoes large-amplitude variation (\ref{phi-largeamp}) over the smaller length scale $l_{\rm ms}$ in the $x$ direction, the component of the electric field normal to the target satisfies the ordering 
\begin{align} \label{E-sheath}
\frac{ \partial_x \phi }{B} \sim \frac{T_{\rm e}}{eB l_{\rm ms}} \sim \frac{\rho_{\rm S}}{l_{\rm ms}} c_{\rm S} \text{ for }  x \sim l_{\rm ms} \rm . 
\end{align}
Notably, this field causes an $\vec{E} \times \vec{B}$ drift in the $y$ direction $ \sim \left( \partial_x \phi / B \right) \cos \alpha$.
When $l_{\rm ms} \sim \rho_{\rm S} \cos \alpha$, this drift of the same order as the sound velocity $c_{\rm S}$ and varies on the length scale $\rho_{\rm S} \cos \alpha$ in the $x$ direction.

We anticipate that ions enter the magnetised sheath region with a velocity component parallel to the magnetic field comparable to the sound speed $c_{\rm S}$ \cite{Chodura-1982}.
Then, the time taken by an ion to reach the target is the width of the region divided by the component $\sim c_{\rm S} \sin \alpha$ of the parallel velocity normal to the wall, giving $l_{\rm ms} / (c_{\rm S} \sin \alpha) \sim (l_{\rm ms} / \rho_{\rm S}) (\Omega \sin \alpha)^{-1} $.
The perpendicular velocity does not contribute significantly to the time taken by an ion to reach the target, as it brings the ion to the target over the fast timescale $\Omega^{-1}$ only after the ion gyro-orbit has itself moved close enough to the target.
The typical time scale for an ion crossing the magnetised sheath is therefore given by
\begin{align} \label{t-scale}
t_X \sim \frac{l_{\rm ms}}{\rho_{\rm S}} \frac{1}{\Omega \sin \alpha} \rm .
\end{align}
Note that $ 1 \lesssim l_{\rm ms}/\rho_{\rm S} \sim \cos \alpha + \lambda_{\rm D}/ \rho_{\rm S} \ll 1/\delta$, so that $l_{\rm ms} / \rho_{\rm S}$ does not impact the expansion in $\delta \ll 1$ or $\alpha \ll 1$.
During this time, the ion moves tangentially to the target: in the $z$ direction due to the projection $\sim c_{\rm S} \cos \alpha$ of its parallel streaming, and in the $y$ direction due to its $\vec{E} \times \vec{B}$ drift $\sim c_{\rm S} \cos \alpha (\rho_{\rm S} / l_{\rm ms})$.
The characteristic displacement of an ion in traversing the magnetised sheath is thus $l_{\rm ms} / \tan \alpha$ in the $z$ direction and $\rho_{\rm S} / \tan \alpha$ in the $y$ direction.
The characteristic length scale of the bulk plasma outside the magnetised sheath in the $z$ direction is $L / \sin \alpha $ (see figure \ref{fig-lzscale}), and is thus much larger than the $z$-displacement of an ion within the magnetised sheath by virtue of (\ref{orderings-length}). 
The bulk length scale in the $y$ direction, $L \sim \rho_{\rm S} / \delta $, is only larger than the displacement of the ion in this direction in the more stringent ordering $\tan \alpha \gg \delta$, which is only a special case of the more general ordering (\ref{alpha-delta}) assumed here.
Hence, for small angles $\alpha \sim \delta \ll 1$, the effect of the variation of the bulk electrostatic potential along an ion trajectory in the thin magnetised sheath must be accounted for in the $y$ direction but not in the $z$ direction.

The analysis of this paper deals with the region far enough away from the target that the electric field is still small compared to the sheath electric field at a distance $ \sim l_{\rm ms}$ from the target, but close enough to the target that the electric field is much larger than the electric fields arising in the bulk plasma.
This region corresponds to $  l_{\rm ms}  \ll x \ll L / (\delta \sin \alpha + \cos \alpha) $, and always exists provided that $l_{\rm ms} \ll L / (\delta \sin \alpha + \cos \alpha) $.
This paper assumes the ordering $l_{\rm ms} \ll L$ (see (\ref{orderings-length})), which is sufficient to satisfy $l_{\rm ms} \ll L / (\delta \sin \alpha + \cos \alpha) $ (given that $\delta \ll 1$).
Furthermore, we take the asymptotic limits
\begin{align} \label{sheath-scale}
\frac{l_{\rm ms}}{L} \rightarrow 0 \text{ and } \frac{x}{L} \rightarrow 0 \text{,}
\end{align}
which correspond to considering the magnetised sheath scale, where $x/l_{\rm ms}$ is finite.
On this scale, we consider large but finite distances from the wall such that $x/l_{\rm ms} \gg 1$.
In taking the limit $l_{\rm ms} / L \rightarrow 0$ in (\ref{sheath-scale}), we are constrained to take $\delta \rightarrow 0$ unless $\alpha$ is close to $90^{\circ}$ and $\lambda_{\rm D} / \rho_{\rm S} \ll 1$, since $l_{\rm ms} / L  \sim \delta \left( \cos \alpha + \lambda_{\rm D} / \rho_{\rm S} \right)$.
Even so, for simplicity, we use the limit $\delta \rightarrow 0$ in all cases, thus considering a drift-kinetic bulk plasma.
For the very small angles where $\alpha \sim \delta \ll 1$, we will additionally consider the limit $\alpha \rightarrow 0$.

We refer to the point infinitely far from the target on the sheath scale, $x/l_{\rm ms} \rightarrow \infty$, as the magnetised sheath entrance.
A necessary condition for the magnetised sheath to be a boundary layer of the bulk plasma is that gradients normal to the wall vanish at its entrance,
\begin{align} \label{dphidx-infty}
\partial_x^n \phi \rvert_{x /l_{\rm ms} \rightarrow \infty} = 0 \text{ for } n > 0 \rm .
\end{align}
We consider the plasma behaviour at large but finite distances from the target, $x / l_{\rm ms} \gg 1$.
From here on, $\phi(x,y)$ will be used only to denote the electrostatic potential calculated on the magnetised sheath scale defined by (\ref{sheath-scale}), with 
\begin{align} \label{phi-infty}
\phi_{\infty} (y) = \phi(x,y)\rvert_{x/l_{\rm ms} \rightarrow \infty}
\end{align}
representing the electrostatic potential at the magnetised sheath entrance.
Note that we have neglected the $z$ dependence of $\phi$ because the ion displacement in the $z$ direction has been shown to be smaller than the characteristic length scale of $\phi$ in this direction.
Since we consider a region near the magnetised sheath entrance, the electrostatic potential difference with respect to its value at the magnetised sheath entrance is small in magnitude,
\begin{align} \label{phihat-def}
\frac{e}{T_{\rm e}} \left( \phi_{\infty}(y) - \phi(x,y) \right) = - \frac{e\phi_1 (x,y)}{T_{\rm e}} = \hat{\phi}(x,y) \ll 1 \rm .
\end{align}
The parameter $\hat \phi$ quantifies ``closeness'' to the magnetised sheath entrance, which corresponds by definition to $\hat \phi = 0$ for $x/l_{\rm ms} \rightarrow \infty$.
The function $\phi_1(x,y) = \phi(x,y) - \phi_{\infty}(y)$ is the magnetised sheath electrostatic potential \emph{relative} to a point at the magnetised sheath entrance with the same value of $y$.
We assume that the potential decreases towards the wall, making $\hat{\phi}$ positive.
Note that variations in the bulk are large in amplitude, implying that the function $\phi_{\infty}$ varies significantly in the directions tangential to the target,
\begin{align} \label{phiinfty-order}
\frac{ e (\phi_{\infty}(y + L) - \phi_{\infty}(y))}{T_{\rm e}} \sim 1 \rm .
\end{align}

The ordering required for the electric field to be weak relative to its characteristic value in the magnetised sheath and large relative to the values in the bulk is
\begin{align} \label{ordering-1}
 \frac{ \partial_y \phi_{\infty}}{B} \sim  c_{\rm S} \delta \ll  \frac{ \partial_x \phi}{B} \sim c_{\rm S} \epsilon \hat{\phi} \frac{\rho_{\rm S}}{l_{\rm ms}} \ll c_{\rm S} \frac{\rho_{\rm S}}{l_{\rm ms}}   \rm .
\end{align}
Here, we introduced the parameter $\epsilon \sim l_{\rm ms}   \hat \phi^{-1} \partial_x \hat \phi$, which for the moment we maximally order to be unity, $\epsilon \sim 1$.
Note that the size of the electric field in the $y$ direction is obtained from the large-amplitude characteristic electrostatic potential variations \emph{in the bulk}, equation (\ref{phiinfty-order}): $\partial_y \phi_{\infty} / B \sim T_{\rm e} / (LeB) \sim c_{\rm S}^2 / (\Omega L) \sim \delta c_{\rm S} $.
The electric field in the $z$ direction is negligible. 
From (\ref{ordering-1}), we extract the ordering $\left(l_{\rm ms} / \rho_{\rm S}\right) \delta \ll \epsilon \hat{\phi} \ll 1 $, which can always be taken since $\left(l_{\rm ms} / \rho_{\rm S}\right) \delta = l_{\rm ms} / L \rightarrow 0$ by virtue of our primary expansion (\ref{sheath-scale}).
The $\vec{E}\times \vec{B}$ drift in the $x$ direction caused by the bulk electric field is $\left( \partial_y \phi_{\infty} / B \right) \cos \alpha \sim c_{\rm S} \delta \cos \alpha$, and is thus comparable to the $x$-component of the parallel streaming $c_{\rm S} \sin \alpha$ if $\tan \alpha \sim \delta$.
This confirms that the gradients (of electrostatic potential, ion distribution function, etc.) in the $y$ direction must be included in any treatment of the magnetised sheath when $\alpha \sim \delta \ll 1$.

We assume that the magnetised sheath is in steady state relative to the bulk plasma.
The time scale over which the magnetised sheath equilibrates is expected to be determined by the crossing time of the ions (the slowest species), given by equation (\ref{t-scale}). 
The expected turbulent time scale over which the bulk plasma evolves is the time taken for an ion to drift across a turbulent structure, $L/ (\delta c_{\rm S}) \sim \rho_{\rm S} / \delta^2 c_{\rm s} \sim (\delta^2 \Omega)^{-1}$.
Here, we have used that the characteristic $\vec{E}\times \vec{B}$ drift of an ion is $\sim \delta c_{\rm S}$ from (\ref{bulkE-ordering}).
For the turbulent (bulk) time scale to be much longer than the magnetised sheath time scale, the ordering $l_{\rm ms} \ll \rho_{\rm S} \sin \alpha / \delta^2 \sim L \sin \alpha / \delta$ is required.
By virtue of the orderings (\ref{orderings-length}) and (\ref{alpha-delta}), the turbulent time scale is always longer than the characteristic magnetised sheath time scale, so that a steady state sheath can be assumed.

This work also assumes that the magnetised sheath is collisionless.
Denoting the ion mean free path as $\lambda_{\rm mfp}$, an ion streaming along the field line can traverse the magnetised sheath without colliding only if the distance it travels along the magnetic field line is shorter than the mean free path for a collision, $l_{\rm ms} / \sin \alpha \ll \lambda_{\rm mfp} $.
Accounting for this consideration, we consider the magnetised sheath in the collisionless limit 
\begin{align} \label{alpha-order}
\frac{l_{\rm ms}}{\lambda_{\rm mfp} \sin \alpha} \rightarrow 0   \rm .
\end{align}


Some readers might be either less interested in the case where the magnetic field is at a grazing angle with the target or less familiar with the derivation of kinetic sheath entrance conditions.
In either case, we suggest to initially focus on the special case where the bulk tangential gradients are negligible, thus neglecting terms of order $\delta / \tan \alpha$ and effectively setting $\partial_y = 0$ everywhere. 
We emphasise that, even in this limit, our derivation of the kinetic Bohm-Chodura condition is the most general one to our knowledge. 

\section{Ion trajectories and velocity distribution} \label{sec-traj}

In this section we solve for the ion trajectories and the ion distribution function in the magnetised sheath far from the target.

Given the orderings and assumptions in section~\ref{sec-orderings}, the ion velocity distribution in the magnetised sheath, denoted $f(\boldsymbol{\xi}) = f(x,y, v_x, v_y, v_z)$ with
\begin{align}
\boldsymbol{\xi} = (x, y, v_x, v_y, v_z) \rm ,
\end{align}
satisfies the collisionless and steady-state Vlasov equation
\begin{align} \label{kineqv1}
\frac{df}{dt} = \frac{dx}{dt} \partial_x f + \frac{dy}{dt} \partial_y f + \frac{dv_x}{dt} \partial_{v_x} f + \frac{dv_y}{dt} \partial_{v_y} f + \frac{dv_z}{dt} \partial_{v_z} f = 0 \rm .
\end{align}
As in (\ref{dphidx-infty}) for $\phi$, for the magnetised sheath to be a boundary layer, we require no gradients normal to the target at the sheath entrance,
\begin{align} \label{dfdx-infty}
\partial_x^n f (\boldsymbol \xi ) \rvert_{x/l_{\rm ms} \rightarrow \infty} = 0 \text{ for } n> 0 \rm .
\end{align}
We impose the boundary conditions 
\begin{align} \label{fv-inf}
|\vec{v}|^5 f (\boldsymbol{\xi}) \rvert_{|\vec{v}| \rightarrow \infty} = 0 \rm ,  
\end{align}
\begin{align} \label{fx-infty}
f (\boldsymbol{\xi}) \rvert_{x/l_{\rm ms} \rightarrow \infty} = f_{\infty}(y,v_x, v_y, v_z) \rm ,
\end{align}
\begin{align} \label{fx-0}
f(\boldsymbol{\xi}) \rvert_{x/l_{\rm ms} = 0, v_x > 0} = 0 \rm ,
\end{align}
and either
\begin{align} \label{fy-infty-alt}
f (\boldsymbol{\xi}) \rvert_{y / L \rightarrow \pm \infty} = 0 \rm ,
\end{align} 
or
\begin{align} \label{fy-infty}
f (\boldsymbol{\xi}) \rvert_{y = L_y} = f (\boldsymbol{\xi}) \rvert_{y = -L_y} \rm ,
\end{align} 
on the ion distribution function.
In (\ref{fv-inf}) we assume that there are a sufficiently small number of ions with large kinetic energy that the average energy in the system is finite.
In (\ref{fx-infty}), we assume no ions coming back from the magnetised sheath, such that $f_{\infty}$ is zero in some parts of velocity space.
In (\ref{fx-0}), we assume no ions to be reflected or re-emitted by the wall.
The boundary condition (\ref{fy-infty-alt}) corresponds to cases where the SOL width $\sim L$, while  (\ref{fy-infty}) corresponds to cases where the SOL width $\gg L$ and we may choose $L_y \sim L$. 
Taken together, the boundary conditions (\ref{fx-infty})-(\ref{fy-infty}) imply that ions enter the system only via the sheath entrance $x/l_{\rm ms} \rightarrow \infty$.
 
The time derivatives of $(x, y, v_x, v_y, v_z)$ in (\ref{kineqv1}) are given by the equations of motion of a single ion moving in an electrostatic field $\vec{E}$ and a magnetic field $\vec{B} = B \uvec{b}$ with $\uvec{b} =  -\sin \alpha ~\uvec{e}_x + \cos \alpha ~ \uvec{e}_z $,
\begin{align} \label{x-EOM}
\frac{dx}{dt} = v_x \rm ,
\end{align}
\begin{align} \label{y-EOM}
\frac{dy}{dt} = v_y \rm ,
\end{align}
\begin{align} \label{vx-EOM}
\frac{dv_x}{dt} = \Omega v_y \cos \alpha - \frac{\Omega \partial_x \phi(x,y)}{B}  \rm ,
\end{align}
\begin{align} \label{vy-EOM}
\frac{dv_y}{dt} = - \Omega v_x \cos \alpha - \Omega v_z \sin \alpha - \frac{\Omega \partial_y \phi(x,y)}{B} \rm ,
\end{align}
\begin{align} \label{vz-EOM}
\frac{dv_z}{dt} = \Omega v_y \sin \alpha  \rm .
\end{align}
The unit vector $\uvec{e}_x \equiv \nabla x$ is normal to the target, $\uvec{e}_y \equiv \nabla y = \uvec{b} \times \uvec{e}_x / | \uvec{b} \times \uvec{e}_x | $ is tangential to the target and perpendicular to the magnetic field, and $\uvec{e}_z \equiv \nabla z = \uvec{e}_x \times \uvec{e}_y$ is the other unit vector tangential to the target and orthogonal to $\uvec{e}_y$.
These unit vectors satisfy $\uvec{e}_x\cdot (\uvec{e}_y \times \uvec{e}_z ) = 1$, and are consistent with the Cartesian axes shown in figure~\ref{fig-lzscale}.
Denoting the gradient operator as $\nabla \equiv \uvec{e}_x \partial_x + \uvec{e}_y \partial_y + \uvec{e}_z \partial_z $, the electric field $\vec{E} \equiv - \nabla \phi$ has been taken to be $\vec{E} \simeq - \uvec{e}_y \partial_y \phi - \uvec{e}_x \partial_x \phi $, since $\partial_z \phi$ is neglected as explained in section~\ref{sec-orderings}. 

The distribution function has no explicit time dependence (steady state).
The individual trajectories, however, can be solved in time and their solution can be used to solve for $f$.
We may integrate (\ref{x-EOM})-(\ref{vz-EOM}) --- in practice this can only be done numerically for a general form of $\phi(x,y)$ --- to find the backwards ion trajectories from a reference time $t=0$ to the time $t = t_{\rm enter} < 0$ at which the ion had crossed the boundary to enter the system (assuming no trapped particles are present in the system).
We can then express the solution to the kinetic equation as
\begin{align} \label{f-sol-xi}
f(\boldsymbol \xi(t) )\rvert_{t=0} = f(\boldsymbol \xi(t) )\rvert_{t=t_{\rm enter}} \rm , 
\end{align}
where $\boldsymbol{\xi}(t)$ are the ion trajectories calculated by solving the equations of motion (\ref{x-EOM})-(\ref{vz-EOM}).
Note that $t_{\rm enter} / t_{X} \rightarrow  -\infty$ for past trajectories that reach the boundary $x/l_{\rm ms} \rightarrow \infty$.

The rest of the section is structured as follows.
In section~\ref{subsec-traj-change} we change to a more convenient set of variables 
\begin{align}
\vec G = (X,Y, \mu, \theta, v_\parallel) \rm ,
\end{align}
obtaining the equations of motion and the kinetic equation in the new variables.
In section~\ref{subsec-traj-perturb} we describe a perturbative approach to solve the equations of motion for the time-dependent variables $\vec G$, and carry out the explicit calculations first to zeroth order and then to first order in $\hat \phi \ll 1$.
We assume that ions never penetrate significantly into the sheath (reaching $\hat \phi \sim 1$) and then get reflected, since the perturbed ($\hat \phi \ll 1$) past trajectory calculation for such ions would not be valid for such a case.
In section~\ref{subsec-traj-noreflect}, we derive the necessary condition for ions not to be reflected back out of the magnetised sheath (equivalently, a sufficient condition for the absence of ion reflection locally near the magnetised sheath entrance).
Finally, in section~\ref{subsec-traj-F} we exploit the trajectory solutions to solve perturbatively for the ion distribution function.

\subsection{Change of variables in the equations of motion and kinetic equation} \label{subsec-traj-change}

The lowest order trajectory far away from the target, solved by neglecting the sheath electrostatic potential variation $\phi_1$, is helical as shown in figure~\ref{fig-geometry}.
Hence, the variables $\boldsymbol{\xi}$ all have a component that is time-periodic.
We change to a more convenient set of variables $ \vec G = (X, Y, \mu, \theta, v_\parallel )$, composed of: the guiding center $x$ coordinate
\begin{align} \label{X-def}
X = x + \frac{v_y}{\Omega} \cos \alpha = O\left( l_{\rm ms} \right) \rm ;
\end{align}
the guiding center $y$ coordinate
\begin{align} \label{Y-def}
Y = y - \frac{1}{\Omega} \left( v_x \cos \alpha + \frac{\partial_y \phi}{B} + v_z \sin \alpha \right) = O(L) \rm ;
\end{align}
the magnetic moment $\mu = v_{\perp}^2 / (2\Omega )$,
\begin{align} \label{mubar-def}
\mu & = \frac{1}{2\Omega} \left( \vec{v} + \frac{\partial_y \phi(x,y)}{B} \uvec{e}_y \times \uvec{b} - v_{\parallel} \uvec{b} \right) \cdot \left( \vec{v} + \frac{\partial_y \phi(x,y)}{B} \uvec{e}_y \times \uvec{b} - v_{\parallel} \uvec{b}  \right)  \nonumber \\
& = \frac{1}{2\Omega} \left[ \left( v_x \cos \alpha + \frac{\partial_y \phi(x, y)}{B} + v_z \sin \alpha \right)^2 + v_y^2 \right] = O(c_{\rm S}\rho_{\rm S}) \rm ,
\end{align}
such that $v_{\perp}(\mu) = \sqrt{2\mu \Omega}$; the phase angle 
\begin{align} \label{theta-def}
\theta = \arctan \left( \frac{ v_x \cos \alpha + \frac{\partial_y \phi(x,y)}{B} + v_z \sin \alpha }{ v_y } \right) = O(1) \rm;
\end{align}
the parallel velocity
\begin{align} \label{vpar-def}
v_{\parallel} = \vec{v} \cdot \uvec{b} = v_z \cos \alpha - v_x \sin \alpha = O(c_{\rm S}) \rm .
\end{align}
Note that the orderings in (\ref{X-def}) and (\ref{Y-def}) refer to the \textit{changes} in these coordinates, as ions traverse the magnetised sheath, that correspond to a significant variation $\sim T_{\rm e} / e$ of the potential $\phi$.

The old variables $\boldsymbol{\xi} = (x, y, v_x, v_y, v_z) $ can be re-expressed as a function of the new ones:
\begin{align} \label{x(GK)}
x =  X + \rho_x ( \mu, \theta)  \rm ,
\end{align}
\begin{align} \label{y(GK)}
y = Y + \rho_y ( \mu, \theta)   \rm ,
\end{align}
\begin{align} \label{vx(GK)}
v_x = v_{\perp}(\mu) \cos \alpha \sin \theta  - \cos \alpha \frac{\partial_y \phi (X + \rho_x ( \mu, \theta),Y + \rho_y ( \mu, \theta))}{B} - v_{\parallel} \sin \alpha \rm ,
\end{align}
\begin{align} \label{vy(GK)}
v_y = v_{\perp}(\mu) \cos \theta \rm ,
\end{align}
\begin{align} \label{vz(GK)}
v_z = v_{\perp}(\mu) \sin \alpha \sin \theta  - \sin \alpha \frac{1}{B}\partial_y \phi (X + \rho_x ( \mu, \theta),Y + \rho_y ( \mu, \theta)) + v_{\parallel} \cos \alpha \rm ,
\end{align}
where we defined the functions
\begin{align} \label{rhox-def}
\rho_x ( \mu, \theta) = - \frac{v_{\perp}(\mu)}{\Omega} \cos \alpha \cos \theta \rm ,
\end{align}
\begin{align} \label{rhoy-def}
\rho_y ( \mu, \theta) =   \frac{v_{\perp}(\mu)}{\Omega} \sin \theta \rm .
\end{align}

Consider an ion that at a reference time $t= - \tau = 0$ has values $\vec{G}(\tau) \rvert_{\tau = 0} = \vec{G}_{\rm f} = (X_{\rm f}, Y_{\rm f}, \mu_{\rm f}, \theta_{\rm f}, v_{\parallel, \rm f})$ with $X_{\rm f} \gg l_{\rm ms}$, where the subscript f stands for final.
We seek to calculate $\vec{G}(\tau) $ at past times $t= - \tau$ with $\tau \geqslant 0$ --- effectively solving for the past particle trajectory.
To obtain the equations of motion in the new variables $\vec G$ and $\tau$, we differentiate the definitions (\ref{X-def})-(\ref{vpar-def}) with respect to time using (\ref{x-EOM})-(\ref{vz-EOM}) and insert (\ref{x(GK)})-(\ref{vz(GK)}),
\begin{align} \label{Xdot-long}
& \frac{dX}{d\tau} = v_{\parallel} \sin \alpha + \cos \alpha \frac{1}{B} \partial_y \phi = O\left( \frac{l_{\rm ms}}{t_{X}} \right) \rm ,
\end{align}
\begin{align} \label{Ydot-long}
\frac{dY}{d\tau} = & - \frac{1}{B} \cos \alpha \partial_x \phi  + \frac{1}{\Omega B} v_{\perp} (\mu) \cos \theta \partial_y^2 \phi \nonumber \\
& + \frac{1}{\Omega B}  \left( v_{\perp}(  \mu) \cos \alpha \sin \theta - \cos \alpha \frac{1}{B} \partial_y \phi  -  {v}_{\parallel} \sin \alpha \right)  \partial_x \partial_y \phi  = O\left( \epsilon \hat \phi \frac{\delta}{\sin \alpha} \frac{L}{t_{X}} \right) \rm ,
\end{align}
\begin{align} \label{mudot-long}
\frac{d\mu}{d\tau} = & v_{\perp}(\mu) \sin \theta \left[ \cos \alpha \frac{1}{B} \partial_x \phi  - \frac{v_\perp ( \mu) }{\Omega B} \cos \theta \partial_{y}^{2} \phi  \right. \nonumber \\
& \left. -   \frac{1}{\Omega B} \partial_x \partial_y \phi   \left( v_{\perp}(  \mu) \cos \alpha \sin   \theta  - \cos \alpha \frac{1}{B} \partial_y \phi  -  {v}_{\parallel} \sin \alpha \right)  \right] = O \left( \frac{\epsilon \hat \phi}{\sin \alpha} \frac{\rho_{\rm S} c_{\rm S} }{t_{X}} \right) \rm ,
\end{align}
\begin{align} \label{vpardot-long}
& \frac{dv_\parallel}{d\tau} = - \sin \alpha \frac{\Omega}{B}  \partial_x \phi = O\left( \epsilon \hat \phi \frac{c_{\rm S}}{t_{X}} \right) \rm ,
\end{align}
\begin{align} \label{thetadot}
& \frac{d  \theta }{d\tau}  = - \Omega + \frac{\Omega}{v_{\perp} (  \mu) B} \cos \alpha  \cos   \theta ~ \partial_x \phi - \cos^2   \theta \frac{1}{B} \partial_{y}^{2} \phi  \nonumber \\ 
& - \frac{1}{B}\partial_x \partial_y \phi \cos   \theta \left( \cos \alpha \sin   \theta  - \frac{\cos \alpha}{v_{\perp}(\mu) B}  \partial_y \phi  - \frac{v_{\parallel} \sin \alpha }{v_{\perp}(  \mu)}  \right) = O\left( \frac{1}{\sin \alpha} \frac{l_{\rm ms}}{\rho_{\rm S}} \frac{1}{t_{X}} \right)  \rm .
\end{align}
The arguments of $\phi(x,y)$ in (\ref{Xdot-long})-(\ref{thetadot}) are $x = X + \rho_x (  \mu,   \theta)$ and $y = Y + \rho_y (  \mu,   \theta)$.

We expand equations (\ref{Xdot-long})-(\ref{thetadot}) in $\delta \ll 1$ while keeping simultaneously $\alpha \sim 1$ and $\delta / \alpha \sim 1$.
These three orderings can never be simultaneously satisfied, but they allow us to retain the terms necessary to treat the range of possible angles $\delta \lesssim \alpha \lesssim 1$.
We keep terms $\gtrsim (\delta/\sin \alpha) \hat \phi G t_{X}^{-1}$, while we neglect any higher order terms in $\delta$, such as $O(\delta (l_{\rm ms}/\rho_{\rm S}) G t_{X}^{-1})$ and $O(\delta \hat \phi G t_{X}^{-1})$. 
Taylor expanding $\phi(  X + \rho_x,   Y + \rho_y)$ for small $\rho_y$ in (\ref{Ydot-long}), (\ref{mudot-long}) and (\ref{thetadot}) cancels out the terms containing $\sin   \theta \cos \alpha  \frac{1}{\Omega B} \partial_x \partial_y \phi (  X + \rho_x, Y) \sim \delta \hat \phi$, leaving
\begin{align} \label{Ydot-long-3}
\frac{dY}{d\tau} = & - \frac{1}{B} \cos \alpha \partial_x \phi (X + \rho_x, Y)  + O\left( \delta^2 \frac{\delta}{\sin \alpha} \frac{l_{\rm ms}}{\rho_{\rm S}} \frac{L}{t_X}, \delta^2 \hat \phi \frac{\delta}{\sin \alpha} \frac{L}{t_X}, \delta^2 \hat \phi \frac{L}{t_X} \right) \rm ,
\end{align}
\begin{align} \label{mudot-long-3}
\frac{d   \mu}{d\tau} =  & v_{\perp}(  \mu) \sin   \theta  \cos \alpha \frac{1}{B} \partial_x \phi(  X + \rho_x(  \mu,   \theta),  Y )  \nonumber \\
& + O \left( \delta \frac{\delta }{ \sin \alpha} \frac{l_{\rm ms}}{\rho_{\rm S}} \frac{c_{\rm S} \rho_{\rm S}}{t_X}, \delta  \hat \phi \frac{c_{\rm S} \rho_{\rm S}}{ t_{X} }, \delta  \hat \phi \frac{\delta}{\sin \alpha} \frac{c_{\rm S} \rho_{\rm S}}{ t_{X} } \right) \rm ,
\end{align}
\begin{align} \label{thetadot-long-3}
\frac{d  \theta }{d\tau} = & - \Omega + \frac{\Omega}{v_{\perp}(  \mu) B} \cos \alpha  \cos   \theta ~ \partial_x \phi (X + \rho_x (  \mu,   \theta),   Y ) ) \nonumber \\
& + O\left( \delta \frac{\delta}{\sin \alpha} \frac{l_{\rm ms}}{\rho_{\rm S}} \frac{1}{t_X},\delta  \hat \phi \frac{1}{t_X}, \delta  \hat \phi \frac{\delta}{\sin \alpha} \frac{1}{ t_{X} } \right) \rm .
\end{align}
Similarly expanding for small $\rho_y$ in (\ref{Xdot-long}) and (\ref{vpardot-long}) gives
\begin{align} \label{Xdot-exact}
\frac{d  X}{d\tau} =   v_{\parallel} \sin \alpha + \cos \alpha \frac{1}{B} \partial_y \phi (  X + \rho_x (  \mu , \theta),   Y ) + O\left( \delta \frac{\delta }{ \sin \alpha } \frac{l_{\rm ms}}{t_X} \right)  \rm ,
\end{align}
\begin{align} \label{vpardot-exact}
\frac{d {v}_\parallel}{d\tau} = - \sin \alpha \frac{\Omega}{B}  \partial_x \phi(  X + \rho_x(  \mu,   \theta ),   Y ) + O\left( \delta \hat \phi \frac{c_{\rm S}}{t_X} \right)  \rm .
\end{align}

While it is possible to obtain the distribution function directly from equation (\ref{f-sol-xi}), it is useful to discuss the kinetic equation in the new coordinates.
Adopting the change of variables $(t, x,y, v_x, v_y, v_z) \rightarrow (\tau, X, Y, \mu, v_{\parallel}, \theta)$  in (\ref{kineqv1}) and denoting $F(\vec{G}) \equiv f(\boldsymbol{\xi}(\vec G))$, we may use the chain rule
\begin{align}
\frac{d\tau}{dt}\frac{d\vec{G}}{d\tau} \cdot \nabla_{\vec{G}} F = \frac{d\boldsymbol{\xi}}{dt} \cdot \nabla_{\boldsymbol{\xi}} f(\boldsymbol \xi )
\end{align}
with $d\tau / dt = -1$, to obtain the kinetic equation
\begin{align} \label{kineqv2}
\frac{d F}{d\tau} = \frac{d  X}{d\tau} \partial_X F + \frac{d  Y}{d\tau} \partial_Y F + \frac{d  \mu}{d\tau} \partial_{\mu} F + \frac{d {v}_{\parallel}}{d\tau} \partial_{v_{\parallel}} F + \frac{d  \theta}{d\tau} \partial_{\theta} F = 0 \rm .
\end{align}
The no-normal gradients condition (\ref{dfdx-infty}) at the sheath entrance can be recast in the new variables as follows.
Using the chain rule we re-express $\partial_X F (\vec G)$ as
\begin{align} \label{dFdX-infty-pre}
(\partial_X F (\vec G) )_{Y, \mu, v_{\parallel}, \theta} =  & (\partial_X x)_{Y, \mu, v_{\parallel}, \theta}  \partial_{x} f (\boldsymbol{\xi}) + (\partial_X y)_{Y, \mu, v_{\parallel}, \theta} \partial_{y} f (\boldsymbol{\xi}) + (\partial_X v_x)_{Y, \mu, v_{\parallel}, \theta} \partial_{v_x} f (\boldsymbol{\xi}) \nonumber \\
 & + (\partial_X v_y)_{Y, \mu, v_{\parallel}, \theta} \partial_{v_y} f (\boldsymbol{\xi}) + (\partial_X v_z)_{Y, \mu, v_{\parallel}, \theta} \partial_{v_z} f (\boldsymbol{\xi})\rm .
\end{align}
From equations (\ref{y(GK)}) and (\ref{vy(GK)}) we obtain  $(\partial_X v_y)_{Y, \mu, v_{\parallel}, \theta} = 0$ and $(\partial_X y)_{Y, \mu, v_{\parallel}, \theta} = 0$, while equation (\ref{x(GK)}) gives $(\partial_X x)_{Y, \mu, v_{\parallel}, \theta} = 1$. From equations (\ref{vx(GK)}) and (\ref{vz(GK)}) we instead obtain $(\partial_X v_x)_{Y, \mu, v_{\parallel}, \theta} / \cos \alpha = (\partial_X x)_{Y, \mu, v_{\parallel}, \theta} / \sin \alpha = - \partial_x \partial_y \phi(X + \rho_x, Y + \rho_y) / B $. 
Hence, (\ref{dFdX-infty-pre}) reduces to
\begin{align} \label{dFdX-infty-pre2}
& (\partial_X F (\vec G) )_{Y, \mu, v_{\parallel}, \theta} =  \partial_{x} f (\boldsymbol{\xi}(\vec{G})) \nonumber \\
& - \frac{1}{B} \partial_x \partial_y \phi(X + \rho_x, Y + \rho_y) (\cos \alpha \partial_{v_x} f (\boldsymbol{\xi} (\vec{G}))) + \sin \alpha \partial_{v_z} f (\boldsymbol{\xi} (\vec{G})))\rm .
\end{align}
Evaluating this at $X \rightarrow \infty$ is equivalent to evaluating at $ X + \rho_x \rightarrow \infty$ unless $\rho_x $ is infinitely large and negative, which requires $\cos \theta > 0$ and $\mu \sim \Omega X^2 / 2 \rightarrow \infty$.
However, boundary condition (\ref{fv-inf}) implies that
\begin{align} \label{Fmu-infty}
F (\vec G) \rvert_{\mu \rightarrow \infty} = 0 \rm .
\end{align}
Imposing conditions (\ref{phi-infty}), (\ref{fx-infty}) and (\ref{Fmu-infty}) on (\ref{dFdX-infty-pre2}) gives $\partial_X F (\vec G) \rvert_{X/l_{\rm ms} \rightarrow \infty} = 0$. 
This can be repeated for the $n$th derivative with respect to $x$, to obtain the condition
\begin{align} \label{dFdX-infty}
\partial_X^n F (\vec G) \rvert_{X/l_{\rm ms} \rightarrow \infty} = 0 \rm .
\end{align}
The boundary conditions (\ref{fx-infty})-(\ref{fx-0}) are equivalent to
\begin{align} \label{FX-infty}
F( \vec G ) \rvert_{X/l_{\rm ms} \rightarrow \infty}  = F_{\infty} (Y, \mu, \theta, v_{\parallel}) \rm ,
\end{align} 
\begin{align} \label{FX-0}
F( \vec G ) \rvert_{X + \rho_x (\mu,\theta) = 0, v_x(\boldsymbol{\vec G}) > 0}  = 0 \rm ,
\end{align} 
and either
\begin{align} \label{FY-infty-alt}
F( \vec G ) \rvert_{Y / L \rightarrow \pm \infty} = 0 \rm ,
\end{align} 
or
\begin{align} \label{FY-infty}
F( \vec G ) \rvert_{Y = L_y} = F( \vec G )  \rvert_{Y = -L_y} \rm ,
\end{align}
where in (\ref{FX-infty}) and (\ref{FY-infty-alt})-(\ref{FY-infty}) we use that $F$ must be exponentially small at large values of $\mu$.
The solution of the kinetic equation (\ref{kineqv2}) is 
\begin{align} \label{F-solution}
F(\vec G (\tau)) \rvert_{\tau = \tau_{\rm enter}} = F(\vec G (\tau)) \rvert_{\tau =0} \rm ,
\end{align}
where $\vec G(\tau)$ are the solutions of the equations of motion (\ref{Xdot-long})-(\ref{thetadot}).
Equation (\ref{F-solution}) is equivalent to equation (\ref{f-sol-xi}), but to derive it, we have obtained equation (\ref{kineqv2}).
We use equation (\ref{kineqv2}) in section \ref{subsec-traj-F} to derive properties of $F$ at the magnetised sheath entrance.

Equations (\ref{mudot-long-3})-(\ref{vpardot-exact}) comprise four nonlinear coupled first order differential equations, which can be compactly re-expressed as
\begin{align} \label{Gdot-exact}
\frac{d\vec G}{d\tau} = \vec h_0(\vec{G}(\tau)) + \vec h_{\phi}(\vec{G}(\tau)) + O \left( \delta\frac{\delta}{\sin \alpha} \frac{l_{\rm ms}}{\rho_{\rm S}} \frac{\vec{G}}{t_X}, \delta \hat \phi \frac{\vec{G}}{t_X}, \delta \hat \phi \frac{\delta}{\sin \alpha} \frac{\vec{G}}{t_X} \right)  \rm .
\end{align}
Here, we introduced the vector functions $\vec h_0 $, comprising all terms that are independent of $\phi_1$, and $\vec h_{\phi} $, comprising all terms that depend on $\phi_1$.
It may appear that we have not made much progress by changing variables. 
However, from the orderings in (\ref{Xdot-long})-(\ref{thetadot}), the variables $\vec{G}$ can be conveniently split into two groups, $\vec{G} = (\boldsymbol{\Gamma}, \boldsymbol{\gamma})$, such that $\boldsymbol{\Gamma} = (Y, \mu, v_{\parallel})$ are slow variables whose time derivatives satisfy
\begin{align} \label{slow-def}
\frac{d\boldsymbol{\Gamma}}{d\tau} \lesssim \frac{\hat \phi}{\sin\alpha} \frac{\boldsymbol{\Gamma} }{t_X} \rm ,
\end{align}
and $\boldsymbol{\gamma} = (X, \theta)$ are fast variables whose time derivatives satisfy
\begin{align}
\frac{d\boldsymbol{\gamma}}{d\tau} \gtrsim \frac{\boldsymbol{\gamma}}{t_X} \rm .
\end{align}
We correspondingly split the time derivatives such that $\vec h_0 = (\vec h^{\Gamma}_0, \vec h^{\gamma}_0)$ and $\vec h_\phi = (\vec h^{\Gamma}_\phi, \vec h^{\gamma}_\phi)$.
By the definition (\ref{slow-def}), the zeroth order time derivative of the slow variables vanishes, $\vec h^{\Gamma}_0 = \vec{0}$, giving
\begin{align} \label{Gslowdot}
\frac{d \boldsymbol{\Gamma}}{d\tau} = \vec h^{\Gamma}_{\phi} (\vec{G}) + O \left( \delta\frac{\delta}{\sin \alpha} \frac{l_{\rm ms}}{\rho_{\rm S}} \frac{\boldsymbol{\Gamma}}{t_X}, \delta \hat \phi \frac{\boldsymbol{\Gamma}}{t_X}, \delta \hat \phi \frac{\delta}{\sin \alpha} \frac{\boldsymbol{\Gamma}}{t_X} \right)  \rm .
\end{align}
The zeroth order time derivative of fast variables is non-zero, but it only depends on the slow variables, $\vec h^{\gamma}_0 = \vec h^{\gamma}_0 (\boldsymbol{\Gamma})$. 
Hence, the time derivative of fast variables satisfies
\begin{align} \label{Gfastdot}
\frac{d \boldsymbol{\gamma}}{d\tau} =  \vec h^{\gamma}_0 (\boldsymbol{\Gamma}) + \vec h^{\gamma}_{\phi} (\vec{G}) + O \left( \delta\frac{\delta}{\sin \alpha} \frac{l_{\rm ms}}{\rho_{\rm S}} \frac{\boldsymbol{\gamma}}{t_X}, \delta \hat \phi \frac{\boldsymbol{\gamma}}{t_X}, \delta \hat \phi \frac{\delta}{\sin \alpha} \frac{\boldsymbol{\gamma}}{t_X} \right) \rm .
\end{align}
Equations (\ref{Gslowdot}) and (\ref{Gfastdot}) can be solved perturbatively in $\hat \phi \ll 1$, as detailed in the next subsection. 
This motivates a perturbative solution of $F(\vec G)$, detailed in subsection~\ref{subsec-traj-F}. 

\subsection{Perturbative calculation of the ion trajectories} \label{subsec-traj-perturb}

We calculate the time-dependence of the variables $\vec{G}$ using the asymptotic expansion $\vec{G}(\tau) = \vec{G}_{\rm f} + \vec{G}_0(\tau) + \vec{G}_1(\tau) + \vec{G}_2(\tau) + O(\hat \phi^3 \vec{G}, \delta \vec{G}) $, explicitly
\begin{align} \label{X-exp}
 X(\tau) = X_{\rm f} +  {X}_0(\tau) + {X}_1(\tau) + {X}_2(\tau) + O\left(\hat \phi^3 l_{\rm ms}, \delta l_{\rm ms} \right) \rm ,
\end{align}
\begin{align} \label{Y-exp}
Y(\tau) = Y_{\rm f} +  {Y}_0(\tau) +  {Y}_1(\tau) +  {Y}_2(\tau) + O\left(\hat \phi^3 L, \delta L \right) \rm ,
\end{align}
\begin{align} \label{vpar-exp}
v_\parallel (\tau) = v_{\parallel, \rm f} +  {v}_{\parallel,0}(\tau) +  {v}_{\parallel,1}(\tau) +  {v}_{\parallel,2}(\tau) + O\left(\hat \phi^3 c_{\rm S}, \delta c_{\rm S} \right) \rm ,
\end{align}
\begin{align} \label{mu-exp}
\mu(\tau) = \mu_{\rm f} +  {\mu}_0(\tau) +  {\mu}_1(\tau) +  {\mu}_2(\tau) + O\left(\hat \phi^3 c_{\rm S} \rho_{\rm S} , \delta c_{\rm S} \rho_{\rm S} \right) \rm ,
\end{align}
\begin{align} \label{theta-exp}
\theta(\tau) = \theta_{\rm f} +  {\theta}_0(\tau) +  {\theta}_1(\tau) +  {\theta}_2(\tau) + O\left(\hat \phi^3 , \delta \right) \rm .
\end{align}
The subscripts indicate orders in $\hat{\phi}$, such that $  \vec{G}_n \sim \hat \phi^n \vec{G}$. 

The time dependence of the $n$th order correction to the variables is obtained from its time derivative via
\begin{align} \label{rule}
\vec G_n (\tau) = \int_0^{\tau} d\tau'  \frac{d\vec G_n (\tau')}{d\tau'} \rm .
\end{align}
The time derivative $d\vec G_n / d\tau$ is calculated perturbatively by considering successive orders in the expansion in $\hat \phi$, and at each order calculating the time dependence of slow variables, $\boldsymbol{\Gamma}_n(\tau)$, first.
At zeroth order, we insert the expansion $\boldsymbol{\Gamma}(\tau) = \boldsymbol{\Gamma}_{\rm f} + \boldsymbol{\Gamma}_0 (\tau) + O(\hat \phi \boldsymbol{\Gamma})$ into (\ref{Gslowdot}) and neglect terms of order $O(\hat \phi t_X^{-1} \boldsymbol{\Gamma})$ to obtain
\begin{align} \label{Gslow0dot}
\frac{d \boldsymbol{\Gamma}_0}{d\tau} = \vec{0}  \rm ,
\end{align}
which integrates to 
\begin{align} \label{Gslow0}
\boldsymbol{\Gamma}_0(\tau) = (Y_0(\tau), \mu_0(\tau), v_{\parallel, 0}(\tau)) = \vec{0}  \rm .
\end{align}
This makes explicit, within the perturbative framework, the ordering (\ref{slow-def}) defining slow variation.
The zeroth order correction to fast variables is then calculated by inserting $\boldsymbol{\Gamma}(\tau) = \boldsymbol{\Gamma}_{\rm f} + O(\hat \phi \boldsymbol{\Gamma})$ (exploiting (\ref{Gslow0})) and $\boldsymbol{\gamma} = \boldsymbol{\gamma}_{\rm f} + \boldsymbol{\gamma}_0 (\tau) + O(\hat \phi \boldsymbol{\Gamma})$ into (\ref{Gfastdot}) and neglecting terms of order $O(\hat \phi t_X^{-1} \boldsymbol{\gamma})$, resulting in
\begin{align} \label{Gfast0dot}
\frac{d \boldsymbol{\gamma}_0}{d\tau} =  \vec h^{\gamma}_0(\bf{\Gamma}_{\rm f}) \rm .
\end{align}
Inserting $\vec{G}(\tau) = \vec{G}_{\rm f} + \vec{G}_0 (\tau) + \vec{G}_1 (\tau) + O(\hat \phi^2 \vec{G})$ in (\ref{Gslowdot}) and neglecting terms of order $O(\hat \phi^2 t_X^{-1} \boldsymbol{\Gamma} )$ gives
\begin{align} \label{Gslow1dot}
\frac{d\bf{\Gamma}_1}{d\tau} = \vec h^\Gamma_{\phi} (\vec{G}_{\rm f} + \vec G_0(\tau))  \rm .
\end{align}
Doing the same with equation (\ref{Gfastdot}) gives
\begin{align} \label{Gfast1dot}
\frac{d\boldsymbol{\gamma}_1}{d\tau} =  \vec h^{\gamma}_0(\boldsymbol{\Gamma}_{\rm f} + \boldsymbol{\Gamma}_1(\tau)) + \vec h^{\gamma}_{\phi} (\vec{G}_{\rm f} + \vec G_0(\tau)) - \frac{d \boldsymbol{\gamma}_0}{d\tau}  \rm .
\end{align}
Inserting $\vec{G}(\tau) = \vec{G}_{\rm f} + \vec{G}_0 (\tau) + \vec{G}_1 (\tau) + \vec{G}_2 (\tau)  + O(\hat \phi^3 \vec{G})$ in (\ref{Gslowdot}) and neglecting terms of order $O(\hat \phi^3 t_X^{-1} \boldsymbol{\Gamma} )$ gives
\begin{align} \label{Gslow2dot}
\frac{d\bf{\Gamma}_2}{d\tau} = \vec h^\Gamma_{\phi} (\vec{G}_{\rm f} + \vec G_0(\tau) + \vec G_1(\tau) ) - \frac{d\bf{\Gamma}_1}{d\tau} \rm .
\end{align}
In this work, the second order correction to fast variables, $\boldsymbol{\gamma}_2(\tau)$, will not be needed.
However, this recursive scheme could be extended to arbitrarily high order: given that the $n$th order correction $\boldsymbol{\Gamma}_n (\tau)$ to slow variables has been calculated, then the $n$th order correction $\boldsymbol{\gamma}_n (\tau)$ to the fast variables can be calculated, at which point the $(n+1)$th order correction to slow variables $\boldsymbol{\Gamma}_{n+1}$ can be calculated, and so on.

By virtue of (\ref{Gslow0}), the zeroth order correction of the slow variables $Y$, $\mu$, and $v_{\parallel}$ is zero.
For the fast variables, we proceed by applying (\ref{Gfast0dot}) to (\ref{thetadot-long-3}) and (\ref{Xdot-exact}) to obtain
\begin{align} \label{Xdot-0th}
\frac{d X_0}{d\tau} =  \cos \alpha \frac{\phi'_{\infty}(Y_{\rm f})}{B} + v_{\parallel, \rm f} \sin \alpha   \rm ,
\end{align}
\begin{align} \label{thetadot-0th}
\frac{d{\theta}_0}{d\tau} = - \Omega \rm .
\end{align}
Hence, the zeroth order correction to the fast variables $X$ and $\theta$ is given by
\begin{align} \label{X-bar-0}
 X_0 (\tau) & = \left( \cos \alpha \frac{\phi'_{\infty}(Y_{\rm f})}{B} + v_{\parallel, \rm f} \sin \alpha \right) \tau  \rm ,
\end{align}
\begin{align} \label{theta-0}
  \theta_0 (\tau) = - \Omega \tau \rm .
\end{align}
Note that the guiding center $X$ also becomes a slow variable if $v_{\parallel, \rm f} +  \frac{\phi_{\infty}'(Y_{\rm f})}{B\tan \alpha} $ is close to zero.
We leave the analysis of such ``slow'' particles to \ref{app-slowtraj}.  

We proceed to calculate $ \boldsymbol{\Gamma}_1 = (Y_1, \mu_1, v_{\parallel, 1})$, and leave the calculation of $\boldsymbol{\gamma}_1 = (X_1, \theta_1)$ to section~\ref{sec-higher} because it will not be required to calculate the ion density at first order.
By applying (\ref{Gslow1dot}) to (\ref{Ydot-long-3}), (\ref{mudot-long-3}) and (\ref{vpardot-exact}), we obtain
\begin{align} \label{mudot-1st}
\frac{d  \mu_1}{d\tau} = & v_{\perp}(\mu_{\rm f}) \sin (\theta_{\rm f} +   \theta_0 (\tau) ) \cos \alpha \frac{1}{B}  \partial_x \phi (X_{\rm f} +   X_0 (\tau) + \rho_x (\mu_{\rm f}, \theta_{\rm f} +   \theta_0 (\tau)), Y_{\rm f})  \rm ,
\end{align}
\begin{align} \label{vpardot-1st}
\frac{d {v}_{\parallel, 1}}{d\tau}  = \Omega \frac{ d Y_1}{d\tau} \tan \alpha = - \sin \alpha \frac{\Omega}{B} \partial_x \phi (X_{\rm f} +   X_0 (\tau) + \rho_x (\mu_{\rm f}, \theta_{\rm f} + \theta_0(\tau)), Y_{\rm f})   \rm .
\end{align}
The correction $\mu_1(\tau) $ is obtained by integrating (\ref{mudot-1st}) according to (\ref{rule}),
\begin{align} \label{mu1-bar}
 &  \mu_1(\tau) = \frac{1}{B} v_{\perp}(\mu_{\rm f}) \cos \alpha \nonumber \\
 & \times  \int_0^{\tau} d\tau'  \sin (\theta_{\rm f} +   \theta_0 (\tau'))  \partial_x \phi (X_{\rm f} +   X_0 (\tau') + \rho_x (\mu_{\rm f}, \theta_{\rm f} +   \theta_0(\tau')), Y_{\rm f})  \rm .
\end{align}
Combining (\ref{mudot-1st}) and (\ref{vpardot-1st}) gives an equation for ${v}_{\parallel, 1}(\tau)$ composed of total derivatives with respect to $\tau$,
\begin{align} \label{combo}
& \Omega \frac{d  \mu_1}{d\tau} + \frac{d  v_{\parallel, 1}}{d\tau} \left( v_{\parallel, \rm f} +  \frac{\phi'_{\infty}(Y_{\rm f})}{B\tan \alpha} \right)  \nonumber \\
&  = - \frac{d}{d\tau} \left[ \frac{\Omega}{B}  \phi (X_{\rm f} +   X_0 (\tau) + \rho_x (\mu_{\rm f}, \theta_{\rm f} +   \theta_0(\tau)), Y_{\rm f}) \right] \rm .
\end{align}
Integrating (\ref{combo}) and rearranging for $  v_{\parallel, 1}(\tau) = \Omega Y_1 (\tau) \tan \alpha $ gives
\begin{align} \label{vparbar1-Ybar1}
  v_{\parallel, 1} (\vec{G}_{\rm f}) = \Omega   Y_1 (\vec{G}_{\rm f}) \tan \alpha = \frac{\Omega}{B \left( v_{\parallel, \rm f} +  \frac{\phi'_{\infty}(Y_{\rm f})}{B\tan \alpha}  \right) } \left[  \phi (X_{\rm f} + \rho_x (\mu_{\rm f}, \theta_{\rm f}), Y_{\rm f} ) \right. \nonumber \\
  \left. - \phi (X_{\rm f} +   X_0 (\tau) + \rho_x (\mu_{\rm f}, \theta_{\rm f} +   \theta_0(\tau)), Y_{\rm f})) - B \mu_1(\tau) \right]  \rm .
  \nonumber \\
\end{align}

\subsection{Necessary criterion for the absence of reflected ions} \label{subsec-traj-noreflect}

In this work, we assume that all ions near the magnetised sheath entrance move towards the wall, with $dX/ d\tau  > 0$.
If the magnetic field angle is large enough that tangential fluctuations of $\phi$ are unimportant, $\alpha \gg \delta$, it has been shown that $\partial_x \phi > 0$ everywhere in the magnetised sheath is sufficient to guarantee that all ions entering the magnetised sheath at $x/l_{\rm ms} \rightarrow \infty$ with $dX/d\tau < 0$ will reach the target \cite{Geraldini-2017}.
For $\alpha \sim \delta \ll 1$, the electric field towards the target ($\partial_x \phi > 0$) accelerates $v_{\parallel}$ but also causes an $\vec{E}\times \vec{B}$ drift in the positive $y$ direction that may increase or decrease $\partial_y \phi$ in expression (\ref{Xdot-exact}) for $dX / d\tau$.
Calculating the time derivative of $d  X / dt$ in (\ref{Xdot-exact}) gives
\begin{align} \label{ddotX-accurate}
\frac{d^2  X}{d\tau^2} \simeq  & \frac{dv_{\parallel}}{d\tau} \sin \alpha + \frac{dY}{d\tau} \cos \alpha  \frac{1}{B} \partial_y^2 \phi (  X + \rho_x (  \mu , \theta),   Y ) \nonumber \\
& + \cos \alpha \left[ \frac{dX}{d\tau} - v_{\perp}(\mu) \cos \alpha \sin \theta \right] \frac{1}{B} \partial_y \partial_x \phi (  X + \rho_x (  \mu , \theta),   Y ) \rm .
\end{align}
Particles just about to reflect must have $dX/d\tau = 0$, such that at that instant the only quantity varying quickly in time is $\theta$ and the final term in (\ref{ddotX-accurate}) averages to zero in $\theta$.
By using (\ref{vpardot-1st}) to re-express the first two terms in (\ref{ddotX-accurate}), and taking $\partial_y^2 \phi \simeq \phi_{\infty}''$, we are left with
\begin{align} \label{ddotX}
\frac{d^2  X}{d\tau^2}  \simeq - \Omega \frac{\partial_x \phi}{B} \sin^2 \alpha \left( 1 + \frac{\phi''_{\infty}(Y)}{\Omega B\tan^2 \alpha} \right) \rm .
\end{align}
Thus, an ion entering the magnetised sheath, with $d   X/ d\tau < 0$, \emph{cannot be reflected} near the magnetised sheath entrance provided that expression (\ref{ddotX}) is negative. 
Presuming that $\partial_x \phi > 0$ remains true due to the electron-repelling nature of the sheath, the resulting condition is  
\begin{align} \label{cond-noreflection}
\phi''_{\infty}(Y) > - \Omega B \tan^2 \alpha \rm .
\end{align}
Condition (\ref{cond-noreflection}) is only local (refers to near the magnetised sheath entrance), though we would expect a similar condition globally across the magnetised sheath to avoid a change in sign of $v_{\parallel} + (B\tan \alpha )^{-1} \partial_y \phi $.

Here, we do not justify why the tangential fluctuations at the magnetised sheath entrance should satisfy (\ref{cond-noreflection}).
One possibility is that the magnetised sheath self-consistently tends to a steady state where tangential gradients are flattened sufficiently that (\ref{cond-noreflection}) is satisfied.
The other possibility is that (\ref{cond-noreflection}) is not necessarily satisfied, and so there are cases in which tangential gradients cause ions to be reflected from the sheath. 
In these cases, the local analysis at the sheath entrance performed here is inaccurate because ions could reach deep into the magnetised sheath where $\hat \phi \sim 1$ before reflecting.
A numerical solution of the magnetised sheath with the inclusion of ions with $dX/d\tau < 0$ is necessary when reflections occur.


\subsection{Perturbative calculation of the ion velocity distribution} \label{subsec-traj-F}

By evaluating the kinetic equation (\ref{kineqv2}) at $X/l_{\rm ms} \rightarrow \infty$, which corresponds to $\hat \phi = 0$, and using the time derivatives (\ref{Gslow0dot}) and (\ref{thetadot-0th}) and the no-normal-gradients condition (\ref{dFdX-infty}), the distribution function at the magnetised sheath entrance satisfies
\begin{align}
\Omega  \partial_{\theta} F \rvert_{X/l_{\rm ms} \rightarrow \infty} = 0 \rm 
\end{align}
to lowest order in $\delta \ll 1$.
Hence, $F_{\infty}$ must be independent of $\theta$.
We proceed to expand the distribution function as the following asymptotic series in increasing orders of $\hat \phi$
\begin{align} \label{F-exp}
F  (\vec G) = & F_{\infty}(Y, \mu, v_{\parallel} ) + F_1 (\vec G) + F_2 (\vec G)  + O\left( \hat \phi^3 n_{\infty} v_{\rm t,i}^{-3} \right) \rm .
\end{align}
The assumption that no ions leave the magnetised sheath at $X/l_{\rm ms} \rightarrow \infty$, and therefore cannot have $dX_0/d\tau < 0$, leads to 
\begin{align} \label{F-noreflected}
F_{\infty} \left( Y, \mu, v_{\parallel}  \right) = 0 ~\text{ for }~ v_{\parallel} < -\frac{\phi'_{\infty}(Y)}{B\tan \alpha}  \rm .
\end{align}
Since we exclude trajectories with $dX_0/d\tau <0$ (ions in such trajectories are assumed to be absent in the system), all considered trajectories with $dX_0/d\tau >0$ must have entered the magnetised sheath at $X/l_{\rm ms} \rightarrow \infty$ at $\tau = \tau_{\rm enter} \rightarrow \infty$. 
Equation (\ref{F-solution}) can thus be re-expressed using $F(\vec G) \rvert_{\tau \rightarrow \infty} = F_{\infty} (Y(\tau), \mu(\tau), v_{\parallel}(\tau)) \rvert_{\tau \rightarrow \infty}$ and $F(\vec G) \rvert_{\tau =0} = F(\vec G_{\rm f})$ to obtain 
\begin{align} \label{F-sol-prequel}
F(\vec G_{\rm f}) = F_{\infty}(Y(\tau), \mu(\tau), v_{\parallel}(\tau)) \rvert_{\tau \rightarrow \infty} \rm .
\end{align} 

The relations (\ref{Y-exp})-(\ref{mu-exp}) can be evaluated at $ \tau \rightarrow \infty$ to obtain $\vec G_{\infty}(\vec G_{\rm f}) =  \vec G \rvert_{\tau \rightarrow \infty}$, where the function $\vec G_{\infty}(\vec G)$ can be expanded to
\begin{align} \label{Y-exp-infty}
Y_{\infty}(\vec G)  = Y +  {Y}_{\infty, 1}(\vec G)  +  Y_{\infty, 2}(\vec G)  + O\left(\hat \phi^3 L, \delta L \right) \rm ,
\end{align}
\begin{align} \label{vpar-exp-infty}
v_{\parallel, \infty} (\vec G)  = v_{\parallel} +  {v}_{\parallel,\infty, 1} (\vec G)  +  {v}_{\parallel,\infty, 2} (\vec G)  + O\left(\hat \phi^3 c_{\rm S}, \delta c_{\rm S} \right) \rm ,
\end{align}
\begin{align} \label{mu-exp-infty}
\mu_{\infty} (\vec G)  = \mu +  {\mu}_{\infty,1} (\vec G)  +  {\mu}_{\infty, 2}(\vec G)  + O\left(\hat \phi^3 c_{\rm S} \rho_{\rm S} , \delta c_{\rm S} \rho_{\rm S} \right) \rm ,
\end{align}
and the functions $\vec G_{\infty, n} (\vec G)$ are obtained from the definition $\vec G_{\infty, n} (\vec G_{\rm f}) =  \vec G_n (\tau) \rvert_{\tau \rightarrow \infty}$.
Note that we have replaced $\vec G_{\rm f}$ by $\vec G$ without ambiguity since the functions (\ref{Y-exp-infty})-(\ref{mu-exp-infty}) no longer have any time dependence.
By evaluating (\ref{mu1-bar}) at $\tau \rightarrow \infty$, we obtain
\begin{align} \label{mu1}
\mu_{\infty, 1} (\vec{G}) = \frac{1}{B} \Phi_{\rm pol} (\vec G)  \rm ,
\end{align}
where we defined the polarisation function
\begin{align} \label{Phipol}
& \Phi_{\rm pol}(\vec{G}) = v_{\perp}(\mu) \cos \alpha   \int_0^{\infty} d\tau  \sin ( \theta +   \theta_0(\tau) ) ~ \partial_x \phi (X +  X_0 (\tau) + \rho_x (\mu, \theta + \theta_0 (\tau)), Y)  \rm .
\end{align}
Equation (\ref{vparbar1-Ybar1}) is evaluated at $\tau \rightarrow\infty$ using (\ref{phi-infty}) to obtain
\begin{align} \label{vpar-Y1}
v_{\parallel, \infty, 1} (\vec{G}) & = \Omega Y_{\infty, 1} (\vec{G}) \tan \alpha = \frac{\Omega \left( \phi_1 (X + \rho_x (\mu, \theta), Y ) - \Phi_{\rm pol}(\vec G)  \right) }{B \left( v_{\parallel} + \frac{\phi'_{\infty}(Y)}{B\tan \alpha} \right)} \rm .
\end{align}
We can re-express $\Phi_{\rm pol} ( \vec G )$ in (\ref{Phipol}) in an alternative form in two steps.
First, we use the relation $v_{\perp}(\mu) \cos \alpha \sin ( \theta +  \theta_0(\tau) ) = - d\rho_x(\mu, \theta + \theta_0(\tau))/ d\tau$, which can be deduced from (\ref{rhox-def}) and (\ref{thetadot-0th}), to obtain
\begin{align} \label{Phipol-rearr}
\Phi_{\rm pol} ( \vec G ) = \int_0^{\infty} d\tau \left[  \frac{d X_0}{d\tau} - \frac{d}{d\tau} \left( X_0 (\tau) + \rho_x (\mu, \theta + \theta_0(\tau)) \right) \right] \nonumber \\
\times \partial_x \phi_1 (X + X_0 (\tau) + \rho_x (\mu, \theta +  \theta_0(\tau)), Y) \rm .
\end{align}
Using the fact that the second term in (\ref{Phipol-rearr}) is an exact integral in $\tau$ which evaluates to $\phi_1(x,y)$, we obtain
\begin{align} \label{Phipol-v2}
\Phi_{\rm pol} ( \vec G ) =  \phi_1 (X + \rho_x (\mu, \theta), Y)  +   \int_0^{\infty} d\tau \frac{d X_0}{d\tau} \partial_x \phi_1 (X + X_0 (\tau) + \rho_x (\mu, \theta +  \theta_0(\tau)), Y) \rm .
\end{align}
Note that $\Phi_{\rm pol}$ depends on $v_{\parallel}$ via $dX_0/d\tau$ and $  X_0 (\tau)$ in (\ref{Xdot-0th}) and (\ref{X-bar-0}).

By inserting the notation introduced in (\ref{Y-exp-infty})-(\ref{mu-exp-infty}) into (\ref{F-sol-prequel}), we obtain the distribution function
\begin{align} \label{F-solVlasov}
F(\vec G) & = F_{\infty}(Y_{\infty}(\vec G), \mu_{\infty}(\vec G), v_{\parallel, \infty}(\vec G)) \simeq F_{\infty} (Y, \mu, v_{\parallel} )  \rm .
\end{align} 
The first order correction in the expansion (\ref{F-exp}) is obtained by Taylor expanding (\ref{F-solVlasov}),
\begin{align} \label{F1}
F_1 (\vec G)  = & Y_{1,\infty}(\vec G) \partial_Y F_{\infty} (Y, \mu, v_{\parallel} ) + \mu_{1, \infty} (\vec G)  \partial_{\mu} F_\infty (Y, \mu, v_{\parallel} ) \nonumber \\
& + v_{\parallel, 1, \infty} (\vec G) \partial_{v_{\parallel}} F_{\infty} (Y, \mu, v_{\parallel} )  \rm ,
\end{align} 
which can be written explicitly upon inserting (\ref{mu1}) and (\ref{vpar-Y1}).
The second order correction is calculated in section~\ref{sec-higher}.

\section{Poisson's equation: Bohm-Chodura and polarisation conditions} \label{sec-Poisson}

In this section we analyse Poisson's equation far from the target in the magnetised sheath, and obtain the necessary conditions for a monotonic decay of the potential at $x \rightarrow \infty$ on the sheath scale.
We first analyse the density of electrons (section~\ref{subsec-deriv-ne}) and ions (section \ref{subsec-deriv-ni}) close to the magnetised sheath entrance using an expansion in a small monotonically decaying potential variation, only writing explicitly terms up to first order in $\hat \phi \ll 1$.
Then, we combine these two expressions in Poisson's equation (section~\ref{subsec-deriv-Poisson}) and we derive the conditions required for the assumption of a monotonic decay to hold: the polarisation condition (section~\ref{subsec-deriv-pol}) and the Bohm-Chodura condition (section~\ref{subsec-deriv-Bohm}).
We then investigate these conditions in the cold ion limit in section~\ref{subsec-deriv-cold}, recovering a result from reference \cite{Loizu-2012}.

\subsection{Electron density as a Taylor expansion in potential variation} \label{subsec-deriv-ne}

Although it is not done here, the electron density can be expressed as an integral of the electron distribution function.
Since the magnetised sheath reflects almost all incoming electrons back to the bulk plasma, the electrons are usually assumed to thermalise in the bulk plasma, so that the density is well-approximated by a Boltzmann distribution close to the magnetised sheath entrance.
In this paper, we allow for the possibility of a different functional form of the electron density as a function of the local electrostatic potential arising from the deviation of the electron distribution from a Maxwellian, but we assume that the electron density can be Taylor expanded in small variations of the electrostatic potential,
\begin{align} \label{ne}
n_{\rm e} (x,y) \simeq  n_{\rm e,\infty} (y) + n_{\rm e,1} (x,y)  + n_{\rm e,2} (x,y) \rm ,
\end{align}
with
\begin{align} \label{ne-1-temp}
n_{\text{e}, 1}(x,y) =  \left. \frac{d n_{\rm e}}{d \phi} (x,y) \right\rvert_{x/l_{\rm ms} \rightarrow \infty}  \phi_1(x, y)  \rm , 
\end{align}
\begin{align} \label{ne-2}
n_{\text{e}, 2} (x,y) =  \left. \frac{1}{2} \frac{d^2 n_{\rm e}}{d \phi^2}(x,y)  \right\rvert_{x/l_{\rm ms} \rightarrow \infty} \phi_1^2(x, y) \rm . 
\end{align}
Here $n_{\rm e,\infty} (y) = n_{\rm e} (\infty, y)$ is the electron density at $x / l_{\rm ms} \rightarrow \infty$.
In the particular case of a Boltzmann distribution of electrons, we would have $n_{\text{e},1} = (e\phi_1/T_{\rm e}) n_{\rm e, \infty}$ and $n_{\text{e},2} = \frac{1}{2} (e\phi_1/T_{\rm e})^2 n_{\rm e, \infty}$.
Without a Boltzmann distribution, we can nonetheless define an effective electron temperature
\begin{align} \label{Te-eff}
T_{\rm e} (y) \equiv e n_{\rm e, \infty}\left( \left. \frac{dn_{\rm e}}{d\phi}(x,y) \right\rvert_{x/l_{\rm ms} \rightarrow \infty}  \right)^{-1} \rm 
\end{align}
to write
\begin{align} \label{ne-1}
n_{\text{e}, 1} =  \frac{e \phi_1(x, y) }{T_{\rm e}(y)} n_{\rm e, \infty} (y)  \rm 
\end{align}
without loss of generality.

\subsection{Ion density as an asymptotic series in potential variation} \label{subsec-deriv-ni}  
 
In this section, we consider an expansion of the ion density up to $O(\hat \phi n_{\rm i, \infty})$ which exploits the results
\begin{align} \label{F-nozero}
\left. F_{\infty} \left( Y, \mu, v_{\parallel}  \right) \right\rvert_{v_{\parallel} = -\frac{\phi'_{\infty}(Y)}{B\tan \alpha}} = 0 \rm ,
\end{align} 
 \begin{align} \label{dF-nozero}
\left. \partial_{v_\parallel} F_{\infty} \left( Y, \mu, v_{\parallel} \right) \right\rvert_{v_{\parallel} = -\frac{\phi'_{\infty}(Y)}{B\tan \alpha}} = 0  \rm ,
 \end{align}
 which in turn rely on $F_{\infty}$ being Taylor expandable in its domain. 
 Note that (\ref{F-nozero}) and (\ref{dF-nozero}) together imply that $\partial_Y F_{\infty}$ evaluated at $v_{\parallel} = -\frac{\phi'_{\infty}(Y)}{B\tan \alpha}$ is also equal to zero.
 In practice, these results allow us to exclude ions for which $v_{\parallel} + \frac{\phi'_{\infty}(Y)}{B\tan \alpha} $ is small, as the number of such ions is higher order than $O(\hat \phi n_{\infty})$. 
 More rigorously, an expansion of the ion density and Poisson's equation which accounts for slow ions leads to the requirement that (\ref{F-nozero}) and (\ref{dF-nozero}) must both be satisfied, as shown in \ref{app-slowdens}.
 A case in which these results are satisfied automatically is the cold ion limit, in which the distribution function in $v_{\parallel}$ is a Dirac delta function centred around $v_{\parallel} = u_{\parallel} \geqslant - \frac{\phi'_{\infty}(y)}{B\tan \alpha}$, with $u_{\parallel}$ the fluid velocity.
 
The ion particle density is, by definition, the velocity integral of the distribution function,
\begin{align} \label{ni-def}
n_{\rm i}(x, y) = \int f (x, y, \vec{v}) d^3\vec{v} \rm .
\end{align}
This can be re-expressed in terms of an integral over the full phase space,
\begin{align} \label{ni-def-re}
n_{\rm i}(x, y) = \int_0^{\infty} dx' \int_{-\infty}^{\infty} dy' \int d^3\vec{v} f (x', y', \vec{v}) \delta_{\rm Dirac} (x' - x) \delta_{\rm Dirac} (y' - y) \rm .
\end{align}
The change of variables $(x',y', v_x, v_y, v_z) \rightarrow (X, Y, \mu, \theta, v_{\parallel})$ has a Jacobian
\begin{align} \label{Jac}
\frac{\partial(x',y', v_x, v_y, v_z)}{\partial (X, Y, \mu, \theta, v_{\parallel})} = \Omega \rm . 
\end{align}
Using (\ref{Jac}) to write the density integral (\ref{ni-def-re}) in the new variables, we obtain
\begin{align} \label{ni-def-new}
n_{\rm i}(x, y) =  \int_0^{\infty} dX \int_{-\infty}^{\infty} dY  \int_0^{\infty} \Omega d\mu \int_{0}^{\infty} d v_{\parallel} \int_0^{2\pi} d\theta F (\vec G)  \nonumber \\
\times \delta_{\rm Dirac} (X + \rho_x(\mu, \theta) - x) \delta_{\rm Dirac} (Y  - y) \rm .
\end{align}
Note that we have taken $\rho_y = 0$ in the last Dirac delta function and have thus neglected some terms small in $\delta$ in the density.
In principle the lower limit of integration in the guiding center coordinate $X$ should depend on the other phase space variables, but since we consider very large distances $x \gg \rho_{\rm S}$ from the target, it can be assumed that there are exponentially few particles with large enough gyro-orbits that intersect the wall.
Inserting (\ref{F-exp}) in (\ref{ni-def-new}), we obtain
\begin{align} \label{ni-exp}
n_{\rm i}(x, y) =  \int_0^{\infty} dX \int_{-\infty}^{\infty} dY  \int_0^{\infty} \Omega d\mu \int_{0}^{\infty} d v_{\parallel} \int_0^{2\pi} d\theta ~ \delta_{\rm Dirac} (X + \rho_x(\mu, \theta) - x)  \nonumber \\
\times \delta_{\rm Dirac} (Y-y)  \left[ F_{\infty}(Y, \mu, v_{\parallel}) + F_1 (\vec G) + F_2 (\vec G ) \right] + O(\hat \phi^3 n_{\infty} ) \rm ,
\end{align}
where $n_{\infty}$ is a characteristic value of the ion density at $x/l_{\rm ms} \rightarrow \infty$.

Re-expressing the ion density far from the wall in the magnetised sheath as an asymptotic series in the smallness parameter $\hat \phi$, 
\begin{align} \label{ni-asymptotic}
n_{\rm i}(x, y) = n_{\rm i, \infty}(y) + n_{\text{i},1}(x,y) + n_{\text{i},2}(x,y) + O(\hat \phi^3 n_{\infty}, \delta n_{\infty} ) \rm , 
\end{align}
we promptly identify
\begin{align} \label{ni-infty}
n_{\rm i, \infty} (y) =  2\pi \int_0^{\infty} \Omega d\mu \int_{0}^{\infty} d v_{\parallel} F_{\infty}  \rm 
\end{align}
and 
\begin{align} \label{ni-1}
n_{\text{i}, 1} (x,y) =  \int_0^{\infty} dX \int_0^{\infty} \Omega d\mu \int_{ - \frac{\phi'_{\infty}(y)}{B\tan \alpha}  }^{\infty} d v_{\parallel} \int_0^{2\pi} d\theta  \delta_{\rm Dirac} (x - X - \rho_x(\mu, \theta))  F_{1} (\vec G) \rvert_{Y=y} \rm .
\end{align}
From here on, the arguments of $F_{\infty}$ and of its derivatives will not be shown explicitly, and it will be understood that these functions are evaluated at $Y=y$. 

Inserting (\ref{mu1}), (\ref{vpar-Y1}) and (\ref{F1}) into (\ref{ni-1}) gives an expression for $n_{\text{i,}1}(x,y)$ which only depends on constants, on the function $F_{\infty}(Y, \mu, v_\parallel)$ and on the yet-to-be-determined function $\phi_1(x,y)$,
\begin{align} \label{ni-1-2}
n_{\text{i}, 1} (x,y) = \int_0^{\infty} dX \int_0^{\infty} \Omega d\mu \int_{ - \frac{\phi_{\infty}'(y)}{B\tan \alpha}  }^{\infty} d v_{\parallel} \int_0^{2\pi} d\theta \delta_{\rm Dirac} (X + \rho_x(\mu, \theta) - x) \nonumber \\
\times \left[  \frac{ \phi_1 (X + \rho_x (\mu, \theta), y ) - \Phi_{\rm pol}(\vec G)  }{ B  \tan \alpha \left( v_{\parallel} + \frac{\phi'_{\infty}(y)}{B\tan \alpha} \right)} \left( \partial_Y  + \Omega \tan \alpha \partial_{v_{\parallel}} \right) F_{\infty} \right. \nonumber \\
\left. + \frac{1}{B} \Phi_{\rm pol} (\vec G)   \partial_{\mu} F_\infty  \right] \rm .
\end{align}
After carrying out the integral over $X$ by using the Dirac delta function $\delta_{\rm Dirac}$ and rearranging, we obtain
\begin{align} \label{ni-1st}
n_{\text{i}, 1} (x,y) = 2\pi \int_0^{\infty} \Omega d\mu \int_{ - \frac{\phi'_{\infty}(y)}{B\tan \alpha}   }^{\infty} d v_{\parallel}  \frac{1 }{B} \overline \Phi_{\rm pol} (x, y, \mu, v_{\parallel} )  \nonumber \\
\times   \left[ \partial_{\mu} F_{\infty} - \frac{\Omega}{v_{\parallel} + \frac{\phi'_{\infty}(y)}{B\tan \alpha} } \left( \partial_{v_{\parallel}} F_{\infty} + \frac{\partial_{Y} F_\infty}{\Omega \tan \alpha} \right) \right]   \nonumber \\ 
+  \frac{ \Omega \phi_1 (x, y )}{B} 2\pi \int_0^{\infty} \Omega d\mu \int_{ - \frac{\phi'_{\infty}(y)}{B\tan \alpha}  }^{\infty} d v_{\parallel}  \frac{1}{ v_{\parallel} + \frac{\phi'_{\infty}(y)}{B\tan \alpha}  } \left( \partial_{v_{\parallel}} F_{\infty} + \frac{\partial_{Y} F_\infty}{\Omega \tan \alpha} \right) \rm ,
\end{align}
where we have defined the gyroaveraged polarisation function at fixed $x$,
\begin{align} \label{barphipol-def}
 \overline \Phi_{\rm pol} (x, Y, \mu, v_{\parallel} ) \equiv \langle \Phi_{\rm pol} (\vec G) \rvert_{X=x - \rho_x(\mu, \theta)}  \rangle_{\theta}  \rm .
\end{align}
and the gyroaveraging operation 
\begin{align} \label{gyroav}
\left\langle \ldots \right\rangle_{\theta} \equiv \frac{1}{2\pi} \int_0^{2\pi} (\ldots ) d\theta \rm .
\end{align}

\subsection{Poisson's equation at first order} \label{subsec-deriv-Poisson}

The electrostatic potential in the magnetised sheath is such that Poisson's equation (\ref{Poisson-3D}) is satisfied. 
Considering only the lowest order terms in $\delta$, and thus neglecting $\varepsilon_0 \partial_{y}^2 \phi_1 $ and $\varepsilon_0 \partial_{z}^2 \phi_1$, Poisson's equation is
\begin{align} \label{Poisson}
\varepsilon_0 \partial_x^2 \phi_1(x,y) = en_{\rm e}(x,y) - Zen_{\rm i} (x,y) + O(\delta^2 n_{\rm i, \infty}) \rm .
\end{align}

To lowest order in $\hat{\phi}$, $O(\hat \phi^0 e n_{\infty})$, we obtain 
\begin{align}
Zn_{\text{i}, \infty} = n_{\text{e}, \infty} \rm ,
\end{align}
which is simply quasineutrality at $x \rightarrow \infty$.
The first order correction to Poisson's equation,
\begin{align} \label{Poisson-1}
\varepsilon_0 \partial_x^2 \phi_1(x,y)  = en_{\text{e},1}(x,y) - Zen_{\text{i},1}(x,y)  \rm ,
\end{align}
is, upon inserting (\ref{ne-1}) and (\ref{ni-1st}) into (\ref{Poisson-1}),
\begin{align} \label{Poisson-1-step2}
& - \lambda_{\rm D}^2 \frac{e}{T_{\rm e}} n_{\rm e, \infty}(y) \partial_x^2 \phi_1  + 2\pi Z \int_0^{\infty} \Omega d\mu \int_{ - \frac{\phi'_{\infty}}{B\tan \alpha}  }^{\infty} d v_{\parallel} \frac{1}{B}  \bar \Phi_{\rm pol} (x, y, \mu, v_{\parallel})   \nonumber \\
& \times  \left[  \frac{( \Omega \partial_{v_{\parallel}}  + \cot \alpha ~ \partial_Y ) F_{\infty} }{  v_{\parallel} + \frac{\phi'_{\infty}(y)}{B\tan \alpha} } - \partial_{\mu} F_\infty   \right]   \nonumber \\
& = - \phi_1 \left[ \frac{en_{\rm e,\infty}(y)}{T_{\rm e}(y)}  - \frac{2\pi Z}{B} \int_0^{\infty} \Omega d\mu \int_{ - \frac{\phi'_{\infty}(y)}{B\tan \alpha}  }^{\infty} d v_{\parallel}  \frac{ (\Omega \partial_{v_{\parallel}} + \cot \alpha \partial_{Y} ) F_{\infty}  }{ v_{\parallel} + \frac{\phi'_{\infty}(y)}{B\tan \alpha}   } \right] \rm .
\end{align}
Here, we introduced the Debye length (at the magnetised sheath entrance)
\begin{align}
\lambda_{\rm D}(y) = \left( \frac{T_{\rm e}(y) \varepsilon_0 }{e^2 n_{\rm e, \infty} (y)}  \right)^{1/2} \rm .
\end{align}

Based on equations (\ref{Phipol-v2}), (\ref{barphipol-def}) and (\ref{gyroav}), the gyroaveraged polarisation function $\overline \Phi_{\rm pol}$ appearing on the left hand side of (\ref{Poisson-1-step2}) consists of a linear integro-differential operator acting on $\phi_1$,
\begin{align} \label{Phipol-init}
& \overline \Phi_{\rm pol} ( x, y, \mu, v_{\parallel} ) = \phi_1 (x, y) \nonumber \\
 & + \frac{1}{2\pi} \int_0^{2\pi} d\theta \int_0^{\infty} d\tau \frac{d X_0}{d\tau} \partial_x \phi_1 (x + X_0 (\tau) + \rho_x (\mu, \theta +  \theta_0(\tau)) - \rho_x (\mu, \theta), y)  \rm .
\end{align}
Poisson's equation (\ref{Poisson-1-step2}) with $\overline \Phi_{\rm pol}$ given by (\ref{Phipol-init}) is a homogeneous linear integro-differential equation that is translationally invariant in $x$, 
and whose eigenfunctions therefore have the form \footnote{Ansatz (\ref{phi-ansatz}) is valid barring any multiplicity $M>1$ in the eigenvalues $k(y)$, which would imply additional eigenfunctions of the form $x^m e^{-k(y)x}$ for integer $m \in [1,M-1]$. The form (\ref{phi-ansatz}) is nevertheless sufficient to solve for all values of $k(y)$.}
\footnote{Note that $\partial_y \phi_1(x,y) = A'(y)e^{-k(y)x} - k'(y)x A(y)e^{-k(y)x}$, where the second term is multiplied by $x$ and thus appears to be large at the magnetic presheath entrance, where $x \gg l_{\rm ms}$.
This term nonetheless remains asymptotically small in $\hat \phi$ relative to the $y$-component of the electric field, since $- k'(y)x A(y)e^{-k(y)x} = \frac{k'(y)}{k(y)} \phi_1 \ln \left( \phi_1 / A(y) \right) = O\left( \frac{T_{\rm e}\delta }{e\rho_{\rm s}} \hat \phi \ln \hat \phi \right) \ll \phi_{\infty}' = O\left( \frac{ T_{\rm e} \delta }{e\rho_{\rm S}} \right)$.} 
\begin{align} \label{phi-ansatz}
\phi_1(x, y)  = A(y) e^{-k(y)x} \text{.}
\end{align}
We note that $\phi_1 < 0$ implies $A(y) < 0$.
Inserting (\ref{phi-ansatz}) into Poisson's equation (\ref{Poisson-1-step2}), we obtain the characteristic equation for $k(y)$.
If no real positive solution for $k(y)$ in (\ref{phi-ansatz}) exists, then no solution for $\phi_1(x,y)$ that is \emph{monotonically decaying} in $x$ can be found. \footnote{In principle one cannot exclude complex solutions for $k(y)$ with a positive real part, corresponding to spatially decaying oscillatory solutions. However, these solutions are inconsistent with our ion density derivation because the peaks of a spatially oscillatory potential profile could reflect ions, while we assumed the absence of reflected ions within the sheath.}

We proceed to re-express $\overline \Phi_{\rm pol}$ in (\ref{Phipol-init}) using Ansatz (\ref{phi-ansatz}) in terms of the modified Bessel function of the first kind,
\begin{align} \label{Bessel-definition}
I_0 (\xi) =   \frac{1}{2\pi} \int_{0}^{2\pi} d\theta \exp\left( \xi\sin(\theta + \zeta )\right) \text{,}
\end{align}
where $\zeta$ is an arbitrary constant phase angle.
Inserting (\ref{phi-ansatz}) into (\ref{Phipol-init}), and changing integration variable from $\tau$ to $s= kX_0(\tau) = \left[ k dX_0 / d\tau \right] \tau$ (where we used that $dX_0/d\tau$ is independent of $\tau$, as shown in (\ref{Xdot-0th})), we obtain
\begin{align} \label{M-reexpressed}
\bar \Phi_{\rm pol} =  -\phi_1 \left[  \frac{1}{2\pi}  \int_0^{2\pi} d\theta \int_0^{\infty} ds  e^{- s - k \left[ \rho_x \left( \mu, \theta - \Omega s / \left[ k dX_0 / d\tau \right] \right)  - \rho_x (\mu, \theta) \right]  } - 1 \right] \rm .
\end{align}
Re-expressing (\ref{M-reexpressed}) using $\theta_0(\tau) = \theta - \Omega \tau$ and the useful relation
\begin{align} \label{rhox-trigrelation}
\rho_x (\mu, \theta - \Omega \tau ) - \rho_x (\mu, \theta) = 2 \frac{v_{\perp}(\mu)}{\Omega} \cos \alpha \sin \left( \frac{\Omega \tau }{2}     \right) \sin \left( \frac{\Omega \tau }{2}   - \theta  \right) \rm ,
\end{align}
obtained from (\ref{rhox-def}) using a trigonometric identity, we obtain
\begin{align} \label{M-reexpressed-alt}
\overline \Phi_{\rm pol} =  -\phi_1(x,Y) \left[ \int_0^{\infty}  e^{- s} I_0 \left( \frac{2kv_{\perp}}{\Omega} \cos \alpha  \sin \left[ \frac{\Omega s}{2k \sin \alpha \left( v_{\parallel} + \frac{\phi'_{\infty}(Y)}{B\tan \alpha} \right) } \right] \right) ds - 1 \right] \rm .
\end{align}
For real values of $k(y)$, $\overline \Phi_{\rm pol}$ in (\ref{M-reexpressed-alt}) is positive. 
This follows from observing that $1 = \int_0^{\infty} e^{-s}ds$ and $I_0(\xi) \geqslant 1$ for real argument $\xi$.

In the limit $\alpha \ll 1$, the period $\Delta S =  2\pi k \sin \alpha \left( v_{\parallel} + \frac{\phi'_{\infty}(y)}{B\tan \alpha} \right) / \Omega$ of the oscillatory term in $s$ in (\ref{M-reexpressed}) is small when compared with the unit decay scale of the exponential term, so that the term $e^{-s}$ is approximately constant over the period of the oscillation.
Thus, the oscillating piece of the integrand in $s$ in (\ref{M-reexpressed}) can be replaced with its average
\begin{align}
\frac{1}{\Delta S} \int_{s}^{s+\Delta S} e^{\frac{kv_{\perp}}{\Omega} \cos \alpha  \cos \left( \theta + \frac{2\pi s}{\Delta S} \right)} ds
= I_0 \left( \frac{kv_{\perp}}{\Omega} \cos \alpha \right) \text{.}
\end{align}
Using $\cos \alpha \simeq 1$, the normalised gyroaveraged polarisation function becomes
\begin{align}
\overline \Phi_{\rm pol}(k) \simeq - \phi_1 \left[ I_0 \left(  \frac{kv_{\perp}}{\Omega}   \right) \int_0^{\infty}   e^{-s} ds  \int_0^{2\pi} d\theta    \exp \left( -\frac{kv_{\perp}}{\Omega} \cos \theta   \right)  - 1 \right]    \rm .
\end{align}
The integral in $s$ evaluates to unity, and the integral in $\theta$ gives another Bessel function, leading to
\begin{align}\label{M-smallalpha}
\overline \Phi_{\rm pol}(k) \simeq -\phi_1 \left[ I_0^2 \left(  \frac{kv_{\perp}}{\Omega} \right)  - 1 \right] \rm .
\end{align}
For small $k$, this further becomes
\begin{align}\label{M-smallk}
\overline \Phi_{\rm pol} \simeq  -  \frac{\mu}{\Omega} \cos^2 \alpha ~ k^2 \phi_1 \rm .
\end{align}

\begin{figure}
\centering
\includegraphics[scale=0.8]{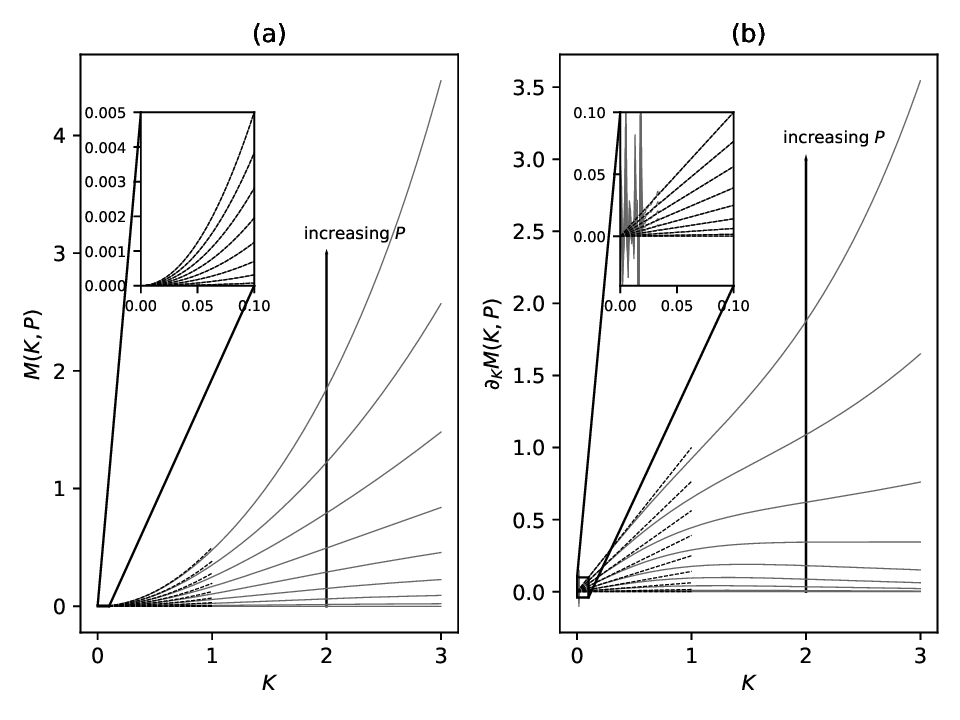}
\caption{\emph{Numerical evaluation} of the functions (a) $M (K, P)$ and (b) $\partial_K M (K, P)$ in (\ref{M-func}) and (\ref{dM-func}) (solid lines) and their analytical small-$K$ limits (\ref{M-func-smallk}) and (\ref{dM-func-smallK}) (dashed lines). Each curve corresponds to a different value of $P$ ranging from $P=0$ to $P=2$ in intervals of $0.25$. The numerical evaluation of $\partial_K M $ is noisy at small values of $K$, as seen in the inset of (b), but the small-$K$ limit accurately represents the function in that portion of the domain.}
\label{fig-mono}
\end{figure}

The gyroaveraged polarisation function in the limit of small positive values of $k$, equation (\ref{M-smallk}), is monotonically increasing in $k$.
We proceed to give strong numerical evidence, supported by the analytical result (\ref{M-smallk}), that $\overline \Phi_{\rm pol}$ is a monotonically increasing function of $k$ for all positive values of $k$, $v_{\perp}$ and $v_{\parallel} + \frac{\phi'_{\infty}(y)}{B\tan \alpha}$.
Consider
\begin{align} \label{M-func}
M (K, P) = \int_0^{\infty}  e^{- s} I_0 \left[ PK  \sin \left( \frac{s}{K} \right) \right] ds - 1 \rm ,
\end{align}
which represents the function $\overline \Phi_{\rm pol} / (-\phi_1)$ in the square brackets of (\ref{M-reexpressed-alt}), after the replacements $K = 2k \sin \alpha \left( v_{\parallel} + \frac{\phi'_{\infty}(Y)}{B\tan \alpha} \right)/\Omega$ and $P = v_{\perp} \cos \alpha / \left[ \sin \alpha \left( v_{\parallel} + \frac{\phi'_{\infty}(Y)}{B\tan \alpha} \right) \right]$.
Its derivative with respect to $K$, while holding $P$ fixed, is
\begin{align} \label{dM-func}
\partial_K M (K, P) = P \int_0^{\infty}  e^{- s} \left[ \sin \left( \frac{s}{K} \right) - \frac{s}{K} \cos \left( \frac{s}{K} \right) \right] I_1 \left[ PK  \sin \left( \frac{s}{K} \right) \right] ds  \rm ,
\end{align} 
where $I_1(\xi) = I_0'(\xi)$ is the modified Bessel function of the first kind.
Since derivatives with respect to $K$ at fixed $P$ are proportional to derivatives with respect to $k$ while keeping $v_{\perp}(\mu)$ and $v_{\parallel} + \frac{\phi'_{\infty}(y)}{B\tan \alpha}$ fixed, we may identify monotonicity in $K$ of the function $M (K, P)$ with monotonicity in $k$ of the function $\overline{\Phi}_{\rm pol}$.
In figure~\ref{fig-mono} we plot a numerical evaluation of (\ref{M-func}) and (\ref{dM-func}) for several values of $P$ in the range $P \in [0, 2]$ as a function of $K \in [0, 3]$, as well as a zoom of the plots in the interval $K \in [0, 0.1]$. 
For small values of $K$, the numerical integration over $s$ \footnote{The integration in $s$ was performed using the scipy.integrate.quad function in Python.} in (\ref{dM-func}) becomes noisy due to the fast oscillation of the integrand in $s$, such that the numerical evaluation of (\ref{dM-func}) has spurious negative values.
We also plot the analytical limits $K \ll 1$ of (\ref{M-func}) and (\ref{dM-func}) which are, from (\ref{M-smallk}),
\begin{align} \label{M-func-smallk}
M (K, P) = \frac{1}{8} P^2 K^2 \rm ,
\end{align}
\begin{align} \label{dM-func-smallK}
\partial_K M (K, P) = \frac{1}{4} P^2 K  \rm .
\end{align}
At small values of $K$, but still large enough that the numerical integration in $s$ is not noisy, (\ref{dM-func-smallK}) overlaps with the numerical evaluation of (\ref{dM-func}).
In the part of the domain where the numerical evaluation of (\ref{dM-func}) is noisy, the plot of (\ref{dM-func-smallK}) is the most accurate evaluation of $\partial_K M $.
With this in mind, the function $ \partial_K M (K, P) $ is deduced to be everywhere positive or zero in the plotted intervals.
It can be seen that for the smaller values of $P$ the function $\partial_K M$ has a maximum and begins to decrease at large values of $K$.
However, taking the limit of large $K \gg 1$ gives $\partial_K M  \simeq \frac{1}{6} K^{-3} P \int_0^{\infty}  e^{- s} s^3  I_1 \left( Ps \right) ds$, which is positive.
Although a formal proof is missing, figure~\ref{fig-mono} is strong evidence that the polarisation function (\ref{M-reexpressed-alt}) is monotonically increasing in $k$.

\subsection{Polarisation condition} \label{subsec-deriv-pol}

We refer to the condition that the left hand side of (\ref{Poisson-1-step2}) be positive for a decaying potential solution as the \emph{polarisation condition}.
The left hand side of (\ref{Poisson-1-step2}) depends on $k(y)$, whereas the right hand side does not.
At normal magnetic field incidence, $\alpha = \pi / 2$, the polarisation function vanishes ($\overline \Phi_{\rm pol} = 0$ for $\cos \alpha = 0$ according to (\ref{Phipol-init})) and $\partial_x^2 \phi_1 < 0$ for a monotonically decaying solution with $\phi_1 < 0$, making the left hand side of (\ref{Poisson-1-step2}) positive and monotonically increasing with $k$.
As a result, for $\alpha = \pi / 2$, (\ref{Poisson-1-step2}) is solvable and gives $k>0$ as long as its right hand side is positive.
For $\alpha \neq \pi/2$, the ion polarisation effects matter.

For a monotonically decaying solution of Poisson's equation (\ref{Poisson-1-step2}), with $\phi_1(x,y) < 0$, we have shown that the gyroaveraged polarisation function $\bar \Phi_{\rm pol}$ is always positive and monotonically increasing with $k$.
Hence, a sufficient (but not necessary) condition for the left hand side of (\ref{Poisson-1-step2}) to be positive and monotonically increasing with $k$ is
\begin{align} \label{polcond-3}
\frac{ \Omega \partial_{v_{\parallel}} F_{\infty} + \cot \alpha ~ \partial_Y F_{\infty} }{  v_{\parallel} + \frac{\phi'_{\infty}(y)}{B\tan \alpha} } - \partial_{\mu} F_\infty  \geqslant 0 \rm 
\end{align}
for all $y$, $\mu$, and $v_{\parallel}$.
The polarisation condition (\ref{polcond-3}) can be re-expressed more concisely as
\begin{align} \label{polcond-4}
(\partial_{\mu} F_{\infty})_{E, Y_{\star}}  \leqslant 0 \rm ,
\end{align}
where the variables 
\begin{align}
E=\frac{1}{2} v_\parallel^2 + \Omega \mu + \frac{\Omega \phi_{\infty}(Y)}{B}  
\end{align} 
and
\begin{align} \label{Ystar}
Y_{\star} = Y - \frac{v_{\parallel}}{\Omega \tan \alpha} 
\end{align}
are held constant in the partial differentiation of (\ref{polcond-4}).
The quantity $Y_{\star}$ is proportional to the canonical momentum of an ion in the direction parallel to the magnetic field.
Equation (\ref{polcond-4}) can be obtained from
\begin{align} \nonumber
(\partial_{\mu} v_{\parallel}) \rvert_{E, Y_{\star}}  = \Omega \tan \alpha (\partial_{\mu} Y )_{E, Y_\star}  =  - \frac{\Omega}{v_{\parallel} +  \frac{\phi'_{\infty}(Y)}{B\tan \alpha}  } \rm ,
\end{align}
which is in turn derived by differentiating the implicit equation 
\begin{align} \nonumber
v_{\parallel}(E, \mu, Y_{\star}) = \sqrt{2\left(E - \Omega \mu - \frac{\Omega}{B} \phi_{\infty}(Y(v_{\parallel}(E, \mu, Y_{\star}), Y_{\star}) \right)} \rm 
\end{align}
with respect to $\mu$ holding $E$ and $Y_{\star}$ fixed.
If condition (\ref{polcond-3}) is not satisfied, the usual condition that the right hand side of (\ref{Poisson-1-step2}) must be positive (see section \ref{subsec-deriv-Bohm}) does not apply.
We leave the study of this situation for future work.

\subsection{Bohm-Chodura condition} \label{subsec-deriv-Bohm}

Supposing that the sufficient polarisation condition (\ref{polcond-3}) is satisfied, and knowing that $\bar \Phi_{\rm pol}$ is positive for $k(y) > 0$, a monotonically decaying solution of Poisson's equation (\ref{Poisson-1-step2}) for the electrostatic potential requires the right hand side of (\ref{Poisson-1-step2}) to also be positive,
 \begin{align} \label{kin-Chod-ymuvpar}
 n_{\rm e\infty}   -   2\pi Z v_{\rm B}^2  \int_0^{\infty} \Omega d\mu  \int_{- \frac{\phi'_{\infty}}{B\tan \alpha} }^{\infty} d v_{\parallel}  \frac{ ( \partial_{v_{\parallel}} + \Omega^{-1} \cot \alpha \partial_{Y} ) F_{\infty} }{ v_{\parallel} + \frac{\phi'_{\infty}(y)}{B\tan \alpha} }   \geqslant 0 \rm .
 \end{align}
We refer to (\ref{kin-Chod-ymuvpar}) as the ``inhomogeneous'' kinetic Bohm-Chodura condition, since it generalises the original Chodura condition (stating that the parallel flow into the target needs to be greater than the sound speed), first derived in reference \cite{Chodura-1982}, by including kinetic physics and small tangential cross-field gradients at small magnetic field angles.
This condition had already been derived in reference \cite{Claassen-Gerhauser-1996b}.
That derivation, however, had been questioned because it made use of charge separation satisfying $\partial_x (n_{\rm e} - Zn_{\rm i}) \geqslant 0$ \cite{Cohen-Ryutov-2004-sheath-boundary-conditions}, and is therefore invalid in the fusion-relevant limit of a quasineutral magnetic presheath.
Indeed, the polarisation condition was absent in reference \cite{Claassen-Gerhauser-1996b}.
Integrating by parts the term containing $\partial_{v_{\parallel}} F_{\infty}$ in (\ref{kin-Chod-ymuvpar}), the boundary terms vanish due to (\ref{F-nozero}) and (\ref{dF-nozero}),
and due to the absence of ions with infinitely high energies, $F_{\infty}\rvert_{v_{\parallel} \rightarrow \infty} = 0$, giving the alternative form
\begin{align} \label{kin-Chod}
Z v_{\rm B}^2 \int \left[ \frac{ f_{\infty} (y, \vec{v}) }{ \left( \vec v\cdot \hat{\vec b} + \frac{\phi'_{\infty}(y)}{B\tan \alpha}  \right)^2 } +   \frac{ \partial_y f_{\infty} (y, \vec{v}) }{ \Omega \tan \alpha  \left(\vec v\cdot \hat{\vec b} + \frac{\phi'_{\infty}(y)}{B\tan \alpha}  \right) } \right] d^3 \vec{v}  \leqslant n_{\rm e,\infty}   \rm .
\end{align}
Here, we denoted $f_{\infty} (y, \vec{v} ) = F_{\infty}(y, \mu, v_{\parallel})$.
  
The Bohm-Chodura condition (\ref{kin-Chod}) can be recast in a coordinate-independent manner as follows.
If $\uvec{n}$ is the unit vector pointing normal to the target and away from the target, and $\uvec{b}$ is the unit vector in the magnetic field direction, (\ref{kin-Chod}) becomes
\begin{align}\label{kin-Chod-coordind-long}
Z v_{\rm B}^2(\vec x) \int \left[ \frac{f_{\infty}(\vec{x}, \vec{v} )}{ \left( \vec{v}\cdot \uvec{b} + \frac{\uvec{n} \times \uvec{b}}{\uvec{n} \cdot \uvec{b}}  \cdot \frac{\nabla \phi(\vec{x})}{B} \right)^2} +  \int  \frac{ \uvec{n} \times \uvec{b}  \cdot \nabla f_{\infty}(\vec{x}, \vec{v}  ) }{ \Omega \uvec{n} \cdot \uvec{b}  \left( \vec{v}\cdot \uvec{b} + \frac{\uvec{n} \times \uvec{b}}{\uvec{n} \cdot \uvec{b}}  \cdot \frac{\nabla \phi(\vec{x})}{B} \right) } \right] d^3 \vec{v}  \leqslant n_{\rm e} \rm .
\end{align}
This form assumes that the electron temperature $T_{\rm e}(\vec x)$ appearing in $v_{\rm B}^2 (\vec x)$ is defined as in (\ref{Te-eff}).
If we denote the gyroaveraged ion velocity as
\begin{align}
\langle \vec{v} \rangle = \vec{v} \cdot \uvec{b} \uvec{b} + \uvec{b} \times \frac{\nabla \phi}{B}
\end{align}
then the Bohm-Chodura condition can be more compactly re-expressed as
\begin{align}\label{kin-Chod-coordind}
Z v_{\rm B}^2(\vec x) \int \left[ \frac{ f_{\rm i, \infty}(\vec{x}, \vec{v} ) \left( \uvec{n} \cdot \uvec{b} \right)^2}{ \left( \langle \vec{v} \rangle \cdot \uvec{n} \right)^2} +  \int  \frac{ \uvec{n} \times \uvec{b}  \cdot \nabla f_{\rm i, \infty}(\vec{x}, \vec{v}  )  }{ \Omega \langle \vec{v} \rangle \cdot \uvec{n} } \right] d^3 \vec{v}  \leqslant  n_{\rm e}  \rm .
\end{align}

A global (weak) form of the inhomogeneous kinetic Chodura condition can be derived by integrating (\ref{kin-Chod-ymuvpar}) over the whole spatial domain in $y=Y$ and changing variables from $(Y, v_{\parallel})$ to $(Y_{\star}, w_{\parallel})$ with $Y_{\star}$ defined in (\ref{Ystar}) and
\begin{align} \label{wpar}
w_{\parallel} = v_{\parallel} + \frac{\phi'_{\infty}(Y)}{B \tan \alpha} \rm .
\end{align}
Using the relations (\ref{Ystar}) and (\ref{wpar}), we obtain the Jacobian
\begin{align} \label{jac-global}
\left| \frac{\partial (Y_{\rm \star}, w_{\parallel})}{ \partial (Y, v_{\parallel}) } \right| =   1 + \frac{\phi''_{\infty}(Y)}{\Omega B \tan^2 \alpha}  \rm .
\end{align}
Using the chain rule, we obtain
\begin{align} \label{app-global-1}
\frac{\partial_Y F_{\infty}}{\Omega \tan \alpha}  = \frac{\phi''_{\infty}(Y)}{\Omega B \tan^2 \alpha} (\partial_{w_{\parallel}} F_{\infty})_{Y_{\star}} +  \frac{(\partial_{Y_{\star}} F_{\infty})_{w_{\parallel}}}{\Omega \tan \alpha}  \rm ,
\end{align}
\begin{align} \label{app-global-2}
\partial_{v_{\parallel}} F_{\infty} =  (\partial_{w_{\parallel}} F_{\infty})_{Y_{\star}} - \frac{(\partial_{Y_{\star}} F_{\infty})_{w_{\parallel}}}{\Omega \tan \alpha}  \rm .
\end{align}
We denote explicitly when we hold a variable that is neither $v_{\parallel}$, $Y$ or $\mu$ constant when taking a partial derivative of $F_{\infty}$, but we have assumed it to be understood that $\partial_{v_{\parallel}} F_{\infty} \equiv (\partial_{v_{\parallel}} F_{\infty})_{Y}$ and $\partial_{Y} F_{\infty} \equiv (\partial_{Y} F_{\infty})_{v_{\parallel}}$.
Adding (\ref{app-global-1}) and (\ref{app-global-2}) results in
\begin{align} \label{change-global}
\partial_{v_{\parallel}} F_{\infty} + \frac{\partial_Y F_{\infty}}{\Omega \tan \alpha} = \left( 1 + \frac{\phi''_{\infty}(Y)}{\Omega B \tan^2 \alpha} \right) (\partial_{w_{\parallel}} F_{\infty})_{Y_{\star}} \rm .
\end{align}
Dividing the local Bohm-Chodura condition (\ref{kin-Chod-ymuvpar}) through by $v_{\rm B}^2 (y)$ and integrating over the whole spatial domain gives (recalling that $y \simeq Y$),
 \begin{align} \label{kin-Chod-ymuvpar-global-1}
 2\pi Z \int_{-\infty}^{\infty} dY \int_0^{\infty} \Omega d\mu  \int_{- \frac{\phi'_{\infty}}{B\tan \alpha} }^{\infty} d v_{\parallel} \frac{ \Omega \partial_{v_{\parallel}} F_{\infty} + \cot \alpha \partial_{Y} F_{\infty} }{\Omega \left( v_{\parallel} + \frac{\phi'_{\infty}}{B\tan \alpha} \right) }  \leqslant \int_{-\infty}^{\infty} dy \frac{n_{\rm e\infty}}{v_{\rm B}^2}  \rm .
 \end{align}
By changing variables from $(Y, v_{\parallel})$ to $(Y_{\star}, w_{\parallel})$, using (\ref{jac-global}) and (\ref{change-global}), integrating by parts in $w_{\parallel}$, and invoking again (\ref{F-nozero}) and (\ref{dF-nozero}),
 \begin{align} \label{kin-Chod-global}
2\pi Z \int_{-\infty}^{\infty} dY_{\star} \int_0^{\infty} \Omega d\mu  \int_{0}^{\infty} d w_{\parallel}  \frac{ F_{\infty}}{w_{\parallel}^2}  \leqslant  \int_{-\infty}^{\infty} dy \frac{ n_{\rm e\infty}(y)}{v_{\rm B}^2(y)} \rm ,
 \end{align}
 where $F_{\infty} = F_{\infty} (Y(Y_{\star}, w_{\parallel}), \mu, v_\parallel(Y_{\star}, w_{\parallel}))$.
 
We proceed to analyse the Bohm-Chodura condition (\ref{kin-Chod}) in some simplified limits.
When $\alpha = 90^{\circ}$, such that $\cos \alpha = 0 = \cot \alpha$, there is no polarisation condition and the kinetic Chodura condition in (\ref{kin-Chod}) recovers the conventional kinetic Bohm condition \cite{Harrison-Thompson-1959, Riemann-review} upon realising that $v_{\parallel} = v_x$,
    \begin{align} \label{kin-Bohm}
Z v_{\rm B}^2 \int  \frac{ f_{\infty}}{ v_x^2 } d^3 \vec{v}  \leqslant n_{\rm e\infty}  \rm .
  \end{align}
In this limit, the only length scale, coming from the Laplacian in Poisson's equation (\ref{Poisson}) is $\lambda_{\rm D}$: the magnetised sheath is analogous to the unmagnetised Debye sheath, with a thickness set by the Debye length $l_{\rm ms} \sim \lambda_{\rm D}$.  
If the $y$ dependences linked to tangential fluctuations are negligible, that is, $\delta \ll \alpha$, the kinetic Bohm-Chodura condition (\ref{kin-Chod}) becomes
 \begin{align} \label{kin-Chod-noy}
2\pi Z v_{\rm B}^2  \int_0^{\infty} \Omega d\mu \int_0^{\infty} dv_{\parallel} \frac{ F_{\infty}}{ v_{\parallel}^2 } = Z v_{\rm B}^2  \int d^3 v \frac{ f_{\infty}}{ |\vec{v} \cdot \uvec{b}|^2 }  \leqslant  n_{\rm e,\infty}(y) \rm .
  \end{align}
  In the limit $\delta \ll \alpha$, we do not need the sufficient polarisation condition (\ref{polcond-3}) to ensure that the left hand side of (\ref{Poisson-1-step2}) is positive.
  Hence, condition (\ref{kin-Chod-noy}) is enough to guarantee a positive solution for $k(y)$ and the presence of a monotonically decaying potential profile (\ref{phi-ansatz}) at large values of $x$ in the magnetised sheath.

\subsection{Cold ion limit} \label{subsec-deriv-cold}

 In the cold ion limit, the ion distribution function at the magnetised sheath entrance tends to a Dirac delta function at $\mu = 0$ and $v_{\parallel} = u_{\parallel}(y)$,
 \begin{align} \label{F-cold}
 F_{\infty}(Y, \mu, v_{\parallel}) = \frac{n_{e, \infty}(Y)}{Z} \frac{1}{2\pi \Omega} \delta_{\rm Dirac}(\mu) \delta_{\rm Dirac} \left( v_{\parallel} - u_{\parallel}(Y) \right) \rm .
 \end{align}
 The gyroaveraged polarisation function $\overline \Phi_{\rm pol}$ in (\ref{Poisson-1-step2}) thus tends to its value at $\mu = 0 = v_{\perp}(\mu)$, which is equal to zero.
 Hence, the integrals of $\partial_{v_{\parallel}} F_{\infty}$ and $\partial_{Y} F_{\infty}$ on the left hand side of (\ref{Poisson-1-step2}) which multiply $\Phi_{\rm pol}$ give zero contribution to the integral.
In the last term on the left hand side of (\ref{Poisson-1-step2}), $\bar \Phi_{\rm pol}$ is multiplied by the partial derivative of the distribution function $F_{\infty}$ with respect to $\mu$, which diverges in the cold ion limit \emph{precisely} at $\mu = 0$.
To establish whether the polarisation condition, i.e. the left hand side of (\ref{Poisson-1-step2}) being positive, is satisfied, we expand the function $\overline \Phi_{\rm pol}$ for small $\mu$ to calculate the non-zero contribution from the term containing $\partial_\mu F_{\infty}$.
Expanding (\ref{M-reexpressed-alt}) in small $v_{\perp}$ using $I_0(\xi) \simeq 1 + \xi^2 / 4 + O(\xi^4)$ for $\xi \ll 1$, we obtain
\begin{align}\label{M-smallmu-1}
\overline \Phi_{\rm pol} = -\phi_1 (x,y) \left( \frac{kv_{\perp}}{\Omega} \cos \alpha \right)^2 \int_0^{\infty}  e^{- s}  \sin^2 \left( \frac{\Omega s}{2k dX_0/d\tau } \right)  ds + O\left( k^{4} \frac{\mu^2}{ \Omega^{2}} | \phi_1 | \right) \rm .
\end{align}
Integrating (\ref{M-smallmu-1}) by parts twice results in
\begin{align}\label{M-smallmu-2}
\overline \Phi_{\rm pol} = -\phi_1 (x,y)  \frac{ \frac{1}{2}  k^2 v_{\perp}^2(\mu) \cos^2 \alpha }{1 + \frac{k^2 \sin^2 \alpha}{\Omega^2} \left( v_{\parallel} + \frac{\phi'_{\infty}}{B\tan \alpha} \right)^2  } + O\left( k^{4} \frac{\mu^2}{ \Omega^{2}} | \phi_1 | \right) \rm .
\end{align}
The term on the left hand side of (\ref{Poisson-1-step2}) involving $\partial_{\mu} F_{\infty}$ can be integrated by parts in $\mu$ using the result (\ref{M-smallmu-2}).
First we exchange the order of integration to carry out the integration in $\mu$ before the one in $v_{\parallel}$, then the integral in $\mu$ becomes, using $v_{\perp}^2 (\mu) = 2\Omega \mu$,
\begin{align} \label{polcold-satis}
 - 2\pi Z \int_0^{\infty} \Omega d\mu \int_{-\frac{\phi'_{\infty}}{B\tan \alpha}}^{\infty} dv_{\parallel} & \frac{\bar \Phi_{\rm pol}}{B} \partial_{\mu} F_{\infty} \nonumber \\ 
 & =  \frac{\Omega \phi_1}{B}  \frac{  k^2  \cos^2 \alpha 2\pi Z \int_{-\frac{\phi'_{\infty}}{B\tan \alpha}}^{\infty} dv_{\parallel} \int_0^{\infty} \Omega d\mu   \mu  \partial_{\mu} F_{\infty} }{\Omega^2 + k^2 \sin^2 \alpha \left( u_{\parallel} + \frac{\phi'_{\infty}}{B\tan \alpha} \right)^2  }  \nonumber \\
  & = - \frac{e \phi_1}{T_{\rm e}}  \frac{ k^2  \cos^2 \alpha ~ \rho_{\rm B}^2  n_{\rm e, \infty} }{1 + \frac{k^2 \sin^2 \alpha}{\Omega^2} \left( u_{\parallel} + \frac{\phi'_{\infty}}{B\tan \alpha} \right)^2  } \rm .
\end{align}
This is the only non-zero term coming from $\overline \Phi_{\rm pol}$ on the left hand side of (\ref{Poisson-1-step2}) because the powers of $\mu^n$ with $n>1$ that were neglected in (\ref{M-smallmu-1}) will lead to terms with $\int d\mu \mu^{n-1} \delta_{\rm Dirac} (\mu)$ after integrating by parts, which vanish for $n>1$.
Since (\ref{polcold-satis}) is manifestly positive, the polarisation condition is automatically satisfied in this cold ion limit.

With the polarisation condition automatically satisfied, we require the Bohm-Chodura condition to be satisfied in order to have monotonically decaying potential solutions in the sheath.
Inserting (\ref{F-cold}) into the right hand side of (\ref{Poisson-1-step2}) and using
  \begin{align} \label{ddy-cold}
  \frac{ \partial_Y F_{\infty} }{\left( v_{\parallel} + \frac{\phi_{\infty}'}{B\tan \alpha} \right)} =  \partial_Y \left(  \frac{ F_{\infty} }{ \left( v_{\parallel} + \frac{ \phi_{\infty}'}{B\tan \alpha} \right)} \right) +  \frac{ F_{\rm \infty}  }{\left( v_{\parallel} + \frac{\phi_{\infty}'}{B\tan \alpha} \right)^2} \frac{\phi_{\infty}''}{ B\tan \alpha}   \rm 
\end{align}
gives, upon bringing the total derivative with respect to $y$ in the first term of (\ref{ddy-cold}) outside of the integrals and then performing all integrals over the Dirac delta functions
\begin{align} \label{RHScold}
- \frac{e\phi_1}{T_{\rm e}} \left[ n_{\rm e,\infty} - 2\pi Z v_{\rm B}^2 \int_0^{\infty} \Omega d\mu \int_{ - \frac{\phi'_{\infty}(y)}{B\tan \alpha}  }^{\infty} d v_{\parallel}  \frac{ (\partial_{v_{\parallel}} + \Omega^{-1}\cot \alpha \partial_{Y} ) F_{\infty}  }{ v_{\parallel} + \frac{\phi'_{\infty}(y)}{B\tan \alpha}   } \right] \nonumber \\
 =  - \frac{e\phi_1}{T_{\rm e}} \left[ n_{\rm e, \infty} - \frac{ v_{\rm B}^2  n_{\rm e, \infty} }{(u_{\parallel} + \frac{\phi_{\infty}'}{B\tan \alpha})^2}  +  \frac{v_{\rm B}^2 }{\Omega \tan \alpha} \frac{d}{dy} \left( \frac{  n_{\rm e,\infty}}{  u_{\parallel} + \frac{\phi_{\infty}'}{B\tan \alpha} } \right) \right. \nonumber \\
 \left. +  \frac{  n_{\rm e \infty}  v_{\rm B}^2  }{   \left( u_{\parallel}  + \frac{\phi_{\infty}'}{B\tan \alpha} \right)^2} \frac{\phi_{\infty}''}{ \Omega B\tan^2 \alpha} \right] \rm .
\end{align}
The Bohm-Chodura condition (\ref{kin-Chod}) constrains (\ref{RHScold}) to be positive, such that
\begin{align}
  \frac{ v_{\rm B}^2  n_{\rm e, \infty} }{(u_{\parallel} + \frac{\phi_{\infty}'}{B\tan \alpha})^2}  +  \frac{v_{\rm B}^2 }{\Omega \tan \alpha} \frac{d}{dy} \left( \frac{  n_{\rm e,\infty}}{  u_{\parallel} + \frac{\phi_{\infty}'}{B\tan \alpha} } \right) +  \frac{  n_{\rm e \infty}  v_{\rm B}^2  }{   \left( u_{\parallel} + \frac{\phi_{\infty}'}{B\tan \alpha} \right)^2} \frac{\phi_{\infty}''}{ \Omega B\tan^2 \alpha} \leqslant n_{\rm e, \infty} \rm .
\end{align}
Explicitly evaluating the derivative with respect to $y$ in the second term and then dividing through by $n_{\rm e, \infty}(y)$ gives
\begin{align}
\frac{v_{\rm B}^2  (\ln n_{\rm e,\infty})'}{\Omega \tan \alpha \left( u_{\parallel} + \frac{\phi_{\infty}'}{B\tan \alpha} \right)}   - \frac{ v_{\rm B}^2 }{\Omega \tan \alpha} \frac{  u_{\parallel}'   - \Omega \tan \alpha }{\left( u_{\parallel} + \frac{\phi_{\infty}'}{B\tan \alpha} \right)^2} \leqslant 1 \rm .
\end{align}
Upon multiplying by $\left( u_{\parallel} + \frac{\phi_{\infty}'}{B\tan \alpha} \right)^2$ and rearranging, we obtain the cold ion limit of the Bohm-Chodura condition (\ref{kin-Chod}) in the presence of tangential gradients,
 \begin{align} \label{fluidChodura-Loizu}
 \left( u_{\parallel} + \frac{\phi_{\infty}'}{B\tan \alpha} \right)^2  -   v_{\rm B}^2 \frac{(\ln n_{\rm e,\infty})'}{\Omega \tan \alpha}  \left( u_{\parallel} +  \frac{\phi_{\infty}'}{B\tan \alpha} \right)  + v_{\rm B}^2 \left( \frac{ u_{\parallel}' }{\Omega \tan \alpha} -1 \right)   \geqslant 0  \rm .
\end{align}
Recall that $\phi_{\infty}$, $u_{\parallel}$, $n_{\rm e, \infty}$ and $v_{\rm B}$ are all functions of $y$.
Note that the marginal (equality) form of (\ref{fluidChodura-Loizu}) recovers equation (12) of reference \cite{Loizu-2012} upon solving for positive $u_{\parallel} + \phi_{\infty}' /(B\tan \alpha)$ (recall $dX_0/d\tau < 0$), 
\begin{align} \label{fluidLoizu}
u_{\parallel} + \frac{\phi'_{\infty}}{B\tan \alpha} = v_{\rm B} \left[ \frac{v_{\rm B} (\ln n_{\rm e \infty})'}{2\Omega \tan \alpha } + \sqrt{ 1 + \left(  \frac{v_{\rm B} (\ln n_{\rm e \infty})'}{2\Omega \tan \alpha } \right)^2 - \frac{u_{\parallel}'}{\Omega \tan \alpha} } \right] \rm .
\end{align}
[To convert our notation to the one in reference \cite{Loizu-2012}, one can use the replacements $u_{\parallel} \sin \alpha + \frac{\phi'_{\infty}}{B} \cos \alpha \rightarrow v_{\rm si}$, $y \rightarrow \rho_{\rm B} x$ such that $\xi'(y) \rightarrow \rho_{\rm B}^{-1} \partial_x \xi$ for any $\xi$, $v_{\rm B} \rightarrow c_{\rm s}$.]
If $y$ dependences are negligible, (\ref{fluidLoizu}) recovers the standard marginal fluid Chodura condition $u_{\parallel} = v_{\rm B} = \sqrt{ZT_{\rm e}/m_{\rm i}}$.
Of course, (\ref{fluidLoizu}) is a differential equation for $u_{\parallel}$ despite the way it is written. 

We remark that 
equation (\ref{polcold-satis}) has an upper bound at $k \rightarrow \infty$, equal to 
\begin{align} \label{LHScold-upper}
- \frac{e \phi_1}{T_{\rm e}}  \frac{v_{\rm B}^2  n_{\rm e, \infty} }{\tan^2 \alpha \left( u_{\parallel} + \frac{\phi'_{\infty}}{B\tan \alpha} \right)^2  } = O\left( \hat \phi \frac{n_{\rm e, \infty}}{\tan^2 \alpha} \right) \rm .
\end{align}
If $\lambda_{\rm D} \ll \rho_{\rm B}$, the first term on the left hand side of (\ref{Poisson-1-step2}) should be negligible at the magnetised sheath entrance, due to the expected ordering $k(y) = O(\rho_{\rm B}^{-1}) $. 
A positive solution for $k(y)$ satisfying this ordering, found by equating (\ref{polcold-satis}) and (\ref{RHScold}), can only exist if (\ref{RHScold}) is smaller than (\ref{LHScold-upper}).
At small $\alpha$, (\ref{RHScold}) is constrained to be of order $O(\hat \phi n_{\rm e, \infty})$ by the ordering $u_{\parallel} + \frac{\phi'_{\infty}}{B\tan \alpha} = O(v_{\rm B})$, where we have used that $c_{\rm S} = v_{\rm B}$ for cold ions.
Since (\ref{LHScold-upper}) is of order $O(\alpha^{-2} \hat \phi n_{\rm e, \infty})$, a positive solution for $k(y)$ is guaranteed to exist if the Bohm-Chodura condition (\ref{fluidChodura-Loizu}) is satisfied and the assumed orderings hold.
In the limit $\delta \ll \alpha \sim 1$, setting (\ref{RHScold}) to be smaller than (\ref{LHScold-upper}) leads to the inequality $u_{\parallel} \sin \alpha < v_{\rm B}$.
If $u_{\parallel} \sin \alpha \geqslant v_{\rm B}$, no solution $k = O(\rho_{\rm B}^{-1}) $ exists: indeed, we expect no electrostatic potential variation on the magnetic presheath scale $\rho_{\rm B}$, since $u_x = - u_{\parallel} \sin \alpha $ implies that the cold-ion Bohm condition $|u_x| \geqslant v_{\rm B}$, required at the Debye sheath entrance, is satisfied. 
A positive solution for $k = O ( \lambda_{\rm D}^{-1} )$ exists due to the first term on the left hand side of (\ref{Poisson-1-step2}).
The magnetic presheath is unnecessary and does not form when $u_{\parallel} \sin \alpha \geqslant v_{\rm B}$.

\section{Marginally satisfied Bohm-Chodura condition: higher order analysis} \label{sec-higher}

Since this section contains extensive calculations that lead to some secondary results, we suggest skipping it during a first read and jumping straight to the conclusions instead.

In this section, we assume that the Bohm-Chodura condition (\ref{kin-Chod-ymuvpar}) is marginally satisfied,
   \begin{align} \label{kin-Chod=}
 2\pi Z v_{\rm B}^2  \int_0^{\infty} \Omega d\mu  \int_{- \frac{\phi'_{\infty}}{B\tan \alpha} }^{\infty} d v_{\parallel}  \frac{ \Omega \partial_{v_{\parallel}} F_{\infty} + \cot \alpha \partial_{Y} F_{\infty} }{ \Omega \left( v_{\parallel} + \frac{\phi'_{\infty}(y)}{B\tan \alpha}  \right) }  =  n_{\rm e\infty}  \rm ,
 \end{align}
 which makes the right hand side of (\ref{Poisson-1-step2}) zero.
For solutions $\phi_1$ of the form (\ref{phi-ansatz}), this case corresponds to $k = 0$.
For $\lambda_{\rm D} / \rho_{\rm S} \rightarrow 0$, when the deviation from quasineutrality at the magnetised sheath entrance is negligible, the mathematical alternative that (\ref{polcond-3}) is precisely an equality for all $y$, $\mu$ and $v_{\parallel}$ gives
  \begin{align} \label{kin-Chod=-pol=-impossible}
\int_0^{\infty} \frac{\Omega d\mu}{ v_{\parallel} + \frac{\phi'_{\infty}}{B\tan \alpha} } \left[  \partial_{ v_{\parallel}} F_{\infty} + \frac{   \partial_{Y} F_{\infty}}{ \Omega \tan \alpha }  \right]  = \int_0^{\infty} \partial_{\mu} F_{\rm i\infty}  d\mu = - F_{\infty}\rvert_{\mu=0} \leqslant 0  \rm ,
  \end{align}
which, upon integrating in $v_\parallel$, is manifestly incompatible with (\ref{kin-Chod=}).
Poisson's equation to order $\sim \hat \phi n_{\infty}$ no longer contains information about the potential variation. 
From (\ref{phi-ansatz}), $k \rightarrow 0$ is seen to correspond to $\epsilon = l_{\rm ms} \partial_x \ln \hat \phi \rightarrow 0$, where $\epsilon$ was defined in (\ref{ordering-1}).
This motivates a subsidiary expansion in $\epsilon \ll 1$ when (\ref{kin-Chod=}) is satisfied.
By Taylor expanding (\ref{Phipol-init}) for small $\rho_x$, or alternatively by making the replacement $k \rightarrow \partial_x$ in (\ref{M-smallk}), we obtain
\begin{align} \label{Phipol-smallepsilon}
& \bar \Phi_{\rm pol} = - \frac{\mu}{\Omega} \cos^2 \alpha ~ \partial_x^{2} \phi_1 \left( x , Y \right) + O(\epsilon^4 \phi_1 ) \text{ if } \epsilon \ll 1 \rm .
\end{align}
The left hand side of Poisson's equation (\ref{Poisson-1-step2}) is $\sim \epsilon^2 \hat \phi n_{\infty}$ and can therefore balance with higher order terms in $\hat \phi$ that have been neglected so far.
Since $\epsilon \ll 1$, the electrostatic potential variation can be obtained by calculating the terms $\sim \hat \phi^2 n_{\infty}$ in Poisson's equation to lowest order in $\epsilon$.
The subsidiary expansion in $\epsilon$ of such terms essentially involves neglecting all appearances of $\rho_x$ in the argument of the electrostatic potential, and is thus akin to drift-kinetics.

Here, we assume that the distribution function at $v_{\parallel} = -\frac{\phi'_{\infty}(Y)}{B\tan \alpha}$ satisfies
 \begin{align} \label{d2F-nozero}
\left. \partial_{v_\parallel}^2  F_{\infty} \left( Y, \mu, v_{\parallel} \right) \right\rvert_{v_{\parallel} = -\frac{\phi'_{\infty}(Y)}{B\tan \alpha}} = 0  \rm ,
 \end{align}
  \begin{align} \label{d3F-nozero}
\left. \partial_{v_\parallel}^3  F_{\infty} \left( Y, \mu, v_{\parallel} \right) \right\rvert_{v_{\parallel} = -\frac{\phi'_{\infty}(Y)}{B\tan \alpha}} = 0 \rm ,
 \end{align}
 so that any divergences appearing in the integrand of the higher order expansion of the ion density at $v_{\parallel} + \frac{\phi'_{\infty}}{B\tan \alpha} \rightarrow 0$ are always integrable.
In \ref{subapp-slow-d2F}, we analyse Poisson's equation at higher order when (\ref{d2F-nozero}) is \emph{not} satisfied.
In this case, the $O(\epsilon^2 \hat \phi)$ term in Poisson's equation must balance with a fractional higher order term not considered here, $O(\hat \phi^{3/2})$, giving $\epsilon \sim \hat \phi^{1/4}$.
In \ref{subapp-slow-d2F} we also analyse Poisson's equation at higher order if (\ref{d2F-nozero}) is satisfied but (\ref{d3F-nozero}) is not, thus including terms of order $O(\hat \phi^{2} \ln (1/\hat \phi))$ which are excluded here.

The rest of this section is structured as follows.
In section~\ref{subsec-higher-DKvpar1}, we expand ion trajectories to higher order in $\hat \phi$, calculating the drift-kinetic ($\sim \epsilon^0$) first order corrections to the fast variables $X$ and $\theta$, and the drift-kinetic second order corrections to the slow variables $Y$, $\mu$ and $v_{\parallel}$. 
Then, in section~\ref{subsec-higher-ni2}, we use the ion trajectory corrections to calculate the $O(\hat \phi^2 n_{\infty})$ correction to the ion density.
In section~\ref{subsec-higher-Poisson}, we collect all higher order terms in Poisson's equation corresponding to the marginal Bohm-Chodura condition (\ref{kin-Chod=}).
In section~\ref{subsec-higher-pol} we derive a necessary polarisation condition, while in section~\ref{subsec-higher-addsheath} we deduce an additional sheath condition required for a monotonic potential spatial decay.

\subsection{Drift-kinetic calculation of the ion trajectory up to second order in $\hat \phi$} \label{subsec-higher-DKvpar1}

Here we consider again the past ion trajectories followed backwards in time $t$ from $t=-\tau = 0$ and $\vec G = \vec G_{\rm f}$ to positive values of $\tau$. 
Recall the expansion in $\hat \phi$ of the time dependence of the variables $\vec G = (X, Y, \mu, \theta, v_\parallel)$ initiated in section~\ref{subsec-traj-perturb}.
We introduce a subsidiary expansion in $\epsilon \ll 1$ of the first and second order terms in that expansion, 
\begin{align} \label{Gn-DK}
\vec G_n (\tau) = \vec G_{n,0}(\tau) + O(\hat \phi^n \epsilon \vec G) \rm ,
\end{align}
with $n =1$ or $n=2$.
In this section, we calculate $v_{\parallel, 2,0}(\tau)$, $Y_{2,0}(\tau)$ and $\mu_{2,0}(\tau)$.
To do so, however, we first calculate $\vec G_{1,0}(\tau)$ for all variables.

Expanding in $\epsilon$ amounts to Taylor expanding the electrostatic potential around the periodic displacements $\rho_x = O( \rho_{\rm S} )$.
Neglecting $\rho_x$ in (\ref{mu1-bar}) thus gives the lowest order term in the $\epsilon$-expansion of $ \mu_1$,
\begin{align} \label{mu10-before}
 \mu_{1,0}(\tau) = \frac{1}{B} v_{\perp}(\mu_{\rm f}) \cos \alpha \int_0^{\tau} d\tau'  \sin (\theta_{\rm f} +  \theta_0 (\tau'))  \partial_x \phi (X_{\rm f} +  X_0 (\tau') , Y_{\rm f}) \rm .
\end{align} 
From $ \theta_0 (\tau') = - \Omega \tau'$, the oscillatory term has a period $\Delta \tau' = \frac{2\pi}{\Omega}$.
Over this time, the change in the term $\partial_x \phi$ is small in $\epsilon$, $ \left[ X_0 (\tau' + \Delta \tau') - X_0 (\tau' ) \right] \partial_x^2 \phi (X_{\rm f} +  X_0(\tau') , Y_{\rm f}) \sim \epsilon \sin \alpha \partial_x \phi (X_{\rm f} +  X_0(\tau')  , Y) $, with the displacement $X_0 (\tau' + \Delta \tau') - X_0 (\tau' ) = \frac{2\pi}{\Omega} \sin \alpha \left( v_{\parallel} + \frac{\phi_{\infty}'(Y_{\rm f})}{B\tan\alpha} \right) = O(\rho_{\rm S} \sin \alpha )$ obtained using (\ref{Xdot-0th}).
Thus, $\partial_x \phi$ is constant to lowest order in $\epsilon$ over the period $\Delta  \tau' $ and we can replace the sinusoidal term by its average over $\Delta \tau'$, which is zero,
\begin{align} \label{mu10}
 \mu_{1,0}(\tau) = 0 \rm .
\end{align}
Similarly, the first term in the $\epsilon$ expansion of the variables $ v_{\parallel}$ and $ Y$ is obtained by setting $\rho_x = 0$ in equation (\ref{vparbar1-Ybar1}), giving
\begin{align} \label{vpar1-Y1-epsilon}
 v_{\parallel, 1, 0}(\tau) = \Omega  Y_{1,0}(\tau) \tan \alpha = & \frac{\Omega \left[  \phi_1 (X_{\rm f} , Y_{\rm f} )  - \phi_1 (X_{\rm f} +  X_0 (\tau), Y_{\rm f})) \right] }{B \left( v_{\parallel, \rm f} +  \frac{\phi'_{\infty}(Y_{\rm f})}{B\tan \alpha}  \right) } \rm .
\end{align}

We proceed to calculate the first order corrections $X_1(\tau)$ and $\theta_1(\tau)$ of the fast variables $\boldsymbol{\gamma} = (X, \theta)$, which were not calculated in section~\ref{sec-traj}, to lowest order in $\epsilon \ll 1$, thus calculating $X_{1,0}(\tau)$ and $\theta_{1,0}(\tau)$.
We use (\ref{Gfast1dot}) to calculate the first order derivatives and then set $\rho_x = 0$ to take the drift-kinetic limit.
We obtain
\begin{align} \label{thetadot10}
\frac{d \theta_{1,0} }{d\tau} = \frac{\Omega }{2v_{\perp} (\mu_{\rm f})} \cos \alpha  \cos ( \theta_{\rm f} +  \theta_0 (\tau) ) \frac{1}{B} \partial_x \phi (X_{\rm f} +  X_0 (\tau), Y_{\rm f})  \rm 
\end{align}
from (\ref{thetadot-long-3}) and (\ref{thetadot-0th}), and
\begin{align} \label{Xdot10}
\frac{d X_{1,0}}{d\tau} =  v_{\parallel,1,0}(\tau) \sin \alpha \left( 1 +  \frac{\phi_{\infty}''(Y_{\rm f})}{\Omega B \tan^2 \alpha } \right) + \cos\alpha \frac{1}{B} \partial_{y} \phi_1 (X_{\rm f}+ X_0(\tau), Y_{\rm f})  \rm 
\end{align}
 from (\ref{Xdot-exact}) and (\ref{Xdot-0th}).
Integrating (\ref{thetadot10}) gives
\begin{align} \label{theta10-before}
 \theta_{1,0}(\tau) = \frac{v_{\perp} (\mu_{\rm f})}{2\mu_{\rm f} B} \cos \alpha \int_0^{\tau} d\tau  \cos \left( \theta_{\rm f} +  \theta_0 (\tau') \right)  \partial_x \phi (X_{\rm f} +  X_0 (\tau') , Y_{\rm f}) \rm .
\end{align} 
The discussion following (\ref{mu10-before}) applies also to (\ref{theta10-before}): to lowest order in $\epsilon$, the oscillatory term in $\tau$ can be replaced by its average, which is zero, giving 
\begin{align}
 \theta_{1,0}(\tau) = 0 \rm .
\end{align}
To integrate (\ref{Xdot10}), we change variables from $\tau'$ to $ X_0'= X_0(\tau')$ in the integration using $d\tau' = d X_0' \left[ \sin \alpha \left( v_{\parallel, \rm f} + \frac{\phi_{\infty}'(Y_{\rm f})}{B\tan\alpha} \right) \right]^{-1}$ and substitute $ v_{\parallel, 1,0}(\tau)$ from (\ref{vpar1-Y1-epsilon}), obtaining
\begin{align} \label{Xbar10}
 X_{1,0}(\tau) = \int_0^{X_0 (\tau)} d  X'_0 \left(  \frac{\Omega \left[ \phi_1 (X_{\rm f} ,Y_{\rm f}) - \phi_1 (X_{\rm f} +  X'_0,Y_{\rm f}) \right]}{B \left( v_{\parallel, \rm f} + \frac{\phi_{\infty}'(Y_{\rm f})}{B\tan\alpha} \right)^2}  \left( 1 +  \frac{\phi_{\infty}''(Y_{\rm f})}{\Omega B \tan^2 \alpha} \right) \right. \nonumber \\
\left. + \frac{\partial_y \phi_1 (X_{\rm f} +  X_0', Y_{\rm f})}{B\tan\alpha \left( v_{\parallel, \rm f} + \frac{\phi_{\infty}'(Y_{\rm f})}{B\tan\alpha} \right)} \right) \rm .
\end{align}

We proceed to calculate the drift-kinetic second order corrections to the slow variables, $\mu_{2,0}(\tau)$, $Y_{2,0}(\tau)$ and $v_{\parallel, 2,0}(\tau)$, using the rule (\ref{Gslow2dot}) to calculate their derivatives with respect to $\tau$ and then setting $\rho_x = 0$ to take the drift-kinetic limit.
From (\ref{mudot-long-3}) and (\ref{mudot-1st}) we obtain
\begin{align} \label{mudot-2-11}
\frac{d  \mu_{2,0}}{d\tau} = & v_{\perp}(\mu_{\rm f} +  \mu_{1,0}(\tau)) \sin (\theta_{\rm f} +  \theta_0(\tau) +  \theta_{1,0}(\tau))  \cos \alpha  \nonumber \\
& \times \frac{1}{B} \partial_x \phi( X_{\rm f} +  X_0(\tau) +  X_{1,0}(\tau), Y_{\rm f} +  Y_{1,0}(\tau) )  \nonumber \\
& - v_{\perp}(\mu_{\rm f}) \sin (\theta_{\rm f} +  \theta_0 (\tau) ) \cos \alpha \frac{1}{B}  \partial_x \phi (X_{\rm f} +  X_0 (\tau) , Y_{\rm f}) \rm ,
\end{align}
while from (\ref{Ydot-long-3}), (\ref{vpardot-exact}) and (\ref{vpardot-1st}) we obtain
\begin{align} \label{dvpardt-2}
\frac{d v_{\parallel,2,0}}{d\tau} = \frac{d  Y_{2,0}}{d\tau} \Omega \tan \alpha = - \sin \alpha \frac{ \Omega}{B} \left[ \partial_x^2 \phi_1 (X_{\rm f} +  X_0(\tau), Y_{\rm f})  X_{1,0}(\tau) \right. \nonumber \\
\left. + \partial_x \partial_y  \phi_1 (X_{\rm f} +  X_0(\tau), Y_{\rm f})  Y_{1,0}(\tau) \right] \rm .
\end{align}
Using that $ \mu_{1,0}(\tau) =  \theta_{1,0}(\tau) = 0$, (\ref{mudot-2-11}) becomes
\begin{align}
\frac{d  \mu_{2,0}}{d\tau} =  \frac{1}{B} v_{\perp}(\mu_{\rm f}) \sin (\theta_{\rm f} +  \theta_0(\tau))  \cos \alpha & \left[  \partial_x^2 \phi( X_{\rm f} +  X_0(\tau) , Y_{\rm f} ) X_1(\tau) \right. \nonumber \\
& \left. + \partial_x \partial_y \phi (X_{\rm f} +  X_0 (\tau) , Y_{\rm f}) Y_1(\tau) \right] \rm .
\end{align}
As before, to lowest order in $\epsilon$ the oscillatory term $\sin (\theta_{\rm f} +  \theta_0(\tau))$ can be replaced with its average, leading to
\begin{align} \label{mubardot20}
 \mu_{2,0}(\tau) = 0 \rm .
\end{align}
Substituting $ X_{1,0}(\tau)$ from (\ref{Xbar10}) and $ Y_{1,0}(\tau)$ from (\ref{vpar1-Y1-epsilon}) in (\ref{dvpardt-2}), and integrating (\ref{dvpardt-2}) using the change of variables from $\tau'$ and $\tau''$ to $ X_0' = X_0(\tau')$ and $X_0'' = X_0(\tau'')$, respectively, in the double integrals 
gives
\begin{align} \label{vparbar20}
v_{\parallel,2,0}(\tau) = & Y_{2,0}(\tau) \Omega \tan \alpha  \nonumber \\
= & - \int_0^{X_0(\tau)} d X_0' \frac{\Omega \partial_x^2 \phi_1 (X_{\rm f} +  X_0', Y_{\rm f})}{B \left( v_{\parallel} + \frac{\phi'_{\infty}(Y_{\rm f})}{B \tan \alpha} \right)} \int_0^{X_0'} d X_0''  \left[  \frac{\partial_y \phi_1 (X +  X_0'',Y_{\rm f}) }{B \tan \alpha \left( v_{\parallel, \rm f} + \frac{\phi'_{\infty}(Y_{\rm f})}{B \tan \alpha} \right) } \right. \nonumber \\
& \left. + \frac{ \Omega \left[   \phi_1 (X_{\rm f}, Y_{\rm f}) - \phi_1 (X_{\rm f} +  X_0'', Y_{\rm f} )  \right] }{B\left( v_{\parallel, \rm f} + \frac{\phi'_{\infty}(Y_{\rm f})}{B \tan \alpha} \right)^2} \left( 1 + \frac{ \phi_{\infty}''(Y_{\rm f})}{\Omega B \tan^2 \alpha} \right)  \right]  \nonumber \\
& - \int_0^{X_0(\tau)} d X_0' \frac{\Omega \partial_x \partial_y  \phi_1 (X_{\rm f} +  X_0' , Y_{\rm f}) \left[   \phi_1 (X_{\rm f}  , Y_{\rm f}) - \phi_1 (X_{\rm f} +  X_0' , Y_{\rm f})  \right]}{B^2 \tan \alpha \left( v_{\parallel, \rm f} + \frac{\phi'_{\infty}(Y_{\rm f})}{B \tan \alpha} \right)^2} \rm .
\end{align}

The second order corrections to the fast variables $X$ and $\theta$ do not enter in the ion velocity distribution at second order, and are therefore not calculated.

\subsection{Drift-kinetic calculation of the ion velocity distribution to second order in $\hat \phi$} \label{subsec-higher-ni2}

To calculate the drift-kinetic limit of the second order correction to the ion distribution function in (\ref{F-exp}), we must first calculate the functions $\vec G_{\infty, 1,0}(\vec G_{\rm f}) = \vec G_{1,0}(\tau) \rvert_{\tau \rightarrow \infty}$ and $\vec G_{\infty, 2,0}(\vec G_{\rm f}) = \vec G_{2,0}(\tau) \rvert_{\tau \rightarrow \infty}$ for the slow variables (recall the expansions (\ref{Y-exp-infty})-(\ref{mu-exp-infty})). 
From (\ref{mu10}) we obtain
\begin{align}
\mu_{\infty, 1,0}(\vec G) = 0 \rm ,
\end{align}
while from (\ref{vpar1-Y1-epsilon}) evaluated at $\tau \rightarrow \infty$, corresponding to $X_0(\tau) \rightarrow \infty$, we obtain
\begin{align} \label{vpar1-Y1-epsilon}
 v_{\parallel, \infty, 1, 0}(\vec G)  = \Omega  Y_{\infty, 1,0}(\vec G)  \tan \alpha = & \frac{\Omega  \phi_1 (X , Y )  }{B \left( v_{\parallel, \rm f} +  \frac{\phi'_{\infty}(Y_{\rm f})}{B\tan \alpha}  \right) } \rm .
\end{align}
Setting $\tau \rightarrow \infty$ and $X_0(\tau) \rightarrow \infty$ in (\ref{vparbar20}) leads to
\begin{align}
v_{\parallel,\infty,2,0}(\vec G) = - \int_0^{\infty} d  X_0 \frac{\Omega \partial_x^2 \phi_1 (X +  X_0, Y)}{B \left( v_{\parallel} + \frac{\phi'_{\infty}(Y)}{B \tan \alpha} \right)} \int_0^{ X_0} d X'_0 \left[  \frac{\partial_y \phi_1 (X +  X_0',Y) }{B \tan \alpha \left( v_{\parallel} + \frac{\phi'_{\infty}(Y_{\rm f})}{B \tan \alpha} \right) } \right. \nonumber \\
\left. + \frac{ \Omega \left[   \phi_1 (X  , Y ) - \phi_1 (X +  X_0' , Y )  \right] }{B\left( v_{\parallel} + \frac{\phi'_{\infty}(Y)}{B \tan \alpha} \right)^2} \left( 1 + \frac{ \phi_{\infty}''(Y)}{\Omega B \tan^2 \alpha} \right)  \right]  \nonumber \\
- \int_0^{\infty} d X_0 \frac{\Omega \partial_x \partial_y  \phi_1 (X +  X_0 , Y ) \left[   \phi_1 (X , Y) - \phi_1 (X +  X_0 , Y )  \right]}{B^2 \tan \alpha \left( v_{\parallel} + \frac{\phi'_{\infty}(Y)}{B \tan \alpha} \right)^2} \rm .
\end{align}
Both terms in this equation can be integrated by parts in $ X_0$, with all boundary terms vanishing, giving
\begin{align}
v_{\parallel,2,0}(\vec G) = \int_0^{\infty} d  X_0 \frac{\Omega \partial_{x} \phi_1 (X +  X_0, Y)}{B \left( v_{\parallel} + \frac{\phi'_{\infty}(Y)}{B \tan \alpha} \right)} \left[ \frac{\partial_y \phi_1 (X +  X_0,Y) }{B \tan \alpha \left( v_{\parallel} + \frac{\phi'_{\infty}(Y)}{B \tan \alpha} \right) }    \right. \nonumber \\
\left. +  \frac{ \Omega \left[   \phi_1 (X , Y ) - \phi_1 (X +  X_0 , Y )  \right]}{B\left( v_{\parallel} + \frac{\phi'_{\infty}(Y)}{B \tan \alpha} \right)^2} \left( 1 + \frac{ \phi_{\infty}''(Y)}{\Omega B \tan^2 \alpha} \right) \right] \nonumber \\
 - \int_0^{\infty} d X_0 \frac{\Omega \partial_{y} \phi_1 (X +  X_0 , Y ) \partial_x \phi_1 (X +  X_0 , Y )}{B^2 \tan \alpha \left( v_{\parallel} + \frac{\phi'_{\infty}(Y)}{B \tan \alpha} \right)^2} \rm .
\end{align}
Noticing that the terms involving $\partial_y \phi_1 \partial_x \phi_1$ cancel and re-expressing the remaining term gives
\begin{align} \label{vpar-2-finstep}
v_{\parallel, \infty, 2,0}(\vec G) = & \frac{\left( 1 + \frac{\phi_{\infty}''(Y)}{\Omega B \tan^2 \alpha} \right)  \Omega^2 }{B^2 \left( v_{\parallel} + \frac{\phi'_{\infty}(Y)}{B \tan \alpha} \right)^3} \nonumber \\
& \times \int_0^{\infty} \partial_{x} \phi_1 (X +  X_0, Y) \left[ \phi_1 (X , Y) - \phi_1 (X +  X_0, Y) \right] d X_0 \rm .
\end{align}
Using the identity $ v_{\parallel,\infty,2,0} =  Y_{\infty,2,0} \Omega \tan \alpha$ from (\ref{dvpardt-2}) and evaluating the integral in (\ref{vpar-2-finstep}), we obtain
\begin{align} \label{vpar2}
 v_{\parallel,\infty,2,0} (\vec G) =  Y_{\infty, 2,0}(\vec G) \Omega \tan \alpha = - \frac{\left( 1 + \frac{\phi_{\infty}''(Y)}{\Omega B \tan^2 \alpha} \right)  \Omega^2 \phi_1^2(X,Y)}{2B^2 \left( v_{\parallel} + \frac{\phi'_{\infty}(Y)}{B \tan \alpha} \right)^3} \rm .
\end{align}
From (\ref{mubardot20}) we also obtain
\begin{align}
 \mu_{\infty,2,0} (\vec G) = 0 \rm .
\end{align}

To lowest order in $\epsilon$, the second order correction $F_2$ of the ion distribution function in (\ref{F-exp}) is denoted $F_{2,0}$ and is independent of gyrophase due to the neglect of $\rho_x$,
\begin{align}
F_2 (X, Y, \mu,  v_{\parallel}, \theta) = F_{2,0} (X, Y, \mu,  v_{\parallel}) + O(\epsilon \hat \phi^2 F) \rm .
\end{align}
Considering the drift-kinetic limit of the second order terms in the Taylor expansion of (\ref{F-solVlasov}) leads to (also making use of (\ref{mu10}))
\begin{align} \label{F20}
F_{2,0} (X, Y, \mu,  v_{\parallel}) =  v_{\parallel, \infty, 2,0}(\vec G) \partial_{v_{\parallel}} F_\infty  +  Y_{\infty,2,0}(\vec G)  \partial_{Y} F_\infty  + \frac{1}{2} v_{\parallel,\infty, 1,0}^2 (\vec G)   \partial_{v_{\parallel} v_{\parallel} } F_\infty \nonumber \\
+ \frac{1}{2} Y_{\infty, 1,0}^2 (\vec G)   \partial_{YY} F_\infty  + v_{\parallel, \infty, 1,0}(\vec G)  Y_{\infty, 1,0}(\vec G)  \partial_{v_{\parallel} Y} F_\infty \rm .
\end{align}
Using $  Y_{n,0} = v_{\parallel,n,0} / (\Omega \tan \alpha )$ for $n=1$ and $n=2$ (see (\ref{vpar1-Y1-epsilon}) and (\ref{vpar2})), equation (\ref{F20}) can be conveniently re-expressed as
\begin{align} \label{F20-reexp}
F_{2,0} (X, Y, \mu,  v_{\parallel}) =  & v_{\parallel, \infty, 2,0} (\vec G)  \left[ \partial_{v_{\parallel}} F_\infty  +  \frac{ \partial_{Y} F_\infty }{\Omega \tan \alpha } \right] \nonumber \\
& + \frac{1}{2} v_{\parallel, \infty, 1,0}^2 (\vec G)  \left[ \partial_{v_{\parallel}} + \frac{\partial_{Y}}{\Omega \tan \alpha} \right] \left[ \partial_{v_{\parallel} } F_\infty + \frac{\partial_Y F_\infty}{\Omega \tan \alpha } \right] \rm .
\end{align}

\subsection{Poisson's equation at second order} \label{subsec-higher-Poisson}

The $O(\hat \phi^2 n_{\infty})$ correction to the ion density at order $\epsilon^0$ is obtained by integrating $F_{2,0}$ as in (\ref{ni-def-new}) and neglecting $\rho_x$ in the Dirac delta function,
\begin{align} \label{ni-2}
n_{\text{i}, 2,0} (x,y) =  2\pi \int_0^{\infty} dX \int_0^{\infty} \Omega d\mu \int_{ - \frac{\phi'_{\infty}(y)}{B\tan \alpha}  }^{\infty} d v_{\parallel} F_{2,0} (X, y, \mu,  v_{\parallel})  \delta_{\rm Dirac} (x - X) \rm .
\end{align}
Computing this explicitly using equations (\ref{vpar1-Y1-epsilon}), (\ref{vpar2}) and (\ref{F20-reexp}) gives
\begin{align}
n_{\text{i},2, 0}(x,y) = &  \frac{\Omega^2 \phi_1^2 (x,y)}{2B^2}  2\pi \int_0^{\infty} \Omega d\mu \int_{ - \frac{\phi'_{\infty}(y)}{B\tan \alpha}  }^{\infty} dv_{\parallel} \nonumber \\
& \times  \left[ - \frac{ 1 + \frac{\phi_{\infty}''(y)}{\Omega B \tan^2 \alpha} }{\Omega \left( v_{\parallel} + \frac{\phi'_{\infty}(y)}{B\tan \alpha} \right)^3}  \left( \Omega \partial_{v_{\parallel}} F_{\infty} + \cot \alpha \partial_Y F_{\infty}  \right) \right. \nonumber \\
& \left.  + \frac{ \left(  \Omega \partial_{v_{\parallel} } + \cot \alpha \partial_{Y} \right)  \left( \Omega \partial_{v_{\parallel}} F_\infty +  \cot \alpha \partial_Y F_\infty  \right)}{ \Omega^2 \left( v_{\parallel} + \frac{\phi'_{\infty}(y)}{B\tan \alpha} \right)^2}    \right] \rm ,
\end{align}
where we take $F_{\infty}$ and its derivatives to be evaluated at $X=x$ and $Y=y$.
Applying the product rule on the last term gives one combination which is twice the size of the first term and has the opposite sign, effectively resulting in the sign of the first term switching, and another term which is an integral of partial derivatives,
\begin{align} \label{ni20-prelim}
& n_{\text{i},2, 0}(x,y)  =  \frac{\Omega^2 \phi_1^2 (x,y)}{2B^2}  2\pi \int_0^{\infty} \Omega d\mu \int_{ - \frac{\phi'_{\infty}(y)}{B\tan \alpha}  }^{\infty} dv_{\parallel} \nonumber \\
& \times  \left[ 
\frac{ 1 + \frac{\phi_{\infty}''(y)}{\Omega B \tan^2 \alpha} }{\left( v_{\parallel} + \frac{\phi'_{\infty}(y)}{B\tan \alpha} \right)^3}  \left( \frac{ \partial_Y F_{\infty} }{\Omega \tan \alpha }  +  \partial_{v_{\parallel}} F_{\infty} \right)  + \left(  \partial_{v_{\parallel} } + \frac{\partial_{Y}}{\Omega \tan \alpha} \right) \left[ \frac{  \partial_{v_{\parallel}} F_\infty + \frac{ \partial_Y F_\infty}{\Omega \tan \alpha }  }{ \left( v_{\parallel} + \frac{\phi'_{\infty}(y)}{B\tan \alpha} \right)^2} \right] \right] \rm .
\end{align}
For the second term in (\ref{ni20-prelim}), the integral of the partial derivative with respect to $v_{\parallel}$ vanishes upon using (\ref{F-nozero}), (\ref{dF-nozero}), (\ref{d2F-nozero}), (\ref{d3F-nozero}) and $F_{\infty} \rvert_{v_\parallel \rightarrow \infty} = 0$, while the partial derivative with respect to $Y$ can be taken outside of the integral to obtain 
\begin{align} \label{ni-2-0-final}
n_{\text{i},2, 0}(x,y) =  \left[ 2\pi \int_0^{\infty} \Omega d\mu \int_{ - \frac{\phi'_{\infty}(y)}{B\tan \alpha}  }^{\infty} dv_{\parallel} 
\frac{ 1 + \frac{\phi_{\infty}''(y)}{\Omega B \tan^2 \alpha} }{\left( v_{\parallel} + \frac{\phi'_{\infty}(y)}{B\tan \alpha} \right)^3}  \left( \partial_{v_{\parallel}} F_\infty + \frac{ \partial_Y F_\infty}{\Omega \tan \alpha }  \right) \right.  \nonumber \\
 +  \left. \frac{\partial_{Y}}{\Omega \tan \alpha} \left( 2\pi   \int_0^{\infty} \Omega d\mu \int_{ - \frac{\phi'_{\infty}(y)}{B\tan \alpha}  }^{\infty} dv_{\parallel}    \frac{  \partial_{v_{\parallel}} F_\infty + \frac{ \partial_Y F_\infty}{\Omega \tan \alpha }  }{ \left( v_{\parallel} + \frac{\phi'_{\infty}(y)}{B\tan \alpha} \right)^2} \right) \right] \frac{\Omega^2 \phi_1^2 (x,y)}{2B^2}   \rm .
\end{align}

The ion density at first order in $\hat \phi$ is given by (\ref{ni-1st}) \emph{to all orders in} $\epsilon$.
Hence, Poisson's equation at higher order in $\hat \phi$ becomes, upon inserting (\ref{Phipol-smallepsilon}) into the left hand side of (\ref{Poisson-1-step2}) and integrating by parts the term ith $\partial_{\mu} F_{\infty}$, and adding $Zn_{\rm i, 2,0} - n_{\rm e, 2}$ to the right hand side using (\ref{ne-2}) and (\ref{ni-2-0-final}),
\begin{align} \label{Poisson-highorder-final}
- \frac{e}{T_{\rm e}} \partial_x^2  \phi_1 (x,y)  \left[ ( \lambda_{\rm D}^2 + \rho_{\rm B}^2 \cos^2 \alpha ) n_{\rm e, \infty} \phantom{\int_0^{\infty}} \right. \nonumber \\
  \left. + 2\pi Z \rho_{\rm B}^2 \cos^2 \alpha  \int_0^{\infty} \Omega d\mu \int_{ - \frac{\phi'_{\infty}(y)}{B\tan \alpha}  }^{\infty} d v_{\parallel}  \mu  \frac{ \Omega \partial_{v_{\parallel}} F_\infty + \cot \alpha \partial_Y F_\infty }{ v_{\parallel} +\frac{\phi'_{\infty}(y)}{B\tan \alpha}  } \right] \nonumber \\
= \frac{1}{2} \left[ - \frac{T_{\rm e}^2}{e^2} \frac{d^2n_{\rm e, \infty}}{d\phi^2}  + 2\pi Z v_{\rm B}^4 \int \Omega d\mu \int dv_{\parallel}  \frac{ 1 + \frac{\phi_{\infty}''(y)}{\Omega B \tan^2 \alpha} }{\left( v_{\parallel} + \frac{\phi'_{\infty}(y)}{B\tan \alpha} \right)^3}  \left( \partial_{v_{\parallel}} F_\infty + \frac{ \partial_Y F_\infty}{\Omega \tan \alpha } \right)  \right. \nonumber \\
\left.  +  \frac{Zv_{\rm B}^4 \partial_{y}}{\Omega \tan \alpha} \left( 2\pi   \int_0^{\infty} \Omega d\mu \int_{ - \frac{\phi'_{\infty}(y)}{B\tan \alpha}  }^{\infty} dv_{\parallel}    \frac{  \partial_{v_{\parallel}} F_\infty + \frac{ \partial_Y F_\infty}{\Omega \tan \alpha }  }{ \left( v_{\parallel} + \frac{\phi'_{\infty}(y)}{B\tan \alpha} \right)^2} \right) \right] \frac{e^2 \phi_1^2 (x,y)}{T_{\rm e}^2}  \rm .
\end{align}

\subsection{Polarisation condition in marginal case} \label{subsec-higher-pol}

In the marginally satisfied Bohm-Chodura condition (\ref{kin-Chod=}), the polarisation condition corresponds to the requirement that the left hand side of (\ref{Poisson-highorder-final}) be positive for a monotonically decaying potential such that $\partial_x^2 \phi_1 < 0$.
Imposing this requirement and rearranging gives the condition
\begin{align} \label{pol-marg}
2\pi Z v_{\rm B}^2  \int_0^{\infty} d\mu \int_{- \frac{\phi'_{\infty}}{B\tan \alpha} }^{\infty}  d v_{\parallel}  \mu \frac{ \partial_{ v_{\parallel}} F_{\infty} + \Omega^{-1} \cot \alpha \partial_{Y} F_{\infty}}{  v_{\parallel} + \frac{\phi'_{\infty}}{B\tan \alpha} } > - \frac{ \lambda_{\rm D}^2 + \rho_{\rm B}^2 \cos^2 \alpha }{\cos^2 \alpha} n_{\rm e,\infty} \rm .
\end{align}
This can be re-expressed by integrating by parts the terms $\partial_{v_\parallel} F_{\infty}$ in $v_{\parallel}$,
\begin{align} \label{pol-margv2}
2\pi Z v_{\rm B}^2  \int_0^{\infty} d\mu \int_{- \frac{\phi'_{\infty}}{B\tan \alpha} }^{\infty}  d v_{\parallel}  \mu \left(  \frac{ F_{ \infty}}{ \left( v_{\parallel} + \frac{\phi'_{\infty}}{B\tan \alpha} \right)^2  }   + \frac{  \cot \alpha \partial_{Y} F_{ \infty}}{ \Omega  \left( v_{\parallel} + \frac{\phi'_{\infty}}{B\tan \alpha} \right) } \right) \nonumber \\
> - \frac{ \lambda_{\rm D}^2 + \rho_{\rm B}^2 \cos^2 \alpha }{\cos^2 \alpha} n_{\rm e,\infty} \rm .
\end{align}
The first term on the left hand side is always positive, while the second term may be negative.
It is therefore only the presence of the second term which could make the polarisation condition violated, since the right hand side of (\ref{pol-marg})-(\ref{pol-margv2}) is negative.

To understand under what conditions (\ref{pol-margv2}) can be violated while (\ref{kin-Chod=}) is satisfied, we evaluate the polarisation condition for a model ion distribution function where the perpendicular and parallel velocity dependences are separable,
\begin{align} \label{F-sep}
F_{\infty}(Y, \mu, v_{\parallel}) = \frac{1}{2\pi} n_{\rm i, \infty}(Y) g(Y, \mu ) h (Y, v_{\parallel}) \rm ,
\end{align}
where $g$ and $h$ are chosen to satisfy the normalisations $\int_0^{\infty} g(Y, \mu ) \Omega d\mu = 1$ and $\int_{-\phi'_{\infty}/(B\tan \alpha)}^{\infty}  h(Y, v_{\parallel})  dv_{\parallel} = 1$.
This may not be a physically realizable distribution function, but it helps to illustrate the case in point. 
Inserting (\ref{F-sep}) into the equality form of the Bohm-Chodura condition (\ref{kin-Chod=}), using $\int_0^{\infty} \partial_Y g(Y , \mu) d\mu = \partial_Y \left[ \int_0^{\infty}  g d\mu \right] = 0$ and $\int_0^{\infty}  \Omega g(Y , \mu) d\mu = 1$, we obtain
\begin{align} \label{int-polapp}
v_{\rm B}^2  \int_{-\frac{\phi'_{\infty}}{B\tan \alpha} }^{\infty} \frac{ dv_{\parallel} }{ v_{\parallel} + \frac{\phi'_{\infty}}{B\tan \alpha} } \left( n_{\rm e, \infty} \partial_{v_{\parallel}} h (Y, v_{\parallel}) + \frac{ \partial_Y \left[ n_{\rm e, \infty} h (Y, v_{\parallel}) \right]}{\Omega \tan \alpha} \right) = n_{\rm e, \infty} \rm .
\end{align}
Inserting (\ref{F-sep}) into the polarisation condition (\ref{pol-marg}) instead gives
\begin{align} \label{int-polapp-2}
\rho_{\rm B}^2 \cos^2 \alpha \int_0^{\infty} \Omega^2 \mu g d\mu \int_{-\frac{\phi'_{\infty}}{B\tan \alpha} }^{\infty} \frac{ dv_{\parallel} }{ v_{\parallel} + \frac{\phi'_{\infty}}{B\tan \alpha} } \left( n_{\rm e, \infty} \partial_{v_{\parallel}} h  + \frac{ \partial_Y \left[ n_{\rm e, \infty} h  \right]}{\Omega \tan \alpha} \right) \nonumber \\
+ \rho_{\rm B}^2 \cos^2 \alpha \int_0^{\infty} \Omega^2 \mu \frac{\partial_Y g }{\Omega \tan \alpha} d\mu \int_{-\frac{\phi'_{\infty}}{B\tan \alpha} }^{\infty} \frac{ n_{\rm e, \infty} h  dv_{\parallel} }{ v_{\parallel} + \frac{\phi'_{\infty}}{B\tan \alpha} } 
 > - n_{\rm e, \infty} \left( \lambda_{\rm D}^2 + \rho_{\rm B}^2 \cos^2 \alpha  \right) \rm .
\end{align}
Substituting the result (\ref{int-polapp}) into the first term of (\ref{int-polapp-2}), using $v_{\rm B}^2 = ZT_{\rm e}/ m_{\rm i}$, re-expressing the second term and recalling that $y \simeq Y$ gives
\begin{align} \label{int-polapp-3}
\rho_{\rm B}^2  \frac{T_{\rm i, \perp}}{ZT_{\rm e}} + \frac{\rho_{\rm B}^2  }{\Omega \tan \alpha}  \frac{d}{dy} \left( \int_0^{\infty} \Omega^2 \mu  g d\mu \right) \int_{-\frac{\phi'_{\infty}}{B\tan \alpha} }^{\infty} \frac{ h dv_{\parallel} }{ v_{\parallel} + \frac{\phi'_{\infty}}{B\tan \alpha} }  
 > - \frac{ \lambda_{\rm D}^2 + \rho_{\rm B}^2 \cos^2 \alpha }{\cos^2 \alpha} \rm ,
\end{align}
having defined the perpendicular ion temperature
\begin{align} \label{Tiperp-def}
T_{\rm i, \perp}(y) \equiv \frac{2\pi}{n_{\rm i, \infty}} \int_0^{\infty} \Omega d\mu \int_{- \frac{\phi'_{\infty}}{B\tan \alpha} }^{\infty}  dv_{\parallel} F_{\infty} m_{\rm i} \Omega \mu = \int_0^{\infty} g(y, \mu) m_{\rm i} \Omega^2 \mu d\mu  \rm .
\end{align}
The polarisation condition is thus satisfied provided that
\begin{align} \label{pol-sep-temperature}
v_{\rm B} \frac{[\ln T_{\rm i,\perp}]'}{\Omega \tan \alpha} > - \left[ 1 + \frac{ZT_{\rm e}}{T_{\rm i\perp}} \left( 1 + \frac{\lambda_{\rm D}^2}{\rho_{\rm B}^2 \cos^2 \alpha} \right) \right] \left[ \int_{-\frac{\phi'_{\infty}}{B\tan \alpha}}^{\infty} dv_{\parallel} \frac{ v_{\rm B} h(Y, v_{\parallel})  }{ v_{\parallel} + \frac{\phi'_{\infty}}{B\tan \alpha} } \right]^{-1} \rm .
\end{align}
Hence, a large and negative gradient of $T_{\rm i,\perp}(y)$ can violate the polarisation condition.

\subsection{Additional sheath condition for an electron-repelling sheath} \label{subsec-higher-addsheath}

If the polarisation condition is satisfied, the left hand side of (\ref{Poisson-highorder-final}) is positive and we therefore require the right hand side to also be positive, giving 
\begin{align} \label{addsheath}
 - \frac{1}{2} \frac{d^2n_{\rm e, \infty}}{d\phi^2}  + 2\pi Z \int \Omega d\mu \int dv_{\parallel} 
\frac{\Omega^2  \left( 1 + \frac{\phi_{\infty}''(Y)}{\Omega B\tan^2 \alpha} \right) }{2 B^2 \left( v_{\parallel} + \frac{\phi'_{\infty}(y)}{B\tan \alpha} \right)^3}  \left( \partial_{v_{\parallel}} F_\infty + \frac{ \partial_Y F_\infty}{\Omega \tan \alpha } \right)   \nonumber \\
+ \frac{\partial_{y}}{\Omega \tan \alpha} \left( 2\pi Z  \int_0^{\infty} \Omega d\mu \int_{ - \frac{\phi'_{\infty}(y)}{B\tan \alpha}  }^{\infty} dv_{\parallel} \frac{\Omega^2}{2B^2}  \frac{  \partial_{v_{\parallel}} F_\infty + \frac{ \partial_Y F_\infty}{\Omega \tan \alpha }  }{ \left( v_{\parallel} + \frac{\phi'_{\infty}(y)}{B\tan \alpha} \right)^2} \right) > 0 \rm .
\end{align}

Assuming $\delta \ll \alpha$, tangential gradients can be neglected.
Then, with Boltzmann electrons, it can be proven \cite{Geraldini-2018}, as follows, that (\ref{addsheath}) is automatically satisfied. 
Eliminating the tangential gradients and integrating by parts using (\ref{F-nozero}), condition (\ref{addsheath}) becomes
\begin{align} \label{addsheath-noy}
\frac{1}{2} \frac{d^2n_{\rm e, \infty}}{d\phi_{\infty}^2} - 2\pi Z \int \Omega d\mu \int dv_{\parallel} 
 \frac{3\Omega^2 }{2 B^2 v_{\parallel}^4} F_{\infty} < 0 \rm .
\end{align}
By applying Schwarz's inequality, we obtain the relation
\begin{align} \label{Schwarz}
\left(\int_0^{\infty} \Omega d\mu \int_0^{\infty} dv_{\parallel} F_{\infty} \right) \left( \int_0^{\infty} \Omega d\mu \int_0^{\infty} dv_{\parallel} \frac{F_{\infty}}{v_{\parallel}^4} \right) \geqslant \left( \int_0^{\infty} \Omega d\mu \int_0^{\infty} dv_{\parallel}  \frac{F_{\infty}}{v_{\parallel}^2} \right)^2 \rm ,
\end{align}
Recall the marginal form of the Bohm-Chodura condition (\ref{kin-Chod=}), from which we obtain, by taking $\partial_y = 0$ and integrating by parts,
\begin{align} \label{kin-Chod-noy-IBP}
 2\pi  \int_0^{\infty} \Omega d\mu  \int_{- \frac{\phi'_{\infty}}{B\tan \alpha} }^{\infty} d v_{\parallel}  \frac{ F_{\infty} }{ v_{\parallel}^2  }  =  \frac{n_{\rm e,\infty}}{Z v_{\rm B}^2}  \rm .
\end{align}
Combining (\ref{Schwarz}) and (\ref{kin-Chod-noy-IBP}) results in
\begin{align}
 2\pi \int \Omega d\mu \int dv_{\parallel} \frac{F_{\infty}}{v_{\parallel}^4}  \geqslant \frac{n_{\rm e,\infty}}{Zv_{\rm B}^4} \rm .
\end{align}
Then, (\ref{addsheath-noy}) is automatically satisfied if
\begin{align} \label{addsheath-noy-simp}
\frac{T_{\rm e}^2}{e^2} \frac{d^2n_{\rm e, \infty}}{d\phi^2} <  3 n_{\rm e, \infty} \rm .
\end{align}
Electrons which are Boltzmann distributed at least to the lowest order in some small expansion parameter satisfy (\ref{addsheath-noy-simp}) because $d^2n_{\rm e, \infty} / d\phi^2 \simeq (e/T_{\rm e})^2 n_{\rm e, \infty}$.

It has not been proven here that the more general condition (\ref{addsheath}), including tangential gradients, is automatically satisfied locally at each $y$ with Boltzmann electrons.
In the absence of such a proof, we must view (\ref{addsheath}) as an additional sheath condition emerging when the Bohm-Chodura condition (\ref{kin-Chod=}) is satisfied marginally and the ion distribution function satisfies (\ref{d2F-nozero})-(\ref{d3F-nozero}).
In \ref{app-slowdens} it is shown that no additional sheath condition emerges in the case that (\ref{d2F-nozero})-(\ref{d3F-nozero}) are not both satisfied.

\section{Conclusions} \label{sec-conc}

The generalisation of the Bohm condition to a magnetised plasma with a magnetic field at an arbitrary angle $\alpha$ with the target, known as the Bohm-Chodura condition, has been derived in the kinetic framework.
This was done by imposing a spatial monotonic decay of the electrostatic potential far from the target on the length scale of the magnetised plasma sheath, $l_{\rm ms} = \max (\rho_{\rm S} \cos \alpha, \lambda_{\rm D})$ (typically $\rho_{\rm S}$), in the limit $l_{\rm ms} / L \rightarrow 0$. 
The effect of spatial fluctuations aligned with the magnetic field, whose characteristic length scale is taken to be $L \sim \rho_{\rm S} / \delta \gg l_{\rm ms}$ across the magnetic field and $L / \delta \gg L$ along the magnetic field, has been included. 
Such fluctuations play a central role at small magnetic field angles, $\alpha \sim \delta$, strongly altering the Bohm-Chodura condition and introducing further constraints.
Collisions, which may become important at small angles, have been neglected.

Without tangential gradients, the generalised criterion is somewhat intuitive.
In the fluid picture, flow towards the sheath must be sonic.
In unmagnetised plasmas, the natural direction of this flow is the wall normal.
In magnetised plasmas, the direction is constrained by the magnetic field.
It thus appears logical that the kinetic Chodura condition must have exactly the same form as the kinetic Bohm condition, with the velocity normal to the wall $v_x$ replaced by the component of the velocity parallel to the magnetic field $v_{\parallel}$.
This is well-known to be the case for the fluid condition \cite{Chodura-1982}.
All the past derivations of the kinetic criterion, however, made simplifying assumptions.
For instance, although reference \cite{Sato-1994} correctly identified the ion polarisation drift as a key ingredient determining the potential profile in the cold ion case, they did not derive this drift in the general case.
Their derivation thus assumes cold ions while inconsistently retaining the ion distribution function.
To our knowledge, this paper provides the most general and rigorous derivation of the kinetic Chodura condition without tangential gradients (see (\ref{kin-Chod-noy})).

The sheath condition is substantially altered at small magnetic field angles, such that $\alpha \sim \delta$. 
In this limit, the gradients from the spatial fluctuations in the bulk, normally negligible on the small sheath length scale, affect the ions in the sheath in two ways:
the $\vec{E} \times \vec{B}$ drifts caused by the bulk gradients move ions towards or away from the wall across the magnetic field line, thus competing with the transport of ions towards the wall by their streaming parallel to the magnetic field;
the $\vec{E}\times\vec{B}$ drifts caused by the sheath electric field move ions in the direction parallel to the wall where there are bulk gradients, and since the ions spend a long time in the sheath (owing to the shallow magnetic field angle) they move large distances (comparable to the bulk length scale when $\alpha \sim \delta$) in this direction.
In a fluid picture, the first effect modifies the sheath condition for the flow velocity parallel to the magnetic field, making it sonic in a frame of reference where the $\vec{E} \times \vec{B}$ drift from the tangential bulk electric field (locally) vanishes \cite{Hutchinson-1996}.
The second effect introduces a further dependence of the sheath condition on tangential gradients of density, temperature and ion parallel flow velocity \cite{Loizu-2012}: more generally, the tangential gradient of the ion velocity distribution.
In the kinetic framework developed here, we recover the fluid condition (\ref{fluidChodura-Loizu}) of reference \cite{Loizu-2012} in the cold ion limit (see (\ref{fluidLoizu})), naturally accounting for $\vec{E} \times \vec{B}$ and diamagnetic flows.
We remark that the boundary condition on the ion flow velocity at the sheath entrance typically used in fluid simulations of a fusion device is a simplified version of (\ref{fluidChodura-Loizu}) which assumes the effect of fluctuations to be subdominant.
To our knowledge, the full implications and complexity of the fluid Bohm-Chodura condition (\ref{fluidChodura-Loizu}) with $\alpha \sim \delta$ have not yet been studied. 

The kinetic Bohm-Chodura condition (\ref{kin-Chod}) is identical to that derived in reference \cite{Claassen-Gerhauser-1996b}.
However, their derivation ignores the ion polarisation density developing in the magnetised sheath, relying solely on $\partial (n_{\rm e} - n_{\rm i} ) / \partial x \geqslant 0$ \cite{Cohen-Ryutov-2004-sheath-boundary-conditions} which is only valid for $\rho_{\rm S} \cos \alpha \ll \lambda_{\rm D}$.
The polarisation density is a key mechanism underlying the upkeep of the monotonous potential profile for $\rho_{\rm S} \cos \alpha \gtrsim \lambda_{\rm D}$.
It can change sign due to the displacement (via the sheath $\vec{E}\times \vec{B}$ drift) of the ions in the tangential direction if the distribution function changes in this direction.
Therefore, the left hand side of (\ref{Poisson-1-step2}) can also change sign for a monotonically decaying electron-repelling ($\phi_1 < 0$) potential.
This invalidates the Bohm-Chodura condition, which imposes that the right hand side of (\ref{Poisson-1-step2}) be positive.
A so-called polarisation condition, which constrains the left hand side of (\ref{Poisson-1-step2}) to be positive (for a monotonically decaying potential profile with $\phi_1 < 0$), is required to further ensure the validity of the Bohm-Chodura condition.
We have also shown that if the polarisation condition is satisfied and the Bohm-Chodura condition is satisfied with the equality sign, an additional sheath condition (\ref{addsheath}) emerges for a specific set of incoming distribution functions. 
Only if tangential gradients are negligible has this additional sheath condition been shown to be automatically satisfied.

This work provides a solid theoretical background to the scale separation between a strongly magnetised plasma and its sheath. 
However, it also raises some questions relevant to the formulation of sheath boundary conditions for a gyrokinetic or drift-kinetic description of a turbulent plasma in contact with a target when the magnetic field angle at the target is shallow, $\alpha \sim \delta \ll 1$.
A first question regards the polarisation condition, whose possible violation effectively reverses the sign of the Laplacian in Poisson's equation, as seen in (\ref{Poisson-highorder-final}) (and more generally in (\ref{Poisson-1-step2})).
As far as we are aware, in most formulations of the open-field-line gyrokinetic equations this sign change is not contemplated.
Therefore, such gyrokinetic equations may need to be modified to account for a more accurate polarisation term near the magnetised sheath entrance. 
Moreover, the spatial variation across the sheath of the tangential electric field $\partial_y \phi$ and the tangential $\vec{E} \times \vec{B}$ drift of ions (caused by the wall-normal sheath electric field $\partial_x \phi$) may cause the net velocity of some ions towards the wall, $v_{\parallel} \alpha + \phi'(y)/ B$, to reverse even if the electric field in the sheath is always directed to the wall (and thus always increases $v_{\parallel}$), $\partial_x \phi > 0$.
For there to be no reflected ions at the sheath entrance, it is necessary that $\left( 1 + \partial_{y}^{2} \phi / (\Omega B \alpha^2 ) \right) \partial_x \phi > 0$ (see (\ref{cond-noreflection})). 
If there are reflected ions, a local analysis of the magnetised sheath entrance becomes harder, and the generalised Bohm-Chodura condition derived here becomes inapplicable.
In this case, numerically solving for the electrostatic potential profile on the length scale $\rho_{\rm S}$ in the magnetic presheath becomes the only obvious way to self-consistently calculate the reflected ions and the full ion distribution function.
In this region a modified gyrokinetic treatment exploiting $\alpha \sim \delta \ll 1$ \cite{Geraldini-2017, Geraldini-2018} can nonetheless be further used to account for the strong distortion of closed ion orbits from non-circular and for the open part of the ion trajectory striking the target. 
From an experimental perspective, accurate measurements of the ion velocity distribution in plasmas near a target with a shallow magnetic field angle using Laser Induced Fluorescence \cite{Siddiqui-Hershkowitz-2016, Caron-2021-exp} would enable the verification of the kinetic Chodura condition derived herein.

~

\ack{Alessandro Geraldini would like to acknowledge helpful discussions with Justin Ball, Joaquim Loizu, Michael Hardman, Micol Bassanini, and members of the EUROfusion TSVV-4 collaboration.
This work has been carried out within the framework of the EUROfusion Consortium, via the Euratom Research and Training Programme (Grant Agreement No 101052200 — EUROfusion) and funded by the Swiss State Secretariat for Education, Research and Innovation (SERI). Views and opinions expressed are however those of the author(s) only and do not necessarily reflect those of the European Union, the European Commission, or SERI. Neither the European Union nor the European Commission nor SERI can be held responsible for them.
This work was supported by the U.S. Department of Energy under contract number DE-AC02-09CH11466. The United States Government retains a non-exclusive, paid-up, irrevocable, world-wide license to publish or reproduce the published form of this manuscript, or allow others to do so, for United States Government purposes.}

\appendix

\section{Slow ion trajectories and velocity distribution} \label{app-slowtraj}

The analysis of Poisson's equation near the magnetised sheath entrance requires considering also the small number of ions which are moving so slowly towards the target, i.e. with $d X / d\tau \ll c_{\rm S} \sin \alpha $, that $d X / d\tau$ itself significantly changes and can no longer be considered constant to lowest order.
From (\ref{Xdot-0th}), $d X / d\tau \simeq  w_{\parallel} \sin \alpha $ where we used the definition of the effective parallel velocity $w_{\parallel}$ in (\ref{wpar}).
From the expansions (\ref{Y-exp}) and (\ref{vpar-exp}) for $ Y$ and $ v_{\parallel}$ we introduce the same asymptotic expansion for the time-dependent $ w_{\parallel}(\tau)$ following an ion trajectory,
\begin{align}
 w_{\parallel} (\tau) = w_{\parallel, \rm f} +  w_{\parallel, 1} (\tau) +  w_{\parallel, 2} (\tau) + O\left( \hat{\phi}^3 c_{\rm S} \right) \rm ,
\end{align}
with $ w_{\parallel, n} (\tau) \rvert_{\tau = 0} = 0$ for all $n$, and $w_{\parallel, \rm f} = w_{\parallel}(\tau) \rvert_{\tau = 0}$.
For $w_{\parallel, \rm f} \sim v_{\parallel, 1}(\tau) \sim w_{\parallel, 1}(\tau)$ in (\ref{vpar-Y1}), the perturbative calculation in section~\ref{subsec-traj-perturb} fails because it requires $w_{\parallel, 1}(\tau) \ll w_{\parallel, \rm f}$.
The failure of the perturbative calculation thus corresponds to the ordering
\begin{align} \label{ordering-slow}
w_{\parallel, \rm f} \sim \hat{\phi}^{1/2} c_{\rm s} \rm .
\end{align}
Ion trajectories satisfying (\ref{ordering-slow}) are called slow.

In the rest of this appendix, we proceed to expand slow ion trajectories (\ref{subapp-slowtraj-1}) and then define and expand the slow ion velocity distribution (\ref{subapp-slowtraj-2}).

\subsection{Alternative expansion for slow ion trajectories} \label{subapp-slowtraj-1}

The time scale over which a slow trajectory, satisfying the ordering (\ref{ordering-slow}), crosses a distance $x \sim l_{\rm ms}$ is much longer than for a bulk trajectory. 
Given the new ordering $d X / d\tau \simeq w_{\parallel} \sin \alpha = O\left(  \hat \phi^{1/2} c_{\rm S} \sin \alpha \right)$, we obtain a time scale $\sim \hat \phi^{-1/2} (l_{\rm ms} / \rho_{\rm S} ) \left(\Omega \sin \alpha \right)^{-1} \sim \hat \phi^{-1/2} t_X \gg t_X$.
While the integration time is longer, the ordering of the time derivatives (\ref{Ydot-long})-(\ref{thetadot}) of the variables $Y$, $\mu$, $v_{\parallel}$ and $\theta$ is unchanged.
Hence, the lowest order variation over the longer time scale of these variables is larger by $\hat \phi^{-1/2}$.
Therefore, the expansions in (\ref{Y-exp})-(\ref{theta-exp}) require the replacements $\boldsymbol{\Gamma}_1 \rightarrow \boldsymbol{\Gamma}_{1/2}$ for the slow variables $\boldsymbol{\Gamma} = (Y, \mu, v_{\parallel})$, where $\boldsymbol{\Gamma}_{1/2} = O(\hat \phi^{1/2} \boldsymbol{\Gamma})$, and (\ref{theta-exp}) requires the replacement $\theta_0 \rightarrow \theta_{-1/2}$, where $\theta_{-1/2} = O(\hat \phi^{-1/2})$,
\begin{align} \label{X-exp-slow}
X(\tau) = X_{\rm f} + X_0(\tau) + O\left( \hat \phi^{1/2} l_{\rm ms} , \delta \hat \phi^{-1/2} l_{\rm ms} \right) \rm ,
\end{align}
\begin{align} \label{theta-exp-slow}
 \theta (\tau) =  \theta_{-1/2} (\tau) + O\left( \hat \phi^{1/2}, \delta \hat \phi^{-1/2} \right) \rm ,
\end{align}
\begin{align}
 \mu (\tau) = \mu_{\rm f} +  \mu_{1/2} (\tau) + O\left( \hat{\phi} \rho_{\rm S} c_{\rm S}, \delta \hat \phi^{-1/2} \rho_{\rm S} c_{\rm S} \right) \rm ,
\end{align}
\begin{align}
 v_{\parallel} (\tau) = v_{\parallel, \rm f} +  v_{\parallel, 1/2}(\tau) + O\left( \hat{\phi} c_{\rm S}, \delta \hat \phi^{-1/2} c_{\rm S} \right) \rm ,
\end{align}
\begin{align} \label{Y-exp-slow}
 Y(\tau) = Y_{\rm f} +  Y_{1/2}(\tau) + O\left( \hat{\phi} L, \delta \hat \phi^{-1/2} L \right) \rm .
\end{align}
We note that (\ref{X-exp-slow})-(\ref{Y-exp-slow}) are expansions in orders of $\hat \phi^{1/2}$.
As it will be convenient to replace $v_{\parallel, 1/2}(\tau)$ and $Y_{1/2}(\tau)$ by $w_{\parallel, 1/2}(\tau)$, we define a corresponding expansion for $w_{\parallel}(\tau)$, 
\begin{align} \label{wpar-exp-slow}
 w_{\parallel} (\tau) = w_{\parallel, \rm f} +  w_{\parallel, 1/2} (\tau) + O\left( \hat{\phi} c_{\rm S}, \delta \hat \phi^{-1/2} c_{\rm S} \right) \rm .
\end{align}
It will be shown in the next subsection and in \ref{app-slowdens} that the slow ion trajectory expansion need not be carried out to higher order than in (\ref{X-exp-slow})-(\ref{wpar-exp-slow}). 

By modifying (\ref{thetadot-0th}) to account for the different ordering for slow ions, the lowest order time derivative of the phase angle is
\begin{align} \label{thetadot-1/2}
\frac{d \theta_{-1/2}}{d\tau} = - \Omega  \rm .
\end{align}
For slow ions, this is the single fastest variable,
\begin{align}
 \theta_{-1/2} (\tau) = - \Omega \tau \rm .
\end{align}
Hence, the time derivatives of the other variables at their lowest order can be replaced by their gyroaverage, i.e. their average over $ \theta$, as in conventional gyrokinetics \cite{Parra-2008}\footnote{Alternatively, we can repeat the steps in section~\ref{subsec-higher-DKvpar1}, where the variation in $\phi_1 (X_0(\tau), Y_{\rm f})$ is very small during the time scale $\tau \sim 1/ \Omega$ over which time-oscillating terms vary; hence, the time-oscillating pieces can be replaced with averages.}
\begin{align} \label{Xdotslow-1st}
\frac{d X_0}{d\tau} = - w_{\parallel, \rm f} \sin \alpha - w_{\parallel, 1/2 }(\tau) \sin \alpha  \rm ,
\end{align}
\begin{align} \label{wpardotslow-1st}
\frac{d  w_{\parallel, 1/2}}{d\tau}  = - \left( 1 + \frac{\phi''_{\infty}(Y_{\rm f})}{\Omega B\tan^2\alpha} \right) \sin \alpha \frac{\Omega}{B} \partial_X \bar \phi_1 (X_{\rm f} +  X_{0}(\tau) , Y_{\rm f} )  \rm ,
\end{align}
\begin{align} \label{mudot-1st-slow}
\frac{d  \mu_{1/2}}{d\tau} = & \left\langle  v_{\perp}(\mu_{\rm f}) \sin \theta \cos \alpha \frac{1}{B}  \partial_x \phi_1 (X_{\rm f} +  X_{0} (\tau) + \rho_x (\mu_{\rm f},  \theta ), Y_{\rm f}) \right\rangle_{ \theta} \nonumber  \\
= &   \frac{\Omega}{B} \left\langle \partial_{\theta} \rho_{x}(\mu_{\rm f}, \theta) \partial_x \phi_1 (X_{\rm f} +  X_{0} (\tau) + \rho_x (\mu_{\rm f},  \theta ), Y_{\rm f}) \right\rangle_{ \theta} = 0 \rm .
\end{align}
To obtain (\ref{wpardotslow-1st}), we substituted (\ref{Ydot-long-3}) and (\ref{vpardot-exact}) into $dw_{\parallel} / d\tau = dv_{\parallel} / d\tau + [\phi_{\infty}''(Y)/(B\tan \alpha)]  dY/d\tau$, obtained from the definition of $w_{\parallel}$ in (\ref{wpar}). 
We then used $\left\langle \partial_x \phi_1 (X_{\rm f} +  X_{0}(\tau)  + \rho_x (\mu_{\rm f},  \theta), Y_{\rm f} ) \right\rangle_{ \theta} = \partial_X \bar \phi_1  (X_{\rm f} +  X_{0}(\tau)  , Y_{\rm f}, \mu_{\rm f} ) $, where we defined the gyroaverage of the electrostatic potential $\phi_1$ (at fixed guiding centre position $X$)
\begin{align}
\bar \phi_1  (X  , Y, \mu ) \equiv \left\langle \phi_1 (X + \rho_x (\mu,  \theta) , Y ) \right\rangle_{ \theta} \rm .
\end{align}

In order to evaluate the correction $ w_{\parallel, 1/2}(\tau)$, we multiply (\ref{wpardotslow-1st}) by $w_{\parallel, \rm f} +  w_{\parallel, 1/2}(\tau)$.
On the left hand side we apply the relation
\begin{align} \label{wpardotsq-rel}
( w_{\parallel, \rm f} +  w_{\parallel, 1/2}(\tau) ) \frac{d}{d\tau} \left( w_{\parallel, \rm f} +  w_{\parallel, 1/2}(\tau) \right) = \frac{1}{2} \frac{d }{d\tau} \left( w_{\parallel, \rm f} +  w_{\parallel, 1/2}(\tau) \right)^2  \rm ,
\end{align}
while on the right hand side we employ
\begin{align} \label{wpar-dphidt}
( w_{\parallel, \rm f} +  w_{\parallel, 1/2}(\tau) ) \sin \alpha  \partial_X \bar \phi_1 (X_{\rm f} +  X_{0}(\tau), Y_{\rm f}, \mu_{\rm f})  = \frac{d }{d\tau} \bar \phi_1 (X_{\rm f} +  X_{0}(\tau)  , Y_{\rm f}, \mu_{\rm f} ) \rm ,
\end{align}
which is obtained from (\ref{Xdotslow-1st}) using the chain rule.
Equation (\ref{wpardotslow-1st}) thus becomes
\begin{align} \label{wpardotslow-1st-1}
\frac{1}{2}  \frac{d }{d\tau} \left( w_{\parallel, \rm f} +  w_{\parallel, 1/2}(\tau) \right)^2 = & - \left( 1 + \frac{\phi''_{\infty}(Y_{\rm f})}{\Omega B\tan^2\alpha} \right) \frac{\Omega}{B} \frac{d}{d\tau} \bar \phi_1 (X_{\rm f} +  X_{0}(\tau), Y_{\rm f}, \mu_{\rm f})  \rm .
\end{align}
Integrating in $\tau$ and using $ w_{\parallel, 1/2} (\tau)\rvert_{\tau = 0} = 0$ gives 
\begin{align} \label{wpardotslow-1st-2}
& \frac{1}{2} \left( w_{\parallel, \rm f} + w_{\parallel, 1/2}(\tau) \right)^2 - \frac{1}{2} w_{\parallel, \rm f}^2  \nonumber \\
& = \left( 1 + \frac{\phi''_{\infty}(Y_{\rm f})}{\Omega B\tan^2\alpha} \right)  \frac{\Omega}{B} \left( \bar \phi_1 (X_{\rm f}, Y_{\rm f}, \mu_{\rm f}) - \bar \phi_1 (X_{\rm f} + X_0(\tau), Y_{\rm f}, \mu_{\rm f}) \right) \rm ,
\end{align}
which is solved by
\begin{align} \label{wparhalf-final}
& w_{\parallel, 1/2}(\tau)   =  -  w_{\parallel, \rm f} \nonumber  \\
& + \left[ w_{\parallel, \rm f}^2 - \left( 1 +  \frac{\phi''_{\infty}( Y) }{\Omega B\tan^2 \alpha} \right) \frac{2\Omega}{B} \left(  \bar \phi_1 (X_{\rm f} + X_0(\tau), Y_{\rm f}, \mu_{\rm f}) - \bar \phi_1 (X_{\rm f}, Y_{\rm f}, \mu_{\rm f})  \right) \right]^{1/2}  \rm .
\end{align} 
Recall that we assume all ions in the system to have $dX/d\tau = w_{\parallel}(\tau) \sin \alpha > 0$, and so we have used $w_{\parallel, \rm f} + w_{\parallel, 1/2}(\tau)  > 0$ to solve (\ref{wpardotslow-1st-2}).
The correction $\mu_{1/2}(\tau)$ is straightforwardly obtained from (\ref{mudot-1st-slow}) using $ \mu_{1/2} (\tau) \rvert_{\tau = 0} = 0$,
\begin{align}
\mu_{1/2}(\tau) = 0 \rm .
\end{align}
The largest non-zero correction to $\mu$ is of order $O(\hat \phi c_{\rm S} \rho_{\rm S})$, and is given by equation (\ref{mu1}) also for the slow ions.
However, as previously hinted this correction will turn out to be unnecessary.

\subsection{Velocity distribution} \label{subapp-slowtraj-2}

The ordering (\ref{ordering-slow}) allows us to identify slow ion trajectories, but it does not allow us to distinguish between slow and bulk trajectories when $\hat \phi^{1/2} c_{\rm S} \ll w_{\parallel, \rm f} \ll c_{\rm S}$.
Such a distinction can be made by introducing an arbitrary split at $w_{\rm cut}$, such that a trajectory is considered to be a slow one if $w_{\parallel, \rm f} < w_{\rm cut}$ and a bulk one if $w_{\parallel, \rm f} > w_{\rm cut}$.
The arbitrary cut-off value $w_{\rm cut}$ must satisfy $\hat \phi^{1/2} c_{\rm S} \ll w_{\rm cut} \ll 1 $ such that
\begin{align} \label{wcuthat}
\hat \phi \ll \hat \phi_{\rm cut} \equiv \frac{w_{\rm cut}^2}{c_{\rm S}^2} \ll 1  \rm ,
\end{align}
with $\hat \phi_{\rm cut}$ being a corresponding dimensionless small parameter.

With this cut-off we can define the full ion distribution function (including slow ions) as
\begin{align} \label{F-full}
F_{\rm full} (\vec G) = \begin{cases} F(\vec G) & \text{ for } w_{\parallel} \geqslant w_{\rm cut} \rm , \\
F_{\rm slow}(\vec G) & \text{ for } w_{\parallel} < w_{\rm cut} \rm .
\end{cases}
\end{align}
The slow ion distribution function $F_{\rm slow}(\vec G)$ is calculated by exploiting the expansion of the ion trajectories and conservation of the distribution function along the ion characteristics.
Similarly to what is done in section~\ref{subsec-traj-F}, we define $w_{\parallel, \infty} (\vec G_{\rm f}) = w_{\parallel}(\tau) \rvert_{\tau \rightarrow\infty}$ and $w_{\parallel, 1/2, \infty} (\vec G_{\rm f}) = w_{\parallel, 1/2}(\tau) \rvert_{\tau \rightarrow \infty}$ such that
\begin{align} \label{vpar-exp-infty-slow}
v_{\parallel, \infty} (\vec G)  = v_{\parallel} +  v_{\parallel,\infty, 1/2} (\vec G)  + O\left(\hat \phi c_{\rm S}, \delta \hat \phi^{-1/2}  c_{\rm S} \right) \rm ,
\end{align}
\begin{align} \label{Ypar-exp-infty-slow}
Y_{ \infty} (\vec G)  = Y +  Y_{\infty, 1/2} (\vec G)  + O\left(\hat \phi L, \delta \hat \phi^{-1/2} L \right) \rm ,
\end{align}
\begin{align} \label{wpar-exp-infty-slow}
w_{\parallel, \infty} (\vec G)  = w_{\parallel} +  w_{\parallel,\infty, 1/2} (\vec G)  + O\left(\hat \phi c_{\rm S}, \delta \hat \phi^{-1/2}  c_{\rm S} \right) \rm ,
\end{align}
are expansions valid for slow ions with $w_{\parallel} \sim \hat \phi^{1/2} c_{\rm S}$.
Evaluating (\ref{wparhalf-final}) at $\tau \rightarrow \infty$, we obtain
\begin{align} \label{wpar-1/2}
w_{\parallel, \infty, 1/2} (\vec G) = \left[ \left( w_{\parallel}^2 - w_{\text{min}, 1/2}^2(X, Y, \mu)  \right)^{1/2} - w_{\parallel} \right]
\end{align}
with
\begin{align} \label{wmin}
w_{\text{min}, 1/2}(X, Y, \mu) = \left[ - \left( 1 +  \frac{\phi''_{\infty}( Y) }{\Omega B\tan^2 \alpha} \right) \frac{2\Omega}{B} \bar \phi_1 (X, Y, \mu)  \right]^{1/2}  \rm .
\end{align}
Given the result (\ref{wpar-1/2}) and the assumption that ions enter the magnetised sheath only at $X/l_{\rm ms} \rightarrow \infty$ and do not reflect within the sheath, the slow ion distribution function $F_{\rm slow} (\vec G)$ must be zero in the interval $[0, w_{\rm min})$,
\begin{align} \label{Fslow-zerobelowmin}
F_{\rm slow} (\vec G) = 0 \text{ for } w_{\parallel} \in [0, w_{\rm min}(X, Y, \mu)) \rm .
\end{align}
An ion entering the magnetised sheath at $x/l_{\rm ms} \rightarrow \infty$ with the minimum possible net parallel velocity just above zero, $w_{\parallel, \infty} = 0^+$, will reach $X/l_{\rm ms} \gg 1$ with $w_{\parallel} = w_{\rm min}(\vec G)$, where we have calculated $w_{\rm min}(\vec G) = w_{\text{min}, 1/2}(X, Y \mu) + O(\hat \phi c_{\rm S})$.

The slow ion distribution function is represented as
\begin{align} \label{Fslow-exp}
F_{\rm slow}(\vec G)   = F_{\infty}(Y, \mu, v_{\parallel}) +  \Delta F_{\text{slow}}(\vec G)  \rm .
\end{align}
From (\ref{F-sol-prequel}) and the expansions (\ref{vpar-exp-infty-slow})-(\ref{Ypar-exp-infty-slow}), we obtain
 \begin{align} \label{F-slow-0}
\Delta F_{\rm slow} (\vec G) = O(\hat \phi^{1/2} n_{\infty} v_{\rm t, i}^{-3} ) \text{}
\end{align}
if $F_{\infty} = O(n_{\infty} v_{\rm t, i}^{-3})$, and
\begin{align} \label{F-slow}
\Delta F_{\rm slow} (\vec G) = & F_{\infty} \left( Y + Y_{\infty,1/2}(\vec G) , v_{\parallel} + v_{\parallel, \infty, 1/2}(\vec G) ,  \mu \right) - F_{\infty}(Y, \mu, v_{\parallel}) \nonumber \\
 & + O(\hat \phi^{(n+1)/2} n_{\infty} v_{\rm t, i}^{-3} )
\end{align}
if $F_{\infty} \propto v_{\parallel}^n$ for small $v_{\parallel}$.
Here, the exponent $n$ is given by $n=1$ if (\ref{F-nozero}) is satisfied, $n=2$ if (\ref{dF-nozero}) is also satisfied, $n=3$ if (\ref{d2F-nozero}) is also satisfied, and $n>3$ if (\ref{d3F-nozero}) is also satisfied.
Hence, $\Delta F_{\rm slow}$ can be obtained for $n > 0$ by Taylor expanding (\ref{F-slow}) up to the relevant order, depending on which of conditions (\ref{dF-nozero}), (\ref{d2F-nozero}) and (\ref{d3F-nozero}) is not satisfied.

From the definition (\ref{wpar}) and from $v_{\parallel, \infty, n} = Y_{\infty, n} \Omega \tan \alpha$, we extract the relations
\begin{align} \label{wparn-slow}
& v_{\parallel, \infty, 1/2}(\vec G) = Y_{\infty, 1/2}(\vec G) \Omega \tan \alpha =  w_{\parallel, \infty, 1/2} \left( 1 + \frac{\phi''_{\infty}(Y) }{\Omega B\tan^2 \alpha} \right)^{-1} \text{ for } w_{\parallel} < w_{\rm cut} \text{, }  
\end{align}
\begin{align} \label{wparn-bulk}
& v_{\parallel, \infty, 1}(\vec G) = Y_{\infty, 1}(\vec G) \Omega \tan \alpha =  w_{\parallel, \infty, 1} \left( 1 + \frac{\phi''_{\infty}(Y) }{\Omega B\tan^2 \alpha} \right)^{-1} \text{ for } w_{\parallel} > w_{\rm cut} \text{, }  
\end{align}
valid for the largest correction terms in the slow and bulk trajectory expansions, respectively.
Combining (\ref{wparn-slow}) and (\ref{wpar-1/2}) gives
\begin{align} \label{vpar1/2}
v_{\parallel, \infty, 1/2}(\vec G)  & = \Omega Y_{\infty, 1/2}(\vec G) \tan \alpha  \nonumber \\
&  = \left[ \left( w_{\parallel}^2 - w_{\text{min}, 1/2}^2 (X, Y, \mu) \right)^{1/2} - w_{\parallel} \right] \left(1 +  \frac{\phi''_{\infty}( Y) }{\Omega B\tan^2 \alpha} \right)^{-1} \rm .
\end{align}
We note that (\ref{vpar1/2}) recovers (\ref{vpar-Y1}) for $w_{\parallel} = w_{\rm cut}$ upon expanding the square root using $w_{\parallel}^2 \gg \hat \phi c_{\rm S}$,
\begin{align} \label{vpar1/2-exp}
v_{\parallel, \infty, 1/2}(\vec G) \rvert_{w_{\parallel} = w_{\rm cut}} = \frac{\Omega \bar \phi_1}{B w_{\rm cut}} + O(\hat \phi^2 \hat \phi_{\rm cut}^{-3/2} c_{\rm S} )  \rm .
\end{align}
Indeed, using
\begin{align} \label{Phipolslow}
\Phi_{\rm pol} ( \vec G ) \rvert_{w_{\parallel} = w_{\rm cut}} & = \Phi_{\rm pol} ( \vec G ) \rvert_{w_{\parallel} = 0} + O(\hat \phi_{\rm cut}^{1/2} |\phi_1| )  
\nonumber \\
& = \phi_1(X+\rho_x(\mu, \theta),Y) - \bar \phi_1 (X, Y, \mu) + O(\hat \phi_{\rm cut}^{1/2} |\phi_1|) \rm ,
\end{align}
equation (\ref{vpar-Y1}) gives
\begin{align}
v_{\parallel, \infty, 1}(\vec G) \rvert_{w_{\parallel} = w_{\rm cut}}  =  \frac{\Omega \bar \phi_1}{B w_{\rm cut}} + O(\hat \phi c_{\rm S} ) \rm .
\end{align}
for $w_{\parallel} = w_{\rm cut}$, which matches (\ref{vpar1/2-exp}).
This verifies that the slow ion correction of $v_{\parallel, \infty, 1/2}(\vec G) \rvert_{w_{\parallel} = w_{\rm cut}}$ is consistent with the bulk ion correction $ v_{\parallel, \infty, 1}(\vec G) \rvert_{w_{\parallel} = w_{\rm cut}}$ at the lowest order $ \sim \hat \phi \hat \phi_{\rm cut}^{-1/2} c_{\rm S}$. 

\section{Poisson's equation including slow ions} \label{app-slowdens}

In this appendix we consider a more complete perturbative analysis of Poisson's equation near the magnetised sheath entrance, including the slow ion density.

By integrating over the full ion distribution function (including slow ions) defined in (\ref{F-full}), the ion density integral can be split in a part corresponding to slow ions and a part corresponding to the bulk ions,
\begin{align} \label{nisplit-def}
n_{\rm i}(x, y) = & \int_0^{\infty} dY \int_0^{\infty} dX \int_0^{\infty} \Omega d\mu \int_0^{2\pi} d\theta \delta_{\rm Dirac} \left( X - x + \rho_x (\mu, \theta ) \right) \delta_{\rm Dirac} (Y - y)  \nonumber \\
& \times \left[ \int_{w_{\rm cut}}^{\infty} d w_{\parallel} F (\vec G)  + \int_{w_{\rm min}(\vec G)}^{w_{\rm cut}} d w_{\parallel}  F_{\rm slow} (\vec G) \right]   \rm .
\end{align}
In (\ref{nisplit-def}) we have used (\ref{Fslow-zerobelowmin}) to change the lower limit of integration in $w_{\parallel}$ in the slow ion integral to $w_{\rm min}(\vec G)$. 
In the main sections of this paper, we calculated the contributions to the ion density expansion coming from the first term in (\ref{nisplit-def}), and simply set $w_{\rm cut} = 0$.
The resulting ion density expansion, equation (\ref{ni-asymptotic}), is only accurate if (\ref{F-nozero})-(\ref{dF-nozero}) and (\ref{d2F-nozero})-(\ref{d3F-nozero}) are satisfied.
Here, we proceed to more carefully expand the ion density as follows:
if (\ref{F-nozero}) is not satisfied,
\begin{align} \label{ni-exp-slow-1}
n_{\text{i}} = n_{\text{i}, \infty} + n_{\text{i},1/2} + O\left( \hat \phi n_{\infty}, \hat \phi \ln \left( 1/\phi \right) n_{\infty} \right) \rm ;
\end{align}
if (\ref{F-nozero}) is satisfied and (\ref{dF-nozero}) is not satisfied,
\begin{align} \label{ni-exp-slow-2}
n_{\text{i}} = n_{\text{i}, \infty} + n_{\text{i},1} + O(\hat \phi \hat \phi_{\rm cut}^{1/2} n_{\infty}, \hat \phi^2 \hat \phi_{\rm cut}^{-1} n_{\infty} ) \rm ;
\end{align}
if (\ref{F-nozero}) and (\ref{dF-nozero}) are satisfied, and (\ref{d2F-nozero}) is not satisfied
\begin{align} \label{ni-exp-slow-3}
n_{\text{i}} = n_{\text{i}, \infty} + n_{\text{i},1} + n_{\text{i},3/2} + O(\hat \phi \hat \phi_{\rm cut} n_{\infty}, \hat \phi^2 \hat \phi_{\rm cut}^{-1/2} n_{\infty}) \rm ;
\end{align}
if (\ref{F-nozero}), (\ref{dF-nozero}) and (\ref{d2F-nozero}) are satisfied, and (\ref{d3F-nozero}) is not satisfied
\begin{align} \label{ni-exp-slow-4}
n_{\text{i}} = n_{\text{i}, \infty} + n_{\text{i},1} + n_{\text{i},2} + O(\hat \phi \hat \phi_{\rm cut}^{3/2} n_{\infty}, \hat \phi^{3} \hat \phi_{\rm cut}^{-1}) \rm .
\end{align}
The density corrections are ordered such that $n_{\text{i}, n} = O( \hat \phi^n n_{\infty} )$, and some of the integer-$n$ terms are $O(\hat \phi^n \ln (1/\hat \phi) n_{\infty})$.
We proceed order by order starting from the unperturbed equation (\ref{ni-infty}) for $n_{\rm i,\infty}$ and using the recursive scheme
\begin{align} \label{n-ordern}
n_{\text{i}, n/2} =  n_{\text{i}} - n_{\text{i}, \infty} - \sum_{m=1}^{n-1} n_{\text{i},m/2} \rm 
\end{align}
to calculate each successive order $n \geqslant 1$.
At every order, we analyse Poisson's equation to check if a dominant balance allowing for a solution to $\phi_1$ exists.
We only proceed to the next order in cases for which such a balance does not exist.
To calculate the errors, we have used the error in equation (\ref{F-slow}) for $w_{\parallel} \sim \hat \phi^{1/2} c_{\rm S}$, and the expansion that led to equation (\ref{vpar1/2-exp}) for $\hat \phi^{1/2} c_{\rm S} \lesssim w_{\parallel} \lesssim \hat \phi_{\rm cut}^{1/2} c_{\rm S}$.
Recall that we are assuming that $F_{\infty} \propto v_{\parallel}^n$ for small $v_{\parallel}$.
We note that the neglected error terms are always subdominant.


The rest of this appendix is structured as follows.
In \ref{subapp-slow-F} we calculate $n_{\text{i},1/2}$ in (\ref{ni-exp-slow-1}) and show that at $O(\hat \phi^{1/2} n_{\infty})$ Poisson's equation only consists of one unbalanced term, which therefore has to be equal to zero, requiring (\ref{F-nozero}).
In \ref{subapp-slow-dF} we calculate $n_{\text{i},1}$ in (\ref{ni-exp-slow-2}) and show that at $O(\hat \phi n_{\infty})$ Poisson's equation cannot have a monotonically decaying potential profile unless (\ref{dF-nozero}) also holds.
Given the results (\ref{F-nozero}) and (\ref{dF-nozero}), both $n_{\text{i}, 1}$ in equation (\ref{ni-1}) and Poisson's equation in the form (\ref{Poisson-1-step2}) are recovered.
Then, in \ref{subapp-slow-d2F} and \ref{subapp-slow-d3F} we analyse Poisson's equation at higher order, which we have shown to be necessary if the kinetic Bohm-Chodura condition (\ref{kin-Chod}) is satisfied marginally as in (\ref{kin-Chod=}).
In \ref{subapp-slow-d2F} we calculate $n_{\text{i},3/2}$ in (\ref{ni-exp-slow-3}) and derive Poisson's equation at $O(\hat \phi^{3/2} n_{\infty})$ in the case that (\ref{d2F-nozero}) is not satisfied.
If (\ref{d2F-nozero}) is satisfied, it is necessary to analyse Poisson's equation to $O(\hat \phi^2 n_{\infty})$, which is done in \ref{subapp-slow-d3F} by first calculating $n_{\text{i},2}$ in (\ref{ni-exp-slow-4}) without assuming (\ref{d3F-nozero}). 
If (\ref{d3F-nozero}) is also satisfied, we show that both $n_{\rm i, 2}$ in (\ref{ni-2}) and the perturbed Poisson's equation (\ref{Poisson-highorder-final}) are recovered.

\subsection{Order $\hat \phi^{1/2} n_{\infty}$: proof of (\ref{F-nozero})} \label{subapp-slow-F}

If (\ref{F-nozero}) is not satisfied, the largest correction to the lowest order ion density in (\ref{ni-infty}) is of order $O(\hat \phi^{1/2} n_{\rm i, \infty} )$ and given by
\begin{align} \label{ni-1/2-final}
n_{\text{i}, 1/2}  =  -\int_0^{\infty} dX  \int_0^{\infty} \Omega d\mu \int_0^{2\pi} d\theta \delta_{\rm Dirac} (X - x + \rho_x (\mu, \theta)) \int^{w_{\text{min},1/2}}_{0} d w_{\parallel}  F_{\infty} \nonumber \\
+ O\left( \hat \phi \ln \left( \frac{\hat \phi_{\rm cut} }{ \hat \phi } \right)  n_{\infty} \right) \rm .
\end{align}
To obtain (\ref{ni-1/2-final}) we took $F = F_{\infty} + O(\hat \phi n_{\infty} v_{\rm t,i}^{-3})$ for bulk ions and $F_{\rm slow} = F_{\infty} + O(\hat \phi^{1/2} n_{\infty} v_{\rm t,i}^{-3})$ for slow ions, and thus rejoined the two integrals in (\ref{nisplit-def}) into a single integral, then subtracted off the lowest order ion density $n_{\text{i},\infty}$ in (\ref{ni-infty}).
We also took $w_{\rm min}(\vec G) = w_{\text{min},1/2}(X, Y, \mu) + O(\hat \phi c_{\rm S})$, making an $ O(\hat \phi n_{\infty})$ error in the density.
The size of the largest neglected terms in (\ref{ni-1/2-final}) is
estimated by integrating the largest neglected terms in the slow ion distribution function in the interval $w_{\parallel} \in [0, w_{\rm cut}]$.
In the interval $w_{\parallel} \in [0, w_{\rm min}]$ we also have $F_{\infty} = F_{\infty} \rvert_{w_{\parallel} = 0} + O(\hat \phi^{1/2} n_{\infty} v_{\rm t,i}^{-3})$, giving
\begin{align}
n_{\text{i}, 1/2} =  - 2\pi \int_0^{\infty} \Omega d\mu \langle w_{\text{min},1/2} \rvert_{X=x-\rho_x (\mu, \theta)} \rangle_{\theta} F_{\infty} \rvert_{w_{\parallel} = 0}  +  O\left( \hat \phi \ln \left( \frac{\hat \phi_{\rm cut} }{ \phi } \right)  n_{\infty} \right) \rm .
\end{align}
Since there is no fractional order in the electron density expansion to balance $n_{\text{i}, 1/2}$, Poisson's equation at $O(\hat \phi^{1/2} n_{\infty})$ is simply
\begin{align} \label{Poisson-1/2}
0 = - 2\pi Z \int_0^{\infty} \Omega d\mu   \langle w_{\text{min},1/2} \rvert_{X=x-\rho_x (\mu, \theta)} \rangle_{\theta} F_{\infty}  \rvert_{w_{\parallel} = 0}   \rm .
\end{align}
Hence, (\ref{Poisson-1/2}) cannot be satisfied at $O(\hat \phi^{1/2} n_{\infty}) $ unless (\ref{F-nozero}) is satisfied, i.e. $F_{\infty} \rvert_{w_{\parallel} = 0} = 0$. 
To solve for $\phi_1$, it is necessary to analyse Poisson's equation at $O(\hat \phi n_{\infty}) $.

Before proceeding further, we note that any function $\mathcal{F}(v_{\parallel}, Y) $ satisfying $\mathcal F \rvert_{w_{\parallel} = 0} = 0$, satisfies the relation
\begin{align} \label{weirdequality}
(\partial_{w_{\parallel}} \mathcal{F} )_{Y_{\star}} \rvert_{w_{\parallel} = 0} = \partial_{v_{\parallel}} \mathcal{F} \rvert_{w_{\parallel} = 0} = \frac{ B\tan\alpha }{ \phi''_{\infty}(Y) } \partial_Y \mathcal{F} \rvert_{w_{\parallel} = 0} \rm .
\end{align}
This can be applied to $\mathcal{F} = F_{\infty} $.

\subsection{Order $\hat \phi n_{\infty}$: proof of (\ref{dF-nozero})} \label{subapp-slow-dF}

Since $n_{\text{i},1/2} = 0$, from (\ref{n-ordern}) we consider again the difference between equation (\ref{nisplit-def}) for the ion density $n_{\rm i}$ and equation (\ref{ni-infty}) for its lowest order asymptotic approximation $n_{\rm i, \infty}$.
This time, however, we take $F \simeq F_{\infty} + F_1$ and $F_{\rm slow} = F_{\infty} + \Delta F_{\rm slow}$ and keep terms up to and including order $O(\hat \phi n_{\infty})$,
\begin{align} \label{ni-1-1}
n_{\rm i,1} = & \int_0^{\infty} dX \int_0^{\infty} \Omega d\mu \int_0^{2\pi} d\theta \delta (X + \rho_x(\mu, \theta) - x)   \left[ \int_{w_{\rm cut}}^{\infty} d w_{\parallel}  F_1 \right. \nonumber \\
& \left.  +  \int_{w_{\rm min}}^{w_{\rm cut}} d w_{\parallel} \Delta F_{\text{slow}} - \int_0^{w_{\rm min}} F_{\infty}  dw_{\parallel} \right]  \rm .
\end{align}
In the following analysis, we will use the relation
\begin{align} \label{wpar-partialF}
v_{\parallel, \infty, n} \partial_{v_{\parallel}} F_{\infty} + Y_{\infty, n} \partial_{Y} F_{\infty} = w_{\parallel, \infty, n}  ( \partial_{w_{\parallel}} F_{\infty} )_{Y_{\star}} \rm , 
\end{align}
with $n = 1/2$ for $w_{\parallel} < w_{\rm cut}$ and $n=1$ for $w_{\parallel} > w_{\rm cut}$, which is obtained by combining (\ref{change-global}) and either (\ref{wparn-slow}) for $n=1/2$ or (\ref{wparn-bulk}) for $n=1$.
Concerning the bulk ion contribution to the density, we re-express $F_1$ in (\ref{F1}) using (\ref{wpar-partialF})  to obtain
\begin{align} \label{F-1-app}
F_1 = \mu_{\infty,1} \partial_{\mu} F_{\infty} + w_{\parallel, \infty, 1} (\partial_{w_{\parallel}} F_{\infty})_{Y_{\star}} \rm .
\end{align}
The correction $w_{\parallel, \infty, 1}$ is obtained explicitly from (\ref{vpar-Y1}) and (\ref{wparn-bulk}) (also using (\ref{wpar})),
\begin{align} \label{wpar1}
w_{\parallel, \infty, 1} = \left( 1 + \frac{\phi''(Y)}{\Omega B \tan^2 \alpha} \right) \frac{\Omega}{B w_{\parallel}}  \left( \phi_1 (X + \rho_x (\mu, \theta), Y ) - \Phi_{\rm pol}(X, Y, \mu, \theta, v_{\parallel})  \right) \text{.}
\end{align}
For slow ions in the interval $w_{\parallel} \in [0, w_{\rm cut}]$, we Taylor expand (\ref{F-slow}) to obtain, using the definition of $\Delta F_{\rm slow}$ in (\ref{Fslow-exp}),
\begin{align} \label{DeltaFslow-1}
\Delta F_{\rm slow} (X, Y, \mu, v_{\parallel}, \theta) =  Y_{\infty,1/2} \partial_Y F_{\infty} + v_{\parallel, \infty, 1/2} \partial_{v_\parallel} F_{\infty} + O( \hat \phi n_{\infty} v_{\rm t,i}^{-3} ) \text{.}
\end{align}
By using (\ref{wpar-partialF}), this is more compactly re-expressed as
\begin{align} \label{DeltaFslow}
\Delta F_{\rm slow} (X, Y, \mu, v_{\parallel}, \theta) =  w_{\parallel, \infty, 1/2} (\partial_{w_{\parallel}} F_{\infty})_{Y_{\star}} + O( \hat \phi n_{\infty} v_{\rm t,i}^{-3} ) \text{.}
\end{align} 
We can further Taylor expand (\ref{DeltaFslow}) around $w_{\parallel} = 0$ and use (\ref{weirdequality}) with $\mathcal{F} = F_{\infty} $ to obtain
\begin{align} \label{Fslow-1/2}
\Delta F_{\text{slow}} =  w_{\parallel, \infty, 1/2}  \partial_{v_{\parallel}} F_{\infty}  \rvert_{w_{\parallel} = 0} + O(\hat \phi_{\rm cut}^{1/2} \hat \phi^{1/2} n_{\infty} v_{\rm t,i}^{-3})  \rm .
\end{align}
The final contribution to the first order perturbation to the ion density is the subtraction of the small piece of $n_{\infty}$ coming from having integrated, in equation (\ref{ni-infty}), also across the interval $w_{\parallel} \in [0, w_{\rm min}]$ where the distribution function $F$ (not $F_{\infty}$) is empty. 
In this interval, we may Taylor expand to obtain, using (\ref{F-nozero}),
\begin{align} \label{Finfty-1/2}
F_{\infty} = w_{\parallel} \partial_{v_{\parallel}} F_{\infty} \rvert_{w_{\parallel} = 0} + O(\hat \phi n_{\infty} v_{\rm t,i}^{-3}) \rm .
\end{align}
Upon inserting (\ref{F-1-app}), (\ref{Fslow-1/2}), (\ref{Finfty-1/2}), (\ref{wpar-1/2}), (\ref{wpar1}) and (\ref{mu1}) into (\ref{ni-1-1}), we obtain
\begin{align} \label{ni-1-star}
n_{\rm i,1} = & \int_0^{\infty} dX  \int_0^{\infty} \Omega d\mu \int_0^{2\pi} d\theta \delta_{\rm Dirac} (X - x + \rho_x (\mu, \theta))   \nonumber \\
& \times \left[ \int_{w_{\rm cut}}^{\infty} d w_{\parallel} \left(  \frac{ \Phi_{\rm pol}}{B} \partial_\mu F_{\infty} +  \left( 1 + \frac{\phi''_{\infty}}{\Omega B\tan\alpha} \right)  \frac{\Omega \left( \phi_1 -  \Phi_{\rm pol} \right)}{B w_{\parallel} } (\partial_{w_{\parallel}} F_{\infty})_{Y_{\star}}   \right) \right.  \nonumber \\
& \left.  + \partial_{v_{\parallel}} F_{\infty} \rvert_{w_{\parallel} = 0} \left( \int_{w_{\text{min},1/2}}^{w_{\rm cut}} d w_{\parallel}  \left( \sqrt{w_{\parallel}^2 - w_{\rm min, 1/2}^2 } - w_{\parallel} \right) - \int_0^{w_{\rm min, 1/2}} w_{\parallel}  dw_{\parallel} \right) \right]  \rm ,
\end{align} 
where we have dropped terms of order $O(\hat \phi^{3/2} n_{\infty}) $.
Note that we have changed the integration limit from $w_{\rm min}$ to $w_{\rm min, 1/2}$ making $O(\hat \phi^{3/2} n_{\infty}) $ errors. 

We proceed to show that (\ref{ni-1-star}) can be reformulated such that it is independent of the arbitrary parameter $w_{\rm cut}$.
Integrating by parts the term $ w_{\parallel}^{-1}$ in (\ref{ni-1-star}) gives
\begin{align} \label{ni-isolate-1-fast}
& \left( 1 +  \frac{\phi''_{\infty} }{\Omega B\tan^2 \alpha} \right) \int_{w_{\rm cut}}^{\infty} d w_{\parallel}   \frac{\Omega (\phi_1 -  \Phi_{\rm pol} )}{Bw_{\parallel}} ( \partial_{w_{\parallel}} F_{\infty} )_{Y_{\star}}
\nonumber \\
& =  \left( 1 +  \frac{\phi''_{\infty} }{\Omega B\tan^2 \alpha} \right)   \int_{w_{\rm cut}}^{\infty} d w_{\parallel}   \ln w_{\parallel} \partial_{v_{\parallel}} \left( \frac{\Omega }{B} (\phi_1 - \Phi_{\rm pol})  (\partial_{w_{\parallel}} F_{\infty})_{Y_{\star}} \right)  \nonumber \\
&  + \frac{1}{2} \left( 1 +  \frac{\phi''_{\infty} }{\Omega B\tan^2 \alpha} \right) \frac{\Omega}{B} \left( \phi_1 - \Phi_{\rm pol} \rvert_{w_{\parallel} = w_{\rm cut}} \right)  \ln w_{\rm cut} \partial_{v_{\parallel}} F_{\infty} \rvert_{w_{\parallel} = w_{\rm cut}} 
  \rm .
\end{align} 
To simplify (\ref{ni-isolate-1-fast}) further, we set $w_{\rm cut} = 0$ everywhere except in $\ln w_{\rm cut}$, thus dropping terms of order $O(\hat \phi \hat \phi_{\rm cut}^{1/2} n_{\infty})$ in the contribution to the density.
In the boundary term of the integration by parts, we also use the limiting result
\begin{align}
- \left( 1 +  \frac{\phi''_{\infty} }{\Omega B\tan^2 \alpha} \right) \frac{\Omega}{B} \left( \phi_1 - \Phi_{\rm pol} \rvert_{w_{\parallel} = 0} \right) = \frac{1}{2} w_{\rm min, 1/2}^2   \rm ,
\end{align}
which is deduced from (\ref{Phipolslow}) and (\ref{wmin}).
Thus, the final result is
\begin{align} \label{ni-isolate-1-fast}
& \left( 1 +  \frac{\phi''_{\infty} }{\Omega B\tan^2 \alpha} \right) \int_{w_{\rm cut}}^{\infty} d w_{\parallel}   \frac{\Omega (\phi_1 -  \Phi_{\rm pol} )}{Bw_{\parallel}} ( \partial_{w_{\parallel}} F_{\infty} )_{Y_{\star}}
\nonumber \\
& =  \left( 1 +  \frac{\phi''_{\infty} }{\Omega B\tan^2 \alpha} \right)   \int_{0}^{\infty} d w_{\parallel}   \ln w_{\parallel} \partial_{v_{\parallel}} \left( \frac{\Omega }{B} (\phi_1 - \Phi_{\rm pol})  (\partial_{w_{\parallel}} F_{\infty})_{Y_{\star}} \right)  \nonumber \\
&  + \frac{1}{2} w_{\rm min, 1/2}^2  \ln w_{\rm cut} \partial_{v_{\parallel}} F_{\infty} \rvert_{w_{\parallel} = 0} + O \left( \hat \phi \hat \phi_{\rm cut}^{1/2} \frac{n_{\infty}}{c_{\rm S}^2} \right)
  \rm .
\end{align} 
A final observation about (\ref{ni-isolate-1-fast}) is that the partial derivative with respect to $w_{\parallel}$ that appears when integrating by parts is performed at fixed $y$ and is therefore denoted equivalently using $\partial_{v_{\parallel}}$ (which throughout this paper is understood to be performed at fixed $y$).
Using
\begin{align} \label{w12wpar-exp}
\int_{w_{\rm min, 1/2}}^{w_{\rm cut}} dw_{\parallel} \left(w_{\parallel}^2 - w_{\rm min,1/2}^2 \right)^{1/2} 
=  \frac{1}{2}  w_{\rm cut} \left( w_{\rm cut}^2 - w_{\rm min,1/2}^2 \right)^{1/2}    \nonumber \\
 - \frac{1}{2} w_{\rm min, 1/2}^2  \ln \left( \frac{w_{\rm cut} + \left( w_{\rm cut}^2 - w_{\rm min,1/2}^2 \right)^{1/2}}{w_{\rm min, 1/2}} \right) \nonumber \\
 =   \frac{1}{2} w_{\rm cut}^2 - \frac{1}{4} w_{\rm min,1/2}^2  - \frac{1}{2} w_{\rm min,1/2}^2  \ln \left( \frac{ 2w_{\rm cut} }{w_{\rm min,1/2}} \right)  + O \left(\frac{w_{\rm min,1/2}^4}{w_{\rm cut}^2} \right)   \rm ,
\end{align}
where in the second equality we expanded in $w_{\rm min,1/2} \sim \hat \phi^{1/2} c_{\rm S} \ll w_{\rm cut} \sim \hat \phi_{\rm cut}^{1/2} c_{\rm S} $, we calculate the terms of (\ref{ni-1-star}) coming from the slow ions, 
\begin{align} \label{bbb}
& \partial_{v_{\parallel}} F_{\infty} \rvert_{w_{\parallel} = 0} \left(  \int_{w_{\rm min,1/2}}^{w_{\rm cut}} d w_{\parallel} \left[  \left(  w_{\parallel}^2 - w_{\rm min,1/2}^2 \right)^{1/2} -  w_{\parallel} \right] - \int_0^{w_{\rm min,1/2}} w_{\parallel} dw_{\parallel} \right) \nonumber \\
& =  - (\partial_{w_{\parallel}} F_{\infty})_{Y_{\star}} \rvert_{w_{\parallel} = 0}  \frac{1}{4} w_{\rm min,1/2}^2  \left( 1 + 2 \ln \left(\frac{ 2w_{\rm cut} }{w_{\rm min,1/2}} \right) \right) + O \left( \hat \phi \hat \phi_{\rm cut}^{1/2} \frac{n_{\infty}}{c_{\rm S}^2},  \frac{\hat \phi^2}{ \hat \phi_{\rm cut}} \frac{n_{\infty}}{c_{\rm S}^2} \right)\rm .
\end{align}
Upon combining (\ref{ni-isolate-1-fast}) and (\ref{bbb}), the terms involving $\ln w_{\rm cut}$ cancel (as hoped), making the expansion at $O(\hat \phi n_{\infty})$ independent of $w_{\rm cut}$,
\begin{align} \label{ni-1-star-final}
n_{\rm i,1}(x, y) = & \int_0^{\infty} dX \int_0^{\infty} \Omega d\mu \int_0^{2\pi} \delta_{\rm Dirac} (X + \rho_x(\mu, \theta ) - x) \left( \int_{0}^{\infty} d w_{\parallel}  \left[ \frac{ \Phi_{\rm pol} }{B} \partial_\mu F_{\infty} \right. \right. \nonumber \\ 
& \left. \left. -  \left( 1 + \frac{\phi''_{\infty}}{\Omega B \tan^2 \alpha} \right) \ln w_{\parallel}  \partial_{v_{\parallel}} \left( \frac{\Omega}{B} (\phi_1 -  \Phi_{\rm pol}) (\partial_{w_{\parallel}} F_{\infty})_{Y_{\star}} \right) \right]  \right. \nonumber \\
& \left. -  \partial_{v_{\parallel}} F_{\infty} \rvert_{w_{\parallel} = 0} \left( 1 + \frac{\phi''_{\infty}}{\Omega B \tan^2 \alpha} \right) \frac{\Omega (\phi_1 - \Phi_{\rm pol})}{B} \ln \left( \frac{ w_{\rm min,1/2} }{2e^{1/2}} \right) \right)  \rm .
\end{align}
The terms that have been neglected to arrive at (\ref{ni-1-star-final}) from (\ref{ni-1-1}) are of order $ O( \hat \phi \hat \phi_{\rm cut}^{1/2} n_{\infty})$ and $ O( \hat \phi^2 \hat \phi_{\rm cut}^{-1} n_{\infty} )$, which are smaller than $ O( \hat \phi n_{\infty})$ by taking $\hat \phi \ll \hat \phi_{\rm cut} \ll 1$. 
Note that the contribution to the error from the neglected term $F_2  \propto c_{\rm S}^4 \hat \phi^2 w_{\parallel}^{-3} \partial_{v_{\parallel}} F_{\infty} $ in the ion distribution function is also of order $O(\hat \phi^2 \hat \phi_{\rm cut}^{-1} n_{\infty})$.

By inserting (\ref{ni-1-star-final}) and (\ref{ne-1}) into (\ref{Poisson-1}), Poisson's equation at $O(\hat \phi n_{\infty})$ is
\begin{align} \label{Poisson-1-star}
& - \lambda_{\rm D}^2 \frac{n_{\rm e, \infty}e}{T_{\rm e}} \partial_x^2 \phi_1 = - n_{\rm e, \infty} \frac{e\phi_1}{T_{\rm e}} + Z\int_0^{\infty} dX \int_0^{\infty} \Omega d\mu \int_0^{2\pi} d\theta ~ \delta_{\rm Dirac} (X + \rho_x(\mu, \theta ) - x)  \nonumber \\
&  \times \left\lbrace \int_{0}^{\infty} d w_{\parallel} \left[ \frac{ \Phi_{\rm pol} }{B} \partial_\mu F_{\infty} -  \left( 1 + \frac{\phi''_{\infty}}{\Omega B \tan^2 \alpha} \right) \ln w_{\parallel}  \partial_{v_{\parallel}} \left( \frac{\Omega}{B} (\phi_1 -  \Phi_{\rm pol}) (\partial_{w_{\parallel}} F_{\infty})_{Y_{\star}} \right) \right]  \right. \nonumber \\
& \left. -  \partial_{v_{\parallel}} F_{\infty} \rvert_{w_{\parallel} = 0} \left( 1 + \frac{\phi''}{\Omega B \tan^2 \alpha} \right) \frac{\Omega (\Phi_{\rm pol} - \phi_1)}{B} \ln \left( \frac{2e^{1/2}}{ w_{\rm min, 1/2} } \right) \right\rbrace  \rm .
\end{align}
The left hand side is order $O(\epsilon^2 \hat \phi n_{\infty} \lambda_{\rm D}^2 / l_{\rm ms}^2 )$ and positive definite for a monotonic electron-repelling potential profile with $\partial_x^2 \phi_1 < 0$.
The first term on the right hand side, coming from the electron density perturbation, is of order $O(\hat \phi n_{\infty} )$.
In the ion density perturbation, the terms in the square bracket contribute to order $O(\hat \Phi n_{\infty} )$, where the small parameter $\hat \Phi = e(\Phi_{\rm pol} - \phi_1)/T_{\rm e} \gtrsim \hat \phi$ is introduced to explicitly allow for the case that $|\phi_1 (x - (v_{\perp}/ \Omega) \cos \alpha, y )| \gg |\phi_1(x,y)|$, which would imply that $\hat \Phi \gg \hat \phi$.
The final ionic term inside the curly bracket is of order $O(\hat \Phi \ln (1/\hat \Phi ) n_{\infty} )$ and is therefore the largest by a factor of $\ln (1/\hat \Phi )$ upon taking the subsidiary limit $\hat \phi \lesssim \hat \Phi \ll [\ln (1/\hat \Phi)]^{-1} \ll 1$.
This term is also negative owing to $\Phi_{\rm pol} > \phi_1$ (see equation (\ref{wmin})) and $\partial_{v_{\parallel}} F_{\infty} \rvert_{w_{\parallel} = 0} \geqslant 0$ (which follows from $F_{\infty} \rvert_{w_{\parallel} = 0} = 0$).
The term on the left hand side and the largest term on the right hand side can be of the same order only if $\epsilon^2 \hat \phi \lambda_{\rm D}^2 / l_{\rm ms}^2 \sim \hat \Phi \ln (1/\hat \Phi )$.
However, they must have opposite signs and so Poisson's equation cannot be satisfied unless both terms vanish at $O(\hat \Phi \ln (1/\hat \Phi))$, implying (\ref{dF-nozero}), i.e. $\partial_{v_{\parallel}} F_{\infty} \rvert_{w_{\parallel} = 0} = 0 = \partial_Y F_{\infty} \rvert_{w_{\parallel} = 0}$.
With this result, we recover (\ref{Poisson-1-step2}) upon integrating by parts (without any cancelling divergent terms appearing) the second term in the square bracket and shifting the terms involving $\Phi_{\rm pol}$ to the left hand side.

\subsection{Order $\hat \phi^{3/2} n_{\infty}$: (\ref{d2F-nozero}) not satisfied} \label{subapp-slow-d2F}

Let us now consider the effect of slow ions when the kinetic Chodura condition is marginally satisfied, and a higher order analysis of Poisson's equation is required.
Although we could exploit the ordering $\epsilon \ll 1$ as in section~\ref{sec-higher}, we perform the analysis at $O(\hat \phi^{3/2})$ without expanding in $\epsilon$, only taking $\epsilon \ll 1$ in the final step.

Following the prescription in (\ref{n-ordern}) and using $n_{\text{i},1/2} = 0$, the ion density at $O(\hat \phi^{3/2} n_{\infty})$ is obtained by subtracting equation (\ref{ni-infty}) for $n_{\rm i, \infty}$ and equation (\ref{ni-1st}) for $n_{\text{i}, 1}$ from equation (\ref{nisplit-def}) for $n_{\rm i}$ with $F \simeq F_{\infty} + F_1$ and $F_{\rm slow} = F_{\infty} + \Delta F_{\rm slow}$,
\begin{align} \label{ni32-pre}
n_{\text{i},3/2} = & 2\pi  \int_0^{\infty} \Omega d\mu  \left( \int_{w_{\text{min}}}^{w_{\rm cut}} d w_{\parallel} \Delta F_{\text{slow}}  -  \int_0^{w_{\text{min}}} d w_{\parallel}  F_{\infty}   - \int_0^{w_{\rm cut}} d w_{\parallel} F_{1} \right)  \rm .
\end{align}
Note that the final term in (\ref{ni32-pre}) corrects for the terms of order $O(\hat \phi \hat \phi_{\rm cut}^{1/2} n_{\infty})$ that were added into $n_{\text{i}, 1}$ by changing the lower limit of integration from $w_{\parallel} = w_{\rm cut} $ to $w_{\parallel} = 0$.
The distribution function $F_1$ in the interval $[0, w_{\rm cut}]$ is given by 
\begin{align} \label{F1-3/2}
F_1 = w_{\parallel, \infty, 1} w_{\parallel} \partial_{v_{\parallel}}^2 F_{\infty} \rvert_{w_{\parallel} = 0} + O\left( \hat \phi^2 n_{\infty} v_{\rm t,i}^{-3} \right)  \rm .
\end{align}
The zeroth order distribution function in the interval $w_{\parallel} \in [0, w_{\text{min},1/2}]$ is 
\begin{align} \label{Finfty-3/2}
F_{\infty} = \frac{1}{2} w_{\parallel}^2 \partial^2_{v_{\parallel}} F_{\infty} \rvert_{w_{\parallel} = 0} + O\left( \hat \phi^{3/2} n_{\infty} v_{\rm t,i}^{-3} \right) \rm .
\end{align}
The perturbed slow ion distribution function defined in (\ref{F-slow}) in the interval $[0, w_{\rm cut}]$ is now given by
\begin{align} \label{Fslow-1}
\Delta F_{\text{slow}} = & Y_{\infty,1/2} w_{\parallel} \partial_{v_{\parallel}} \partial_Y F_{\infty} \rvert_{w_{\parallel} = 0}   + v_{\parallel, \infty, 1/2} w_{\parallel} \partial_{v_{\parallel}}^2 F_{\infty} \rvert_{w_{\parallel} = 0}  + \frac{1}{2} Y_{\infty, 1/2}^2 \partial_Y^2 F_{\infty} \rvert_{w_{\parallel} = 0}  \nonumber \\
& + \frac{1}{2} v_{\parallel, \infty, 1/2}^2 \partial_{v_{\parallel}}^2 F_{\infty} \rvert_{w_{\parallel} = 0}  + v_{\parallel, \infty, 1/2} Y_{1/2} \partial_{v_{\parallel}}^2 F_{\infty} \rvert_{w_{\parallel} = 0} + O(\hat \phi \hat \phi_{\rm cut}^{1/2} n_{\infty} v_{\rm t,i}^{-3})  \rm .
\end{align} 
Owing to (\ref{F-nozero}) and (\ref{dF-nozero}), the terms in the expansion depending on neglected first order slow trajectory corrections $w_{\parallel, \infty, 1}$ and $\mu_{\infty, 1}$ in (\ref{F-slow}) only contribute to (\ref{Fslow-1}) at order $O(\hat \phi \hat \phi_{\rm cut}^{1/2} n_{\infty})$ and $O(\hat \phi \hat \phi_{\rm cut} n_{\infty})$, respectively, and are thus negligible.
To simplify (\ref{Fslow-1}) we use the relations
\begin{align} \label{ggg}
 ( v_{\parallel, \infty, 1/2} \partial_{v_{\parallel}} + Y_{\infty, 1/2} \partial_Y )^n F_{\infty} \rvert_{w_{\parallel} = 0} =  w_{\parallel, \infty, 1/2}^n (\partial_{w_{\parallel}}^n F_{\infty})_{Y_{\star}} \rvert_{w_{\parallel} = 0} \rm  ,
\end{align} 
\begin{align} \label{hhh}
\partial_{v_{\parallel}} (\partial_{w_{\parallel}}^{n-1} F_{\infty})_{Y_{\star}} \rvert_{w_{\parallel} = 0} =  \partial_{v_{\parallel}}^{n} F_{\infty} \rvert_{w_{\parallel} = 0} \rm 
\end{align}
with $n=2$.
Equation (\ref{ggg}) follows from (\ref{change-global}) provided that $\partial_{v_{\parallel}}^{m} F_{\infty} \rvert_{w_{\parallel} = 0}  = (\partial_{w_{\parallel}}^m F_{\infty})_{Y_{\star}} \rvert_{w_{\parallel} = 0} = 0$ for $m < n$.
Equation (\ref{hhh}) is deduced from (\ref{weirdequality}) with $\mathcal F = \partial_{v_{\parallel}} F_{\infty}$ upon exploiting the commutation property $(\partial_{v_{\parallel}} (\partial_{w_{\parallel}} F_{\infty})_{Y_{\star}})_Y = (\partial_{w_{\parallel}} (\partial_{v_{\parallel}} F_{\infty})_Y)_{Y_{\star}}$, which follows from (\ref{change-global}).
Equation (\ref{Fslow-1}) becomes
\begin{align} \label{newFslow-1}
\Delta F_{\text{slow}} = & w_{\parallel, \infty, 1/2} w_{\parallel} \partial_{v_{\parallel}}^2 F_{\infty} \rvert_{w_{\parallel} = 0} + \frac{1}{2} w_{\parallel, \infty, 1/2}^2  \partial_{v_{\parallel}}^2 F_{\infty} \rvert_{w_{\parallel} = 0} + O(\hat \phi \hat \phi_{\rm cut}^{1/2} n_{\infty} v_{\rm t,i}^{-3}) \rm .
\end{align}
Upon inserting (\ref{newFslow-1}), (\ref{F1-3/2}) and (\ref{Finfty-3/2}) into (\ref{ni32-pre}), we obtain
\begin{align} \label{ni32-1}
n_{\text{i},3/2} = & 2\pi  \int_0^{\infty} \Omega d\mu   \partial_{v_{\parallel}}^2 F_{\infty}  \rvert_{w_{\parallel} = 0} \left[ \int_{w_{\text{min},1/2}}^{w_{\rm cut}} d w_{\parallel} \left( w_{\parallel, \infty, 1/2} w_{\parallel} + \frac{1}{2} w_{\parallel, \infty, 1/2}^2   \right) \right. \nonumber \\
& \left. - \frac{1}{2} \int_0^{w_{\text{min}, 1/2}} d w_{\parallel}  w_{\parallel}^2  - \int_0^{w_{\rm cut}} d w_{\parallel} w_{\parallel, \infty, 1} w_{\parallel}  \right]  \rm ,
\end{align}
where we have changed the integration limit from $w_{\rm min}$ to $w_{\rm min, 1/2}$ making a small $O(\hat \phi^2 n_{\infty})$ error in the respective integrals, and we have neglected error terms up to order $O\left( \hat \phi \hat \phi_{\rm cut} n_{\infty} \right)$.
Inserting (\ref{wparn-slow}) for $w_{\parallel, \infty, 1/2}$ and using $w_{\parallel, \infty, 1} = - w_{\text{min},1/2}^2/ (2w_{\parallel})$ (valid in the slow trajectory interval $w_{\parallel} \in [0, w_{\rm cut}]$) into (\ref{ni32-1}) we obtain
\begin{align} \label{ni32-2}
n_{\text{i},3/2} = &  2\pi \int_0^{\infty} \Omega d\mu  \partial_{v_{\parallel}}^2 F_{\infty} \rvert_{w_{\parallel} = 0}   \left[ \int_{w_{\text{min}, 1/2}}^{w_{\rm cut}} d w_{\parallel} \left( \left[ \left(w_{\parallel}^2 - w_{\text{min}, 1/2}^2 \right)^{1/2} - w_{\parallel} \right] w_{\parallel} \right. \right.  \nonumber  \\
& \left. \left.  + \left[ \left(w_{\parallel}^2 - w_{\text{min}, 1/2}^2 \right)^{1/2} - w_{\parallel} \right]^2 \right) - \frac{1}{2} \int_0^{w_{\text{min}, 1/2}} dw_{\parallel}  w_{\parallel}^2 + \frac{1}{2}  \int_0^{w_{\rm cut}} dw_{\parallel}  w_{\text{min}, 1/2}^2   \right) \rm .
\end{align} 
Upon carrying out the integrals in (\ref{ni32-2}), using 
\begin{align}
\int_{w_{\rm min, 1/2}}^{w_{\rm cut}} d w_{\parallel}  \left[w_{\parallel}^2 - w_{\rm min, 1/2}^2 \right]^{1/2} w_{\parallel} & = \frac{1}{3} \left[w_{\rm cut}^2 - w_{\rm min, 1/2}^2 \right]^{3/2} \nonumber \\
& = \frac{1}{3} w_{\rm cut}^3 -  \frac{1}{2} w_{\rm min, 1/2}^2 w_{\rm cut} + O\left( \frac{w_{\rm min, 1/2}^4}{w_{\rm cut}} \right)  \rm ,
\end{align}
and neglecting terms of order $ O(  \hat \phi^2  \hat \phi_{\rm cut}^{-1/2} n_{\infty} )$, we obtain
\begin{align} \label{ni-3/2-final}
n_{\text{i},3/2} = & \frac{1}{3} \int_0^{\infty} dX \int_0^{\infty} \Omega d\mu \int_0^{2\pi} d\theta \delta_{\rm Dirac} (X + \rho_x (\mu, \theta) - x)  \partial_{v_{\parallel}}^2 F_{\infty} \rvert_{w_{\parallel} = 0} w_{\text{min},1/2}^3 \rm .
\end{align}
Note that the error term coming from neglecting $F_2 \sim c_{\rm S}^4 \hat \phi^2 w_{\parallel}^{-3} \partial_{v_{\parallel}} F_{\infty}$ is of order $ O(  \hat \phi^2  \hat \phi_{\rm cut}^{-1/2} n_{\infty} )$, which can be seen upon integrating this term by parts and using (\ref{dF-nozero}).
The left hand side of (\ref{Poisson-1-step2}) is unchanged, while the right hand side is now given by (\ref{ni-3/2-final}) as the $O(\hat \phi n_{\infty})$ terms have vanished due to the marginal Chodura condition,
\begin{align} \label{Poisson-3/2}
& - n_{\rm e, \infty} \lambda_{\rm D}^2  \frac{e}{T_{\rm e}} \partial_x^2 \phi_1  + 2\pi Z \int_0^{\infty} \Omega d\mu \int_{ - \frac{\phi'_{\infty}(y)}{B\tan \alpha}  }^{\infty} d v_{\parallel} \frac{\Omega \overline \Phi_{\rm pol} }{B}  \left[  \frac{ \partial_{v_{\parallel}} F_{\infty} + \frac{\partial_Y F_{\infty}}{\Omega \tan \alpha} }{  v_{\parallel} + \frac{\phi'_{\infty}(y)}{B\tan \alpha} } - \partial_{\mu} F_\infty   \right]  \nonumber \\
& = \frac{1}{3} Z \int_0^{\infty} dX \int_0^{\infty} \Omega d\mu \int_0^{2\pi} d\theta \delta_{\rm Dirac} (X + \rho_x (\mu, \theta) - x)  \partial_{v_{\parallel}}^2 F_{\infty} \rvert_{w_{\parallel} = 0} w_{\text{min}, 1/2}^3 \rm .
\end{align}
Equation (\ref{Poisson-3/2}) implies $\hat \Phi_{\rm pol} = e\Phi_{\rm pol} / T_{\rm e} \sim \hat \phi^{3/2} \ll \hat \phi$.
Since $\Phi_{\rm pol}$ depends on derivatives of $\phi_1$, this ordering requires $\epsilon \ll 1$ (as stated in the beginning of the subsection).
Neglecting $|\Phi_{\rm pol}| \ll |\phi_1|$ in the expression for $w_{\rm min, 1/2}$ and further using the expansion (\ref{Phipol-smallepsilon}) in $\epsilon \ll 1$ of $\overline \Phi_{\rm pol} $, Poisson's equation takes the form
\begin{align} \label{Poisson-3/2-0}
- \frac{e}{T_{\rm e}}  \partial_x^2  \phi_1  \left[ \left( \lambda_{\rm D}^2 + \rho_{\rm B}^2 \cos^2 \alpha \right) n_{\rm e, \infty} + 2\pi Z \rho_{\rm B}^2 \cos^2 \alpha \int_0^{\infty} \Omega d\mu \int_{ - \frac{\phi'_{\infty}(y)}{B\tan \alpha}  }^{\infty} d v_{\parallel} \times  \right. \nonumber \\
 \left. \frac{\Omega \mu }{ v_{\parallel} +\frac{\phi'_{\infty}(y)}{B\tan \alpha}  } \left( \partial_{v_{\parallel}} F_\infty + \frac{ \partial_Y F_\infty}{\Omega \tan \alpha } \right) \right] \nonumber \\
 = \frac{1}{3} Z \left( - \frac{2e \phi_1}{T_{\rm e}}  \right)^{3/2} \left(1 + \frac{\phi_{\infty}''(y)}{\Omega B \tan^2 \alpha} \right)^{3/2} v_{\rm B}^3 \int_0^{\infty} \Omega d\mu \int_0^{2\pi} d\theta  \partial_{v_{\parallel}}^2 F_{\infty} \rvert_{w_{\parallel} = 0} \rm ,
 \end{align}
This equation always has a solution for $\phi_1$ if the polarisation condition (\ref{polcond-3}) is satisfied because $\partial_{v_{\parallel}}^2 F_{\infty} \rvert_{w_{\parallel} = 0} \geqslant 0$ (which follows from (\ref{F-nozero}) and (\ref{dF-nozero})) makes the right hand side positive definite.
 If (\ref{d2F-nozero}) is satisfied, i.e. $\partial_{v_{\parallel}}^2 F_{\infty} \rvert_{w_{\parallel} = 0} = 0$, the right hand side of (\ref{Poisson-3/2}) becomes zero to all orders in $\epsilon$, and determining the form of $\phi_1$ requires analysing $O(\hat \phi^2 n_{\infty})$ terms.
This is done in the next subsection.

\subsection{Order $\hat \phi^2 n_{\infty}$: (\ref{d2F-nozero}) satisfied, (\ref{d3F-nozero}) not satisfied} \label{subapp-slow-d3F}

In this subsection we exploit the ordering $\epsilon \ll 1$ and replace all quantities can by their zeroth order in $\epsilon$.
Hence, we take $\bar \phi_1 \rvert_{X=x- \rho_x(\mu, \theta)} \simeq \phi_1  (x, y)$ such that $w_{\text{min},1/2} = w_{\text{min},1/2, 0}+ O(\epsilon^2 \hat \phi^{1/2} c_{\rm S} )$ with
\begin{align} \label{wmin0}
w_{\text{min},1/2, 0} = \left( - \left( 1 +  \frac{\phi''_{\infty} }{\Omega B\tan^2 \alpha} \right) \frac{2\Omega}{B}  \phi_1  (x, y) \right)^{1/2}   \rm .
\end{align}
We also have $w_{\parallel, \infty, n} \simeq w_{\parallel, \infty, n, 0}$ with
\begin{align} \label{wparhalf0}
w_{\parallel, \infty, 1/2, 0} = \sqrt{ w_{\parallel}^2 - w_{\text{min},1/2,0}^2 } - w_{\parallel} \rm ,
\end{align}
\begin{align} \label{wparhalf0}
w_{\parallel, \infty, 1, 0} =  - \frac{w_{\text{min}, 1/2, 0}^2}{2w_{\parallel}} = \left( 1 + \frac{\phi''_{\infty}}{\Omega B \tan^2 \alpha} \right) \frac{\Omega \phi_1}{B w_{\parallel}} \rm .
\end{align}
Moreover, we recall that $\mu_{\infty, 1} \simeq \mu_{\infty, 1,0} = 0$ from (\ref{mu10}).

Following the prescription (\ref{n-ordern}) and exploiting $n_{\text{i}, 1/2} = n_{\text{i}, 3/2} = 0$ from (\ref{F-nozero}) and (\ref{d2F-nozero}), the second order correction $n_{\text{i}, 2,0}$ to the ion density is obtained by subtracting equation (\ref{ni-infty}) for $n_{\text{i},\infty}$, and equation (\ref{ni-1st}) for $n_{\text{i},1}$ from (\ref{nisplit-def}) with $F \simeq F_{\infty} + F_{1,0} + F_{2,0}$ and $F_{\rm slow} = F_{\infty} + \Delta F_{\rm slow}$,
\begin{align} \label{ni-2-app-1}
n_{\text{i},2, 0}   = & 2\pi  \int_0^{\infty} \Omega d\mu  \left[  \int_{w_{\rm cut}}^{\infty} dw_{\parallel} F_{2,0} + \int_{w_{\text{min}}}^{w_{\rm cut}} d w_{\parallel} \Delta F_{\text{slow}} \right. \nonumber \\
& \left.  - \int_0^{w_{\rm cut}} d w_{\parallel} F_{1,0} - \int_0^{w_{\text{min}}} d w_{\parallel} F_{\infty} \right] \rm .
\end{align}
The first term in (\ref{ni-2-app-1}) is effectively equation (\ref{ni-2-0-final}) with the lower limit of integration in $v_{\parallel}$ raised by $w_{\rm cut}$.
The second, third and fourth terms are exactly as in (\ref{ni32-pre}), but are now order $O(\hat \phi^2 n_{\infty})$. 
Similarly to (\ref{Fslow-1}), we exploit (\ref{ggg}) and (\ref{hhh}) to write the slow ion distribution function defined in the interval $w_{\parallel} \in [0, w_{\rm cut}]$ as
\begin{align} \label{Fslow-3/2}
\Delta F_{\text{slow}} = & \frac{1}{2} w_{\parallel, \infty, 1/2} w_{\parallel}^2 \partial_{v_{\parallel}}^3 F_{\infty} \rvert_{w_{\parallel} = 0} + \frac{1}{2} w_{\parallel, \infty, 1/2}^2 w_{\parallel} \partial_{v_{\parallel}}^3 F_{\infty} \rvert_{w_{\parallel} = 0} + \frac{1}{6} w_{\parallel, \infty, 1/2}^3 \partial_{v_{\parallel}}^3 F_{\infty} \rvert_{w_{\parallel} = 0} \nonumber \\ & + O(\hat \phi \hat \phi_{\rm cut} n_{\infty} v_{\rm t,i}^{-3}) \rm .
\end{align}
In the interval $[0, w_{\rm cut}]$, we may write $F_{1,0}$ as
\begin{align} \label{F10-2}
F_{1,0} = \frac{1}{2} w_{\parallel, \infty, 1,0} w_{\parallel}^2 \partial_{v_{\parallel}}^3 F_{\infty} \rvert_{w_{\parallel}=0} + O(\hat \phi \hat \phi_{\rm cut}^{3/2} n_{\infty} v_{\rm t,i}^{-3}) \rm .
\end{align}
In the interval $[0, w_{\text{min}}]$, we have
\begin{align} \label{Finfty-2}
F_{\infty} = \frac{1}{6} w_{\parallel}^3 \partial_{v_{\parallel}}^3 F_{\infty} \rvert_{w_{\parallel}=0} + O(\hat \phi^2 n_{\infty} v_{\rm t,i}^{-3}) \rm .
\end{align}
Inserting (\ref{F20}) for $F_{2,0}$ into (\ref{ni-2-app-1}) results in the same terms in (\ref{ni20-prelim}) (note that (\ref{ni-2-0-final}) assumes (\ref{d3F-nozero}), not assumed here) but with the lower limit of integration in $v_{\parallel}$ increased by $w_{\rm cut}$.
Inserting also (\ref{Fslow-3/2}), (\ref{F10-2}) and (\ref{Finfty-2}) into (\ref{ni-2-app-1}), and re-expressing the terms coming from $F_{2,0}$ (equation (\ref{ni20-prelim})) using (\ref{change-global}) and (\ref{wmin0}), we obtain
\begin{align} \label{ni-2-0-app}
& n_{\text{i},2, 0}(x,y) =  2\pi \int_0^{\infty} \Omega d\mu \left\lbrace \int_{w_{\rm cut} }^{\infty} dw_{\parallel} \frac{ w_{\text{min},1/2,0}^4 }{ 8w_{\parallel}^3}   (\partial_{w_{\parallel}} F_\infty)_{Y_{\star}} \right.  \nonumber \\
& \left. +  \frac{\Omega^2 \phi_1^2 (x,y)}{2B^2} \int_{ w_{\rm cut} }^{\infty} dw_{\parallel} \partial_{v_{\parallel}} \left( \frac{ (\partial_{w_{\parallel}} F_\infty)_{Y_{\star}} \left(1+\frac{\phi''_{\infty}(y)}{\Omega B \tan^2 \alpha} \right) }{ w_{\parallel}^2}  \right)   \right. \nonumber \\
&  +  \left. \frac{\Omega^2 \phi_1^2 (x,y)}{2B^2} \frac{\partial_{y}}{\Omega \tan \alpha} \left( \int_{ w_{\rm cut} }^{\infty} \frac{dw_{\parallel}}{ w_{\parallel}^2} (\partial_{w_{\parallel}} F_\infty)_{Y_{\star}} \left(1+\frac{\phi''_{\infty}(y)}{\Omega B \tan^2 \alpha} \right) \right)  \right. \nonumber \\
& \left.  + \partial_{v_{\parallel}}^3 F_{\infty} \rvert_{w_{\parallel}=0} \left[ \int_{w_{\text{min},1/2,0}}^{w_{\rm cut}} d w_{\parallel} \left(  \frac{1}{2} w_{\parallel, \infty, 1/2, 0} w_{\parallel}^2  + \frac{1}{2} w_{\parallel, \infty, 1/2, 0}^2 w_{\parallel} + \frac{1}{6} w_{\parallel, \infty, 1/2, 0}^3  \right)  \right. \right. \nonumber \\
& \left. \left. - \int_{0}^{w_{\rm cut}} d w_{\parallel} \frac{1}{2} w_{\parallel, \infty, 1, 0} w_{\parallel}^2 - \int_{0}^{w_{\text{min},1/2,0}} d w_{\parallel} \frac{1}{6} w_{\parallel}^3   \right] \right\rbrace \rm .
\end{align}
We have changed integration limit from $w_{\rm min}$ to $w_{\rm min, 1/2,0}$ making errors small in $O(\hat \phi^2 \epsilon^2 n_{\infty})$ in the respective integrals.
In writing (\ref{ni-2-0-app}) we have already neglected error terms of order $O( \hat \phi \hat \phi_{\rm cut}^{3/2}n_{\infty})$ coming from integrating the error terms in (\ref{Fslow-3/2}) over the interval $w_{\parallel} \in [w_{\rm min}, w_{\rm cut}]$. 
The first term in (\ref{ni-2-0-app}) is integrated by parts three times in $w_\parallel$ to get
\begin{align}
\int_{w_{\rm cut} }^{\infty} dw_{\parallel} \frac{ w_{\text{min},1/2,0}^4 }{ 8w_{\parallel}^3}   (\partial_{w_{\parallel}} F_\infty)_{Y_{\star}} = - \int_{w_{\rm cut} }^{\infty} dw_{\parallel} \frac{1}{16} w_{\text{min},1/2,0}^4 \ln w_{\parallel} \partial_{v_{\parallel}}^3 (\partial_{w_{\parallel}} F_\infty)_{Y_{\star}} \nonumber \\
+ \frac{3}{32} w_{\text{min},1/2,0}^4   \partial_{v_{\parallel}}^2 (\partial_{w_{\parallel}} F_\infty)_{Y_{\star}} \rvert_{w_{\parallel} = 0} - \frac{1}{16} w_{\text{min},1/2,0}^4 \ln w_{\rm cut} \partial_{v_{\parallel}}^2 (\partial_{w_{\parallel}} F_\infty)_{Y_{\star}} \rvert_{w_{\parallel} = 0}  \rm .
\end{align}
The boundary terms in the first two integration by parts have been evaluated using
\begin{align} \label{limit1}
\left. \frac{(\partial_{w_{\parallel}} F_\infty)_{Y_{\star}}}{w_{\parallel}^2} \right\rvert_{w_{\parallel} = w_{\rm cut}} = \frac{1}{2} \partial_{v_{\parallel}}^2 (\partial_{w_{\parallel}} F_\infty)_{Y_{\star}} \rvert_{w_{\parallel} = 0} + O(\hat \phi_{\rm cut}^{1/2} n_{\infty} v_{\rm t,i}^{-6}) \rm ,
\end{align}
\begin{align}
\left. \frac{\partial_{v_{\parallel}} (\partial_{w_{\parallel}} F_\infty)_{Y_{\star}}}{w_{\parallel}} \right\rvert_{w_{\parallel} = w_{\rm cut}} =  \partial_{v_{\parallel}}^2 (\partial_{w_{\parallel}} F_\infty)_{Y_{\star}} \rvert_{w_{\parallel} = 0} + O(\hat \phi_{\rm cut}^{1/2} n_{\infty} v_{\rm t,i}^{-6}) \rm ,
\end{align}
while in the third integration by parts $\partial_{v_{\parallel}}^2 (\partial_{w_{\parallel}} F_\infty)_{Y_{\star}} \rvert_{w_{\parallel} = w_{\rm cut}} = \partial_{v_{\parallel}}^2 (\partial_{w_{\parallel}} F_\infty)_{Y_{\star}} \rvert_{w_{\parallel} = 0} +  O(\hat \phi_{\rm cut}^{1/2} n_{\infty} v_{\rm t,i}^{-6})$ was used.
From (\ref{limit1}), the integral in the second term is
\begin{align}
\int_{ w_{\rm cut} }^{\infty} dw_{\parallel} \partial_{v_{\parallel}} \left( \frac{ (\partial_{w_{\parallel}} F_\infty)_{Y_{\star}}  }{ w_{\parallel}^2}  \right)  = - \frac{1}{2} \partial_{v_{\parallel}}^2 (\partial_{w_{\parallel}} F_\infty)_{Y_{\star}} \rvert_{w_{\parallel} = 0} + O(\hat \phi_{\rm cut}^{1/2} n_{\infty} v_{\rm t,i}^{-6}) \rm .
\end{align}
The third term in (\ref{ni-2-0-app}) has no divergence because (\ref{limit1}) is finite, and so can be left as it is.
Using (\ref{wparhalf0}), the three integrals in (\ref{ni-2-0-app}) that come from the slow ion contribution to the density are combined to
\begin{align} \label{A3}
& \int_{w_{\text{min},1/2,0}}^{w_{\rm cut}} d w_{\parallel} \left(  3 w_{\parallel, \infty, 1/2, 0} w_{\parallel}^2  +  3 w_{\parallel, \infty, 1/2, 0}^2 w_{\parallel} +  w_{\parallel, \infty, 1/2, 0}^3  \right) \nonumber \\
& =  \int_{w_{\text{min},1/2,0}}^{w_{\rm cut}} d w_{\parallel} \left[ \left( w_{\parallel}^2 - w_{\text{min},1/2,0}^2 \right)^{3/2} - w_{\parallel}^3  \right] \nonumber \\
& =  - \frac{3}{4} w_{\rm cut}^2 w_{\rm min, 1/2, 0}^2  + \frac{17}{32} w_{\rm min, 1/2, 0}^4 + \frac{3}{8} w_{\rm min, 1/2, 0}^4 \ln \left(\frac{2w_{\rm cut}}{w_{\rm min, 1/2, 0}} \right) + O\left( w_{\rm min}^4 \frac{w_{\rm min}^2}{w_{\rm cut}^2}  \right)\rm .
\end{align}
Note that the result
\begin{align}
& \int_{w_{\rm min, 1/2, 0}}^{w_{\rm cut}} d w_{\parallel} \left( w_{\parallel}^2 - w_{\text{min},1/2,0}^2 \right)^{3/2} = \frac{3}{8} \ln \left( \frac{w_{\rm cut} + \sqrt{ w_{\rm cut}^2 - w_{\rm min, 1/2, 0}^2}}{w_{\rm min, 1/2, 0}} \right) \nonumber \\
& + \left( w_{\rm cut}^2 - w_{\rm min, 1/2, 0}^2 \right)^{1/2} \left( \frac{1}{4} w_{\rm cut}^3 - \frac{5}{8} w_{\rm min, 1/2,, 0}^2 w_{\rm cut}  \right)  \nonumber \\
= & \frac{1}{4} w_{\rm cut}^4 - \frac{3}{4} w_{\rm cut}^2 w_{\rm min, 1/2, 0}^2  + \frac{9}{32} w_{\rm min, 1/2, 0}^4 + \frac{3}{8} w_{\rm min, 1/2, 0}^4 \ln \left(\frac{2w_{\rm cut}}{w_{\rm min, 1/2, 0}} \right) \nonumber \\
&  + O\left( w_{\rm min}^4 \frac{w_{\rm min}^2}{w_{\rm cut}^2}  \right)
\end{align}
was used in (\ref{A3}).
The integral coming from the piece $F_{1,0}$ of the distribution function is
\begin{align}
\int_{0}^{w_{\rm cut}} d w_{\parallel} w_{\parallel, \infty, 1, 0} w_{\parallel}^2 = - \int_{0}^{w_{\rm cut}} d w_{\parallel} w_{\parallel} \frac{w_{\text{min}, 1/2, 0}^2}{2} = - \frac{1}{4} w_{\text{min},1/2, 0}^2 w_{\text{cut}}^2 \rm .
\end{align}
Finally, the integral coming from the piece $F_{\infty}$ of the distribution function is
\begin{align}
\int_{0}^{w_{\text{min},1/2,0}} d w_{\parallel} w_{\parallel}^3  = \frac{1}{4} w_{\text{min},1/2, 0}^4  \rm .
\end{align}
Inserting these results into (\ref{ni-2-0-app}), the terms containing $w_{\rm cut}$ cancel leaving
\begin{align} \label{ni-2-0-app-2}
& n_{\text{i},2, 0}(x,y) = 2\pi \int_0^{\infty} \Omega d\mu \left[  - \int_{0}^{\infty} dw_{\parallel} \frac{1}{16} w_{\text{min},1/2,0}^4 \ln w_{\parallel} \partial_{v_{\parallel}}^3 (\partial_{w_{\parallel}} F_\infty)_{Y_{\star}} \right. \nonumber \\
& \left. - \frac{\Omega^2 \phi_1^2 (x,y)}{2B^2} \frac{\partial_{y}}{\Omega \tan \alpha} \left( \int_{0}^{\infty} dw_{\parallel} \ln w_{\parallel} \partial_{v_{\parallel}}^2 (\partial_{w_{\parallel}} F_\infty)_{Y_{\star}} \left(1+\frac{\phi''_{\infty}(y)}{\Omega B \tan^2 \alpha} \right)  \right) \right. \nonumber \\
& \left. + \frac{5}{64} w_{\text{min},1/2,0}^4   \partial_{v_{\parallel}}^3 F_{\infty}  \rvert_{w_{\parallel} = 0} + \frac{1}{16} w_{\text{min},1/2,0}^4 \ln \left( \frac{2}{w_{\text{min},1/2,0}} \right)  \partial_{v_{\parallel}}^3 F_{\infty}  \rvert_{w_{\parallel} = 0}  \right] \rm.
\end{align}
To obtain (\ref{ni-2-0-app-2}), we have neglected additional error terms of order $O(\hat \phi^{3} \hat \phi_{\rm cut}^{-1} n_{\infty})$ coming from (\ref{A3}); error terms of the same order also come from the third order correction to the bulk ion distribution function $F_3 \sim c_{\rm S}^6 \hat \phi^3 w_{\parallel}^{-5} \partial_{v_{\parallel}} F_{\infty}$, as seen by integrating by parts twice and using (\ref{dF-nozero}) and (\ref{d2F-nozero}).
Poisson's equation finally takes the form
\begin{align} \label{Poisson-secondorder-app}
& - \partial_x^2  \phi_1 (x,y) \frac{ \Omega}{B} \cos^2 \alpha \left[ \left( \frac{\lambda_{\rm D}^2}{\rho_{\rm B}^2 \cos^2 \alpha} + 1 \right) n_{\rm e, \infty}\right. \nonumber \\
& \left. + 2\pi Z \int_0^{\infty} \Omega d\mu \int_{ - \frac{\phi'_{\infty}(y)}{B\tan \alpha}  }^{\infty} d v_{\parallel}  \frac{\Omega \mu }{ v_{\parallel} +\frac{\phi'_{\infty}(y)}{B\tan \alpha}  } \left( \partial_{v_{\parallel}} F_\infty + \frac{ \partial_Y F_\infty}{\Omega \tan \alpha } \right) \right] \nonumber \\
= &  \left[  - \frac{B^2}{\Omega^2} \frac{d^2n_{\rm e, \infty}}{d\phi_{\infty}^2}  +  2\pi \int_0^{\infty} \Omega d\mu \left[ - \frac{1}{2} \int_{0}^{\infty} dw_{\parallel} \left( 1 + \frac{\phi''_{\infty}}{\Omega B \tan^2 \alpha} \right)^2 \ln w_{\parallel} \partial_{v_{\parallel}}^3 (\partial_{w_{\parallel}} F_\infty)_{Y_{\star}} \right. \right. \nonumber \\
& \left. \left. - \frac{\partial_{y}}{\Omega \tan \alpha} \left( \int_{0}^{\infty} dw_{\parallel} \ln w_{\parallel} \partial_{v_{\parallel}}^2 (\partial_{w_{\parallel}} F_\infty)_{Y_{\star}} \left(1+\frac{\phi''_{\infty}(y)}{\Omega B \tan^2 \alpha} \right)  \right) \right. \right. \nonumber \\
& \left. \left.  + \frac{1}{2} \left( 1 + \frac{\phi''_{\infty}}{\Omega B \tan^2 \alpha} \right)^2 \partial_{v_{\parallel}}^3 F_{\infty}  \rvert_{w_{\parallel} = 0} \ln \left( \frac{2e^{5/4}}{w_{\text{min},1/2,0}} \right) \right] \right] \frac{\Omega^2 \phi_1^2 (x,y)}{2B^2}  \rm .
\end{align}
The positive term $\sim \ln (2e^{5/4}/w_{\rm min, 1/2, 0}) \sim \ln (1/\hat \phi )$ in the coefficient multiplying $\phi_1^2 $ makes the right hand side always positive, thus guaranteeing a monotonically decaying potential profile (provided the polarisation condition (\ref{polcond-3}) is satisfied).
Hence, no additional sheath condition emerges at second order if (\ref{d3F-nozero}) is not satisfied.

We remark that to our knowledge an equation of the form $-\partial_x^2 \phi_1 \sim A \phi_1^2 \ln (1/\phi_1)$ has never been obtained even in the more extensively studied case of the (unmagnetised, one-dimensional) Debye sheath entrance.
This is because the distribution function is usually Taylor expanded in energy, so that its third derivative with respect to velocity is assumed to be zero.
Therefore, the asymptotic expansion of Poisson's equation near the sheath entrance performed in this appendix is also a generalisation of the conventional one for the sheath of an unmagnetised plasma: one simply sets $\Phi_{\rm pol} = 0$ ($\lambda_{\rm D} / \rho_{\rm B} = 0$), $\phi_{\infty} = 0$, $w_{\parallel} = v_{\parallel} = v_x$, and removes all $y$-dependences.
This generalisation addresses criticisms to the kinetic Bohm criterion that question it on the grounds that its derivation assumes a restricted class of ion distribution functions \cite{Baalrud-Hegna-2012-reply}.

\section*{References}

\bibliography{gyrokineticsbibliography}{}

\begin{thebibliography}{10}

\bibitem{Tonks-1929}
L.~Tonks and I.~Langmuir.
\newblock A general theory of the plasma of an arc.
\newblock {\em Physical review}, 34(6):876, 1929.

\bibitem{Bohm-1949}
D.~Bohm.
\newblock {\em The Characteristics of Electrical Discharges in Magnetic
  Fields}, page~77, 1949.

\bibitem{Loizu-2012}
J.~Loizu, P.~Ricci, F.~D. Halpern, and S.~Jolliet.
\newblock Boundary conditions for plasma fluid models at the magnetic presheath
  entrance.
\newblock {\em Physics of Plasmas (1994-present)}, 19(12):122307, 2012.

\bibitem{Krashenninikov-2017}
S.~I. Krasheninnikov and A.~S. Kukushkin.
\newblock Physics of ultimate detachment of a tokamak divertor plasma.
\newblock {\em Journal of Plasma Physics}, 83(5):155830501, 2017.

\bibitem{Shi-2017}
E.~L. Shi, G.~W. Hammett, T.~Stoltzfus-Dueck, and A.~Hakim.
\newblock Gyrokinetic continuum simulation of turbulence in a straight
  open-field-line plasma.
\newblock {\em Journal of Plasma Physics}, 83(3), 2017.

\bibitem{Bender-Orszag}
C.~M. Bender and S.~A. Orszag.
\newblock {\em Advanced mathematical methods for scientists and engineers I:
  Asymptotic methods and perturbation theory}.
\newblock Springer Science \& Business Media, 1999.

\bibitem{Riemann-review}
K.-U. Riemann.
\newblock The {B}ohm criterion and sheath formation.
\newblock {\em Journal of Physics D: Applied Physics}, 24(4):493, 1991.

\bibitem{Chodura-1982}
R.~Chodura.
\newblock Plasma--wall transition in an oblique magnetic field.
\newblock {\em Physics of Fluids (1958-1988)}, 25(9):1628--1633, 1982.

\bibitem{Tskhakaya-2002b}
D.~Tskhakaya, S.~Kuhn, V.~Petr{\v{z}}ilka, and R.~Khanal.
\newblock Effects of energetic electrons on magnetized electrostatic plasma
  sheaths.
\newblock {\em Physics of Plasmas}, 9(6):2486--2496, 2002.

\bibitem{Riemann-1994}
K.-U. Riemann.
\newblock Theory of the collisional presheath in an oblique magnetic field.
\newblock {\em Physics of Plasmas (1994-present)}, 1(3):552--558, 1994.

\bibitem{Kim-1995}
G.-H. Kim, N.~Hershkowitz, D.~A. Diebold, and M.-H. Cho.
\newblock Magnetic and collisional effects on presheaths.
\newblock {\em Physics of Plasmas}, 2(8):3222--3233, 1995.

\bibitem{Tskhakaya-2000}
D.~Tskhakaya and S.~Kuhn.
\newblock Influence of initial energy on the effective secondary-electron
  emission coefficient in the presence of an oblique magnetic field.
\newblock {\em Contributions to Plasma Physics}, 40(3-4):484--490, 2000.

\bibitem{Komm-2020}
M.~Komm, S.~Ratynskaia, P.~Tolias, and A.~Podolnik.
\newblock Space-charge limited thermionic sheaths in magnetized fusion plasmas.
\newblock {\em Nuclear Fusion}, 60(5):054002, 2020.

\bibitem{Ewart-2021}
R.~J. Ewart, F.~I. Parra, and A.~Geraldini.
\newblock Sheath collapse at critical shallow angle due to kinetic effects.
\newblock {\em Plasma Physics and Controlled Fusion}, 64(1):015010, 2021.

\bibitem{Castillo-2024}
A.~M. Castillo and K.~Hara.
\newblock Loss cone effects and monotonic sheath conditions of a partially
  magnetized plasma sheath.
\newblock {\em Physics of Plasmas}, 31(3), 2024.

\bibitem{Sato-1994}
K.~Sato, H.~Katayama, and F.~Miyawaki.
\newblock Effects of an oblique magnetic field on sheath formation in the
  presence of electron emission.
\newblock {\em Contributions to Plasma Physics}, 34(2-3):133--138, 1994.

\bibitem{Pan-2018}
Q.~Pan, D.~Told, E.~L. Shi, G.~W. Hammett, and F.~Jenko.
\newblock Full-f version of {GENE} for turbulence in open-field-line systems.
\newblock {\em Physics of Plasmas}, 25(6):062303, 2018.

\bibitem{Claassen-Gerhauser-1996b}
H.~A. Claa{\ss}en and H.~Gerhauser.
\newblock Generalized {B}ohm's criterion for thermal ions in oblique magnetic
  and electric fields.
\newblock {\em Contributions to Plasma Physics}, 36(2-3):361--365, 1996.

\bibitem{Cohen-Ryutov-2004-sheath-boundary-conditions}
R.~H. Cohen and D.~D. Ryutov.
\newblock Sheath physics and boundary conditions for edge plasmas.
\newblock {\em Contributions to Plasma Physics}, 44(1-3):111--125, 2004.

\bibitem{Geraldini-2017}
A.~Geraldini, F.~I. Parra, and F.~Militello.
\newblock Gyrokinetic treatment of a grazing angle magnetic presheath.
\newblock {\em Plasma Physics and Controlled Fusion}, 59(2):025015, 2017.

\bibitem{Carralero-2015}
D.~Carralero, P.~Manz, L.~Aho-Mantila, G.~Birkenmeier, M.~Brix, M.~Groth, H.~W.
  M{\"u}ller, U.~Stroth, N.~Vianello, E.~Wolfrum, et~al.
\newblock Experimental validation of a filament transport model in turbulent
  magnetized plasmas.
\newblock {\em Physical review letters}, 115(21):215002, 2015.

\bibitem{Mosetto-2015}
A.~Mosetto, F.~D. Halpern, S.~Jolliet, J.~Loizu, and P.~Ricci.
\newblock Finite ion temperature effects on scrape-off layer turbulence.
\newblock {\em Physics of Plasmas}, 22(1):012308, 2015.

\bibitem{Ricci-2012}
P.~Ricci, F.~D. Halpern, S.~Jolliet, J.~Loizu, A.~Mosetto, A.~Fasoli, I.~Furno,
  and C.~Theiler.
\newblock Simulation of plasma turbulence in scrape-off layer conditions: the
  {GBS} code, simulation results and code validation.
\newblock {\em Plasma Physics and Controlled Fusion}, 54(12):124047, 2012.

\bibitem{Harrison-Thompson-1959}
E.~R. Harrison and W.~B. Thompson.
\newblock The low pressure plane symmetric discharge.
\newblock {\em Proceedings of the Physical Society}, 74(2):145, 1959.

\bibitem{Geraldini-2018}
A.~Geraldini, F.~I. Parra, and F.~Militello.
\newblock Solution to a collisionless shallow-angle magnetic presheath with
  kinetic ions.
\newblock {\em Plasma Physics and Controlled Fusion}, 60(12):125002, 2018.

\bibitem{Hutchinson-1996}
I.~H. Hutchinson.
\newblock The magnetic presheath boundary condition with {E}$\times${B} drifts.
\newblock {\em Physics of Plasmas}, 3(1):6--7, 1996.

\bibitem{Siddiqui-Hershkowitz-2016}
M.~U. Siddiqui, D.~S. Thompson, C.~D. Jackson, J.~F. Kim, N.~Hershkowitz, and
  E.~E. Scime.
\newblock Models, assumptions, and experimental tests of flows near boundaries
  in magnetized plasmas.
\newblock {\em Physics of Plasmas (1994-present)}, 23(5):057101, 2016.

\bibitem{Caron-2021-exp}
D.~Caron and E.~E. Scime.
\newblock Influence of magnetic angle on the {E}$\times${B} drift in a magnetic
  presheath.
\newblock {\em Physics of Plasmas}, 28(8), 2021.

\bibitem{Parra-2008}
F.~I Parra and P.~J. Catto.
\newblock Limitations of gyrokinetics on transport time scales.
\newblock {\em Plasma Physics and Controlled Fusion}, 50(6):065014, 2008.

\bibitem{Baalrud-Hegna-2012-reply}
S.~D. Baalrud and C.~C. Hegna.
\newblock Reply to comment on ‘{K}inetic theory of the presheath and the
  {B}ohm criterion’.
\newblock {\em Plasma Sources Science and Technology}, 21(6):068002, 2012.

\end{thebibliography}
\bibliographystyle{unsrt}

\end{document}